\documentclass{elsart}
\usepackage{amssymb}
\usepackage{color}
\usepackage{epsfig}
\usepackage{palatino}
\usepackage{graphicx}  
\usepackage{subfigure}
\usepackage[greek,english]{babel}
\def\ifm#1{\relax\ifmmode#1\else$#1$\fi}  
    \def\x{\ifm{\times}}
     
  \def\to{\ifm{\rightarrow}}

\let\cal=\mathcal   \def\ORD#1!{\ifm{{\cal O}\hbox{(#1)}}}
\def\pt#1,#2,{\ifm{#1\x10^{#2}}} 
\def\blankline{\vskip 12 pt\noindent}
\newcommand{\bea}{\begin{eqnarray}}
\newcommand{\eea}{\end{eqnarray}}
\newcommand{\be}{\begin{equation}}
\newcommand{\eee}{\end{equation}}
\newcommand{\sig}{\sigma}
\newcommand{\nn}{\nonumber}
\newcommand{\veps}{\varepsilon}

\begin{document}

\begin{frontmatter}

\title{\bf Prospects for $e^+e^-$ Physics at Frascati between the $\phi$ and the $\psi$ 
}

\author{F.Ambrosino$^{e}$}
\author{F.Anulli$^{c}$}
\author{D.Babusci$^{c}$}
\author{S.Bianco$^{c}$} 
\author{C.Bini$^{i,*}$} 
\author{N.Brambilla$^{d}$} 
\author{R.De Sangro$^{c}$} 
\author{P.Gauzzi$^{i}$} 
\author{P.M.Gensini$^{h}$}
\author{S.Giovannella$^{c}$}
\author{V.Muccifora$^{c}$}
\author{M.Negrini$^{b}$} 
\author{F.Nguyen$^{l}$} 
\author{S.Pacetti$^{c}$}
\author{G.Pancheri$^{c}$} 
\author{M.Passera$^{f}$} 
\author{A.Passeri$^{l}$} 
\author{A.D.Polosa$^{i}$}
\author{M.Radici$^{g}$}
\author{Y.N.Srivastava$^{h}$}
\author{A.Vairo$^{d}$} 
\author{G.Venanzoni$^{c}$}
\author{G.Violini$^{a}$}

\address{$^{a}$ Universit\'a della Calabria, Cosenza and INFN Gruppo
  collegato di Cosenza}
\address{$^{b}$ Universit\'a di Ferrara and INFN Sezione di Ferrara}
\address{$^{c}$ Laboratori Nazionali di Frascati, INFN}
\address{$^{d}$ Universit\'a di Milano and INFN Sezione di Milano}
\address{$^{e}$ Universit\'a ``Federico II'' di Napoli and INFN Sezione di Napoli}
\address{$^{f}$ Universit\'a di Padova and INFN Sezione di Padova}
\address{$^{g}$ Dip. di Fisica Nucleare e Teorica, Universit\'a di Pavia and INFN Sezione di Pavia}
\address{$^{h}$ Dip. di Fisica dell'Universit\`a di Perugia 
and I.N.F.N., Sezione di Perugia, Italy}
\address{$^{i}$ Universit\'a ``La Sapienza'' e INFN Sezione di Roma}
\address{$^{l}$ Universit\'a Roma3 and INFN Sezione di Roma3}

\begin{abstract}We present a detailed study, done in the framework of
  the INFN 2006 Roadmap, of the prospects for $e^+e^-$ physics at the
  Frascati National Laboratories. The physics case for an $e^+e^-$ collider
  running at high luminosity at the $\phi$ resonance energy and also reaching
  a maximum center of mass energy of 2.5 GeV is discussed, together with the
  specific aspects of a very high luminosity $\tau$-charm factory. Subjects
  connected to Kaon decay physics are not discussed here, being part of
  another INFN Roadmap working group. The significance
  of the project and the impact on INFN are also discussed.
\par\noindent All the documentation related to the activities of the working group can be
  found in {\it http://www.roma1.infn.it/people/bini/roadmap.html}.
\end{abstract}
\end{frontmatter}
{\small $^{*}$ corresponding author, cesare.bini@roma1.infn.it,
    tel.+390649914266}
\pagebreak
\tableofcontents
\section{Introduction}
\label{Intro}
The Frascati National Laboratories (LNF) of INFN have a long and successful 
tradition in $e^+e^-$ accelerators
and in $e^+e^-$ physics. The concept of $e^+e^-$ colliders with the two beams
circulating in the same vacuum chamber in opposite directions was developed
at Frascati by Bruno Touscheck and his collaborators in the early sixties 
with the project AdA. 

In the seventies, the new accelerator Adone worked 
in the center of mass energy region
between 1.5 and 3.1 GeV. The Adone experiments contributed 
to the discovery of the unexpectedly rich multihadronic
production and in 1974 
confirmed the discovery of the $J/\psi$. 
Some of the hadronic cross section measurements done at Adone
are still today the best results in that energy region, and are used
in precision tests of the Standard Model.
     
The tradition continued with {\small DAFNE} that provided 
the first beam collisions in 1999 and that is still working at present. 
{\small DAFNE} is a $\phi$-factory, that is an $e^+e^-$ machine centred at the 
$\phi$ resonance energy $\sqrt{s}=1019.4$ MeV. It has recently reached a
peak luminosity of 1.5 $\times 10^{32}$cm$^{-2}$s$^{-1}$ that is the
highest luminosity ever reached by an 
$e^+e^-$ collider in this energy region.
Many features of this machine are unique. First of all the production 
of $K^0\overline{K^0}$ final states in a pure quantum 
state with the consequent possibility to study quantum interference effects,
and to have pure monochromatic tagged 
$K_S$ and $K_L$ beams. 
In addition a $\phi$-factory is also a source of 
high statistic samples of pseudoscalar and scalar
mesons obtained through the $\phi$ radiative decays, and of
monochromatic charged kaons directly from the $\phi$. These samples 
allow one to obtain relevant results in
hadronic physics, and also in atomic and nuclear physics (the study of 
hypernuclei and exotic atoms). For a comprehensive review of the aspects of 
physics that concern
{\small DAFNE}, we refer to the {\small DAFNE} Physics Handbook that was written in
1995\cite{DPH} before the start-up of the experiments. For the status and the
physics results of the experiments, we refer the reader
to the Web sites of each experiment, namely KLOE
\cite{KLOE}, Finuda \cite{Finuda} and Dear \cite{Dear}.   

The {\small DAFNE} schedule is defined up to the end of year 2008. In the 
last few years a
discussion has started at LNF about  
the $e^+e^-$ future program for the laboratory. 
Two main options have been considered up to now, not necessarily mutually
exclusive. 

The first option, that we call {\small DAFNE}-2
in the following, corresponds to continue a low energy $e^+e^-$
program with a new version of {\small DAFNE} of higher luminosity,
and also by allowing the center of mass energy to span from 
the $\phi$ resonance
energy up to $\sqrt{s}$=2.5 GeV. 
The new machine can be built within the same {\small DAFNE}/Adone building 
and the Frascati Accelerator Division is studying now which are the possible
machine schemes to obtain the required performance. A first
project has been already developed
\cite{notemacchina}.

The second option, that we call the Flavour-Factory, is more ambitious
\cite{ConvegnoGiorgi}. The
idea is to profit from the experience developed by accelerator physicists 
with the linear collider studies, to
build an $e^+e^-$ machine of completely new concept, able to run around 10
GeV center of mass energy (Super B-factory) but also in the 3-4 GeV region
(as a $\tau$-charm factory). This project doesn't fit the present
laboratories size, so that it requires a new site, and a very big financial
and technical effort.  

In this document, that is part of ``Gruppo-1'' section of the INFN Roadmap, we consider in
full detail the physics case for {\small DAFNE}-2 (in Sect.\ref{DAFNE-2} below), 
describing the main particle physics 
issues and showing the possible reach of the project. We do not
discuss the physics of Kaon decays and Kaon interferometry,
since it will be extensively discussed in the document of the
kaon physics working group\cite{morandin}. Moreover we
don't discuss the Super-B factory program, that also belongs to another
working group. Nevertheless Sect.\ref{Tauc} will be devoted to the
presentation of the main physics topics of the $\tau$-charm factory that
could be part of the Flavour-Factory program. Considerations about detector
issues for the case of {\small DAFNE}-2,  
are presented in Sect.\ref{DetAcc}. Finally in
Sect.\ref{Concl} we summarise the relevance of the programs here
outlined. 
\section{The physics case for {\small DAFNE}-2}
\label{DAFNE-2}
\subsection{\bf Overview}
\label{Over}
{\small DAFNE}-2 is planned to be optimised in luminosity at the $\phi$ peak, reaching 
a luminosity of $\sim$8$\times$10$^{32}$cm$^{-2}$s$^{-1}$. In the higher energy
region between 1 and 2.5 GeV is expected to reach a luminosity of
$\sim$10$^{32}$cm$^{-2}$s$^{-1}$, much larger than any previous machine in the
same energy region.  With such a machine one can think to collect an integrated
luminosity of $\sim 50$ fb$^{-1}$ in few years of data taking at the $\phi$
and $\sim 5$ fb$^{-1}$ in the same
running time between 1 and 2.5 GeV. With
respect to {\small DAFNE}, it corresponds to increase by a factor 20 the 
statistics at the $\phi$ and to open a new window on high statistics $e^+e^-$
physics in the 1 - 2.5 GeV energy region.
The only direct competitor project is VEPP-2000 at Novosibirsk \cite{VEPPtot} 
that will cover
the center of mass energy between 1 and 2 GeV with two experiments. This
project is expected to start by year 2007 with a luminosity ranging between 
$10^{31}$cm$^{-2}$s$^{-1}$ at 1 GeV and $10^{32}$cm$^{-2}$s$^{-1}$ at 2 GeV. 
Other ``indirect'' competitors are the higher
energy $e^+e^-$ colliders ($\tau$-charm and B-factories) that in principle
cover the {\small DAFNE}-2 energy range by means of radiative return. Moreover for some
specific issues
experiments at hadron machines are also competitive. A list of the
competitors is reported in Sect.\ref{Concl}.

In the following sections we present the main physics issues of the {\small DAFNE}-2
project. We start with the possibility to improve the knowledge of the
$e^+e^-$ to hadrons cross sections in a very wide center of mass 
energy region, from the
$\pi\pi$ threshold up to 2.5 GeV, and its implications on the precision
tests of the Standard Model ~(\ref{Hadro}), and on the vector meson
spectroscopy ~(\ref{Vecto}). Then we describe the
physics potential of studying radiative decays ~(\ref{Radia}) and
$\gamma\gamma$ collisions ~(\ref{Gamma}). Finally we discuss also the subjects
of the hadron form factors ~(\ref{Bario}) and of the kaon-nucleus interactions
~(\ref{Kaons}).   
\subsection{\bf Hadronic cross section}
\label{Hadro}
%%%%%%%%%%%%%%%%%%%%%%%%%%%%%%%%%%%%%%%%%%%%%%%%%%%%%%%%%%%%%%%%%%%%%%%%%%%%%
\newcommand{ \mysmall}[1]{\scriptscriptstyle #1} % a smaller #
\newcommand{ \mw}{M_{\mysmall{W}}}
\newcommand{ \mz}{M_{\mysmall{Z}}}
\newcommand{ \mh}{M_{\mysmall{H}}}

%---------------------------------------------------------------------------%
%
\subsubsection{Precision 
tests of the Standard Model: overview}
\label{subsubsec:SMTESTS}

The systematic comparison of the Standard Model ({\small SM}) predictions
with very precise experimental data served in the last few decades as an
invaluable tool to test the theory at the quantum level. It has also
provided stringent constraints on many ``new physics'' scenarios.  The (so
far) remarkable agreement between the precise measurements of the
electroweak observables and their {\small SM} predictions is a striking
experimental confirmation of the theory, even if there are a few observables
where the agreement is not so satisfactory.  On
the other hand, the Higgs boson has not yet been observed, and there are
strong theoretical arguments hinting at the presence of physics beyond the
{\small SM}. Future colliders, like the upcoming {\small
LHC} or an $e^+ e^-$ International Linear Collider ({\small ILC}), will
hopefully answer many such questions, offering at the same time great
physics potential and a new challenge to provide even more precise
theoretical predictions.

Precise {\small SM} predictions require precise input parameters. Among the
three basic input parameters of the electroweak ({\small EW}) sector of the
{\small SM} -- the fine-structure constant $\alpha$, the Fermi coupling
constant $G_F$ and the mass of the $Z$ boson -- $\alpha$ is by far the most
precisely known, determined mainly from the anomalous magnetic moment of the
electron with an amazing relative precision of 3.3 parts per billion
(ppb)~\cite{CODATA02}. However, physics at nonzero squared momentum transfer
$q^2$ is actually described by an effective electromagnetic coupling
$\alpha(q^2)$ rather than by the low-energy constant $\alpha$ itself. The
shift of the fine-structure constant from the Thomson limit to high energy
involves non-perturbative hadronic effects which spoil this fabulous
precision. Indeed, the present accuracies of these basic parameters
are~\cite{CODATA02,PDG04,Jegerlehner:2003rx}
\begin{eqnarray}
  \delta \alpha / \alpha \sim 3 \times 10^{-9}, &&
  \delta G_F / G_F \sim 9 \times 10^{-6},  \\
  \delta \mz / \mz \sim 2 \times 10^{-5},  &&
  \delta \alpha(\mz^2)/ \alpha(\mz^2) \sim O(10^{-4}).  
\end{eqnarray}
The relative uncertainty of $\alpha(\mz^2)$ is roughly one order of
magnitude worse than that of $\mz$, making it one of the limiting factors in
the calculation of precise {\small SM} predictions.

\begin{figure}[h]
\begin{center}
\includegraphics[width=11cm]{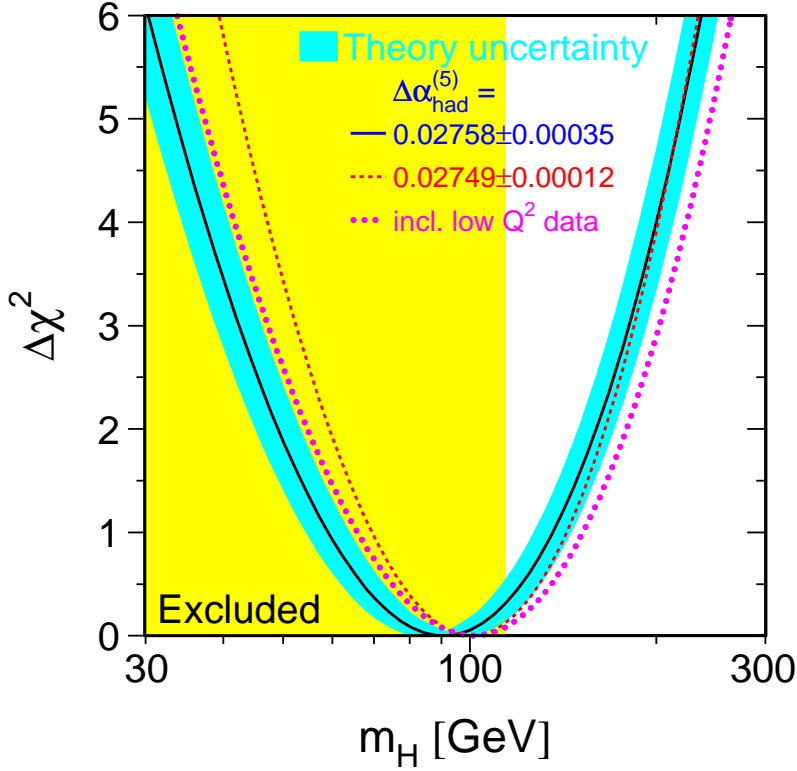}
\vspace{-0.5cm}
\caption{The line is the result of the Electroweak Working Group fit using
  all data~\cite{LEPEWWG}; the band represents an estimate of the
  theoretical error due to missing higher order corrections.  The vertical
  band shows the 95\% CL exclusion limit on $\mh$ from the direct search.}
\label{fig:blueband}
\end{center}
\end{figure}

The effective fine-structure constant at the scale $\mz$, $\alpha(\mz^2) =
\alpha/[1-\Delta \alpha(\mz^2)]$, plays a crucial role in basic {\small EW}
radiative corrections of the {\small SM}. An important example is the
{\small EW} mixing parameter $\sin^2 \!\theta$, related to $\alpha$, $G_F$ and
$\mz$ via the Sirlin relation~\cite{Sirlin}
\begin{equation}
  \sin^2 \!\theta_{\mysmall{S}} \cos^2 \!\theta_{\mysmall{S}} = 
  \frac{\pi \alpha}{\sqrt 2 G_F \mz^2 (1-\Delta r_{\mysmall{S}})},
\label{eq:sirlin}
\end{equation}
where the subscript $S$ identifies the renormalization scheme. $\Delta
r_{\mysmall{S}}$ incorporates the universal correction $\Delta
\alpha(\mz^2)$, large contributions that depend quadratically on the top
quark mass~\cite{Veltman} (which led to its indirect determination before
the discovery of this quark at the Tevatron~\cite{TopQuark}), plus all
remaining quantum effects.
In the {\small SM}, $\Delta r_{\mysmall{S}}$ depends on various physical
parameters such as $\alpha$, $G_F$, $\mz$, $\mw$, $\mh$, $m_f$, etc., where
$m_f$ stands for a generic fermion mass. As $\mh$, the mass of the Higgs
boson, is the only relevant unknown parameter in the {\small SM}, important
indirect bounds on this missing ingredient can be set by comparing the
calculated quantity in Eq.~(\ref{eq:sirlin}) with the experimental value of
$\sin^2 \!\theta_{\mysmall{S}}$. These constraints can be easily derived
using the simple formulae of Refs.~\cite{Formulette}, which relate the
effective {\small EW} mixing angle $\sin^2 \!\theta_{\rm eff}^{\rm lept}$
(measured at {\small LEP} and {\small SLC} from the on-resonance
asymmetries) with $\Delta\alpha(\mz^2)$ and other experimental inputs like
the mass of the top quark. It is important to note that the present error in
the effective electromagnetic coupling constant, $\delta \Delta\alpha(\mz^2)
= 35 \times 10^{-5}$~\cite{BP05}, dominates the uncertainty of the
theoretical prediction of $\sin^2 \!\theta_{\rm eff}^{\rm lept}$, inducing
an error $\delta(\sin^2 \!\theta_{\rm eff}^{\rm lept}) \sim 12 \times
10^{-5}$ which is not much smaller than the experimental value
$\delta(\sin^2 \!\theta_{\rm eff}^{\rm lept})^{\mysmall \rm EXP} = 16 \times
10^{-5}$ determined by {\small LEP-I} and {\small
SLD}~\cite{LEPEWWG}. 
Moreover, as measurements of the effective {\small EW}
mixing angle at a future linear collider may improve its precision by one
order of magnitude~\cite{Weiglein:2004hn}, a much smaller
value of $\delta\Delta\alpha(\mz^2)$ will be required (see next section). 
It is therefore crucial to
assess all viable options to further reduce this uncertainty. The latest
global fit of the {\small LEP} Electroweak Working Group, which employs the
complete set of {\small EW} observables, leads to the value $\mh =
91^{+45}_{-32}$GeV, with a 95\% confidence level upper limit of 186~GeV (see
Fig.~\ref{fig:blueband})~\cite{LEPEWWG}. This limit increases to 219 GeV
when including the {\small LEP-II} direct search lower limit of 114~GeV.

In the next few years the {\small LHC} may delight us with many discoveries,
and the Higgs boson may be close at hand. Once its mass will be known,
precision {\small EW} tests will provide excellent possibilities to
establish new physics contributions beyond the {\small SM}. A high-precision
{\small EW} program will be the natural complement to direct searches of new
particles and will help indicate the directions that future studies must
take. Eventually, if ``new physics'' will be directly uncovered at collider
facilities, precision measurements of its properties will guide our search
for even higher scale phenomena.

%---------------------------------------------------------------------------%
\subsubsection{The effective fine-structure constant at the scale 
\boldmath $M_Z$ \unboldmath}
\label{subsubsec:ALPHAEFF}

Let us examine the determination of the running of the effective
fine-structure constant to the scale $\mz$, that can be defined by
$ \Delta \alpha(\mz^2) = 4\pi\alpha \mbox{Re}[\Pi^{(f)}_{\gamma
  \gamma}(0)- \Pi^{(f)}_{\gamma \gamma}(\mz^2)],
$
where $\Pi^{(f)}_{\gamma\gamma}(q^2)$ is the fermionic part of the photon
vacuum polarisation function (with the top quark decoupled). Its evaluation
includes hadronic contributions where long-distance {\small QCD} dynamics
cannot be calculated analytically.  These contributions cause the
aforementioned dramatic loss of accuracy, by several orders of magnitude,
which occurs moving from the value of $\alpha$ at vanishing momentum
transfer to that at $q^2=\mz^2$. The shift $\Delta \alpha(\mz^2)$ can be
split in two parts: $\Delta\alpha(\mz^2) = \Delta\alpha_{\rm lep}(\mz^2) +
\Delta \alpha_{\rm had}^{(5)}(\mz^2)$. The leptonic contribution is
calculable in perturbation theory and known up to three-loops:
$\Delta\alpha_{\rm lep}(\mz^2) = 3149.7686\times 10^{-5}$~\cite{MS98}. The
hadronic contribution $\Delta \alpha_{\rm had}^{(5)}(\mz^2)$ of the five
light quarks ($u$, $d$, $s$, $c$, and $b$) can be computed from hadronic
$e^+ e^-$ annihilation data via the dispersion relation~\cite{CG61}
\begin{equation} 
  \Delta \alpha_{\rm had}^{(5)}(\mz^2) = -\left(\frac{\alpha \mz^2}{3\pi}
  \right) \mbox{Re}\int_{4m_\pi^2}^{\infty} ds \frac{R(s)}{s(s- \mz^2
  -i\epsilon)},
\label{eq:delta_alpha_had}
\end{equation}
where $R(s) = \sigma^{(0)}(s)/(4\pi\alpha^2\!/3s)$ and $\sigma^{(0)}\!(s)$
is the total cross section for $e^+ e^-$ annihilation into any hadronic
state, with extraneous {\small QED} corrections subtracted off.  In the
1990s, detailed evaluations of this dispersive integral have been carried
out by several authors~\cite{deltaalpha,ADH98}. More recently, some of these
analyses were updated to include new $e^+ e^-$ data -- mostly from {\small
CMD-2}~\cite{Akhmetshin:2003zn} and {\small BES}~\cite{Bai:2001ct} --
obtaining:
$
\Delta\alpha_{\rm had}^{(5)} = 2761 \, (36) \times 10^{-5}
$
\cite{BP01},
$
\Delta\alpha_{\rm had}^{(5)} = 2757 \, (36) \times 10^{-5}
$
\cite{JegerJPG29},
$
\Delta\alpha_{\rm had}^{(5)} = 2755 \, (23) \times 10^{-5}
$
\cite{Hagiwara:2003da}, and
$
\Delta\alpha_{\rm had}^{(5)} = 2749 \, (12) \times 10^{-5}
$
\cite{dTY04}. The reduced uncertainty of the latter result has been obtained
making stronger use of theoretical inputs.
The reduction, by a factor of two, of the uncertainty quoted
in the first article of ref.~\cite{deltaalpha} ($70 \times 10^{-5}$), with
respect to that in~\cite{JegerJPG29} ($36 \times 10^{-5}$), is mainly due to
the data of {\small BES}. The latest update,
$
\Delta\alpha_{\rm had}^{(5)} = 2758 \, (35) \times 10^{-5}
$
\cite{BP05}, includes also the measurements of {\small
KLOE}~\cite{Aloisio:2004bu}.  
Tab.~\ref{tab:future} (from Ref.~\cite{JegerJPG29}) shows that an
uncertainty $\delta \Delta\alpha_{\rm had}^{(5)} \sim 5 \times 10^{-5}$,
needed for precision physics at a future linear collider, requires the
measurement of the hadronic cross section with a precision of $O(1\%)$ from
threshold up to the $\Upsilon$ peak.
\begin{table}[h]
\begin{center}
 \renewcommand{\arraystretch}{1.4}
 \setlength{\tabcolsep}{1.6mm}
\begin{tabular}{|c|c|c|c|}
\hline
$\delta \Delta\alpha_{\rm had}^{(5)} \times 10^{5} $ 
& $\delta(\sin^2 \!\theta_{\rm eff}^{\rm lept}) \times 10^{5}$  
& Request on $R$\\
\hline \hline
35   &  12.5 & Present \\
\hline
7   &   2.5 & $\delta R/R \sim 1\%$ for $\sqrt{s} \leq M_{J/\psi}$\\
\hline   
5   &   1.8 & $\delta R/R \sim 1\%$ for $\sqrt{s} \leq M_{\Upsilon}$ \\
\hline   
\end{tabular}
\caption{\label{tab:future} Values of the uncertainties $\delta
\Delta\alpha_{\rm had}^{(5)}$ (first column) and the errors induced by these
uncertainties on the theoretical {\small SM} prediction for $\sin^2
\!\theta_{\rm eff}^{\rm lept}$ (second column). The third column indicates
the corresponding requirements on the $R$ measurement.}
\end{center}
\end{table}

%---------------------------------------------------------------------------%
\subsubsection{The muon \boldmath $g$$-$$2$ \unboldmath}
\label{subsubsec:GMINUS2}

During the last few years, in a sequence of increasingly precise
measurements, the {\small E821} Collaboration at Brookhaven has determined
$a_{\mu} = (g_{\mu}-2)/2$ with a fabulous relative precision of 0.5 parts
per million (ppm)~\cite{BNL,BNL04}, allowing us to test all sectors of
the {\small SM} and to scrutinise viable alternatives to this
theory~\cite{CM01}. The present world average experimental value 
is
$
    a_{\mu}^{\mbox{$\scriptscriptstyle{\rm EXP}$}}  = 
               116 \, 592 \, 080 \, (63) \times 10^{-11} 
               ~(0.5~\mbox{ppm})
$~\cite{BNL04}.
This impressive result is still limited by statistical errors.  A new
experiment, {\small E969}, has been approved (but not yet funded) at
Brookhaven in 2004~\cite{E969}. Its goal is to reduce the present
experimental uncertainty by a factor of 2.5 to about 0.2 ppm. A letter of
intent for an even more precise $g$$-$$2$ experiment was submitted to
{\small J-PARC} with the proposal to reach a precision below 0.1
ppm~\cite{JPARC}. But how precise is the theoretical prediction?

The {\small SM} prediction $a_{\mu}^{\mysmall \rm SM}$ is usually split into
three parts: {\small QED}, electroweak and hadronic
(see~\cite{Gminus2Reviews} for recent reviews).  The {\small QED}
contribution to $a_{\mu}$ arises from the subset of {\small SM} diagrams
containing only leptons ($e,\mu,\tau$) and photons. First computed by
Schwinger more than fifty years ago~\cite{Sch48}, it is now known up to
terms of order $(\alpha/\pi)^4$, and leading five-loop contributions have
been evaluated. The prediction currently stands at
$
    a_{\mu}^{\mysmall \rm QED} = 
    116 \, 584 \, 719.4 \, (1.4) \times 10^{-11}       
$~\cite{KinoshitaNio0512330},
where the error is due to the uncertainty of the $O(\alpha^4)$ and
$O(\alpha^5)$ terms, and to the uncertainty of $\alpha$.  The {\small EW}
contribution to $a_{\mu}$ is suppressed by a factor $(m_{\mu}/\mw)^2$ with
respect to the {\small QED} effects. Complete one- and two-loop calculations
have been carried out leading, for $\mh=150$~GeV, to
$
    a_{\mu}^{\mysmall \rm EW} = 154(1)(2)\times 10^{-11}
$~\cite{CMV03}.
The first error is due to hadronic loop uncertainties, while the second one
corresponds to an allowed range of $\mh \in [114,250]$~GeV, to the current
top mass uncertainty, and to unknown three-loop effects.  The
leading-logarithm three-loop contribution to $a_{\mu}^{\mysmall \rm EW}$ is
extremely small~\cite{CMV03,DGi98}.

\begin{figure}[h]
\begin{center}
\includegraphics[width=14cm]{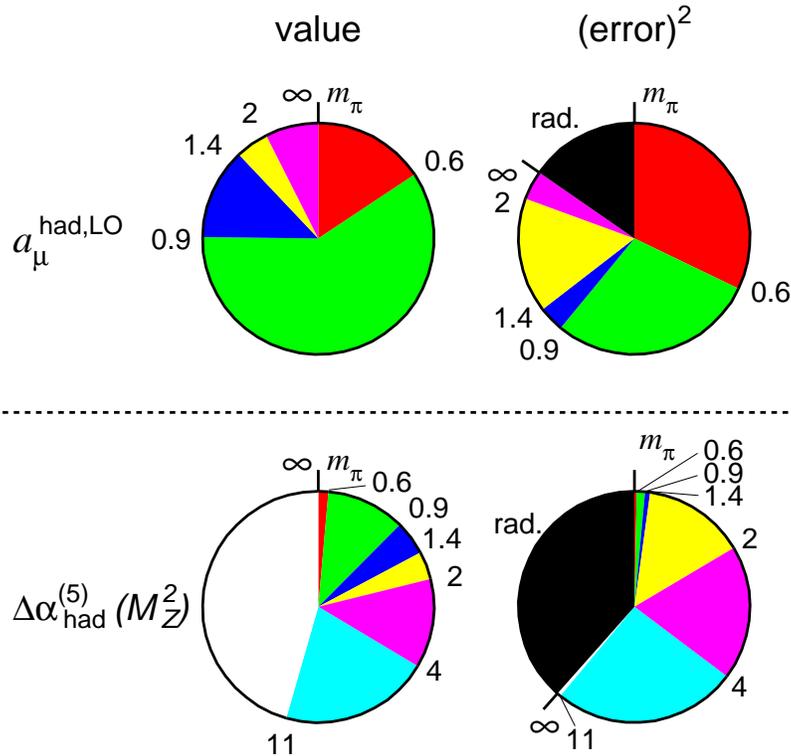}
\vspace{-2.cm}
\caption{The pie diagrams show the fractions of the total contributions and
  the squared errors from various energy intervals in the dispersion
  integrals in Eqs.~(\ref{eq:delta_alpha_had}) and (\ref{eq:amu_had}). The
  diagrams for the leading-order hadronic contribution to the muon $g-2$, shown in the first
  row, correspond to sub-contributions with energy boundaries at 0.6, 0.9,
  1.4, 2 GeV and $\infty$, whereas for the hadronic contribution to the
  effective fine-structure constant, shown in the second row, the boundaries
  are at 0.6, 0.9, 1.4, 2, 4, 11.09 GeV and $\infty$. In the squared
  error diagrams, the contributions arising from the treatment of the
  radiative corrections to the data are also
  included~\cite{Hagiwara:2003da}.}
\label{fig:pies}
\end{center}
\end{figure}

Like the effective fine-structure constant at the scale $\mz$, the {\small
SM} determination of the anomalous magnetic moment of the muon is presently
limited by the evaluation of the hadronic vacuum polarisation and, in turn,
by our knowledge of the low-energy total cross-section for $e^+ e^-$
annihilations into hadrons. Indeed, the hadronic leading-order contribution
$a_{\mu}^{\mbox{$\scriptscriptstyle{\rm HLO}$}}$, due to the hadronic vacuum
polarisation correction to the one-loop diagram, involves long-distance
{\small QCD} effects which cannot be computed perturbatively. However, using
analyticity and unitarity, it was shown long ago that this term can be
computed from hadronic $e^+ e^-$ annihilation data via the dispersion
integral~\cite{DISP_REL_AMU}
\begin{equation}
      a_{\mu}^{\mbox{$\scriptscriptstyle{\rm HLO}$}}= 
      (1/4\pi^3)
      \int^{\infty}_{4m_{\pi}^2} ds \, K(s) \sigma^{(0)}\!(s) =
      (\alpha^2/3\pi^2)
      \int^{\infty}_{4m_{\pi}^2} ds \, K(s) R(s)/s.
\label{eq:amu_had}
\end{equation}
The kernel function $K(s)$ decreases monotonically for increasing~$s$. This
integral is similar to the one entering the evaluation of the hadronic
contribution $\Delta \alpha_{\rm had}^{(5)}(\mz^2)$ in
Eq.~(\ref{eq:delta_alpha_had}). Here, however, the weight function in the
integrand gives a stronger weight to low-energy data. Figure~\ref{fig:pies}
(from ref.~\cite{Hagiwara:2003da}) shows the fractions of the total
contributions and the squared errors from various energy intervals in the
dispersion integrals for $a_{\mu}^{\mbox{$\scriptscriptstyle{\rm HLO}$}}$
and $\Delta \alpha_{\rm had}^{(5)}(\mz^2)$.

An important role among all $e^+ e^-$ annihilation measurements is played by
the precise data collected in 1994-95 by the {\small CMD-2} detector at the
{\small VEPP-2M} collider in Novosibirsk for the $e^+e^-\rightarrow
\pi^+\pi^-$ cross section at values of $\sqrt{s}$ between 0.61 and 0.96
GeV~\cite{Akhmetshin:2003zn} (quoted systematic error 0.6\%,
dominated by the uncertainties in the radiative corrections). 
Recently~\cite{CMD2-05} the {\small CMD-2} Collaboration released its
1996-98 measurements for the same cross section in the full energy range
$\sqrt{s} \in [0.37,1.39]$ GeV. The part of these data for $\sqrt{s} \in
[0.61,0.96]$ GeV (quoted systematic error 0.8\%) agrees with their earlier
result published in~\cite{Akhmetshin:2003zn}.
In 2005, also the {\small SND} Collaboration (at the {\small VEPP-2M}
collider as well) released its analysis of the $e^+e^-\rightarrow
\pi^+\pi^-$ process for $\sqrt{s}$ between 0.39 and 0.98 GeV, with a
systematic uncertainty of 1.3\% (3.2\%) for $\sqrt{s}$ larger (smaller) than
0.42 GeV~\cite{Achasov:2005rg}. However, a recent preliminary reanalysis of
these data~\cite{Achasov:2006} uncovered an error in the treatment of the
radiative corrections, reducing the value of the measured cross-section.
The new preliminary result appears to be in good agreement with the
corresponding one from {\small CMD-2}. Further significant progress is
expected from the $e^+e^-$ collider VEPP-2000 \cite{VEPPtot} under
construction in Novosibirsk.
In 2004 the {\small KLOE} experiment at the {\small DAFNE} collider in
Frascati presented a precise measurement of $\sigma(e^+e^-\rightarrow
\pi^+\pi^-)$ via the initial-state radiation ({\small ISR}) method at the
$\phi$ resonance~\cite{Aloisio:2004bu} (see later).  This cross section was
extracted for $\sqrt{s}$ between 0.59 and 0.97 GeV with a systematic error
of 1.3\% and a negligible statistical one.  There are some discrepancies
between the {\small KLOE} and {\small CMD-2} results, although their
integrated contributions to $a_{\mu}^{\mbox{$\scriptscriptstyle{\rm HLO}$}}$
are similar. The data of {\small KLOE} and {\small
SND}~\cite{Achasov:2005rg} disagree above the $\rho$ peak, where the latter
are significantly higher. However, the values of the latter appears to be
lower after the new preliminary reanalysis presented in~\cite{Achasov:2006}.
The study of the $e^+e^-\rightarrow \pi^+\pi^-$ process via the {\small ISR}
method is also in progress at {\small BABAR}~\cite{BABAR-ISR} and
Belle~\cite{BELLE-ISR}.On the theoretical side, analyticity, unitarity and
chiral symmetry provide strong constraints for the pion form factor in the
low-energy region~\cite{Colangelo03}.
Recent evaluations of the dispersive integral based on the {\small CMD-2}
analysis of ref.~\cite{Akhmetshin:2003zn} are in good agreement:
$
      a_{\mu}^{\mbox{$\scriptscriptstyle{\rm HLO}$}} =  
      6934 \, (53)_{exp} (35)_{rad} \times 10^{-11}
$\cite{Hoecker04},
$
      a_{\mu}^{\mbox{$\scriptscriptstyle{\rm HLO}$}} =
      6948 \, (86) \times 10^{-11} 
$\cite{Jegerlehner:2003rx,JegerJPG29},
$
      a_{\mu}^{\mbox{$\scriptscriptstyle{\rm HLO}$}} = 
      6924 \, (59)_{exp} (24)_{rad} \times 10^{-11}
$\cite{Hagiwara:2003da},
$
      a_{\mu}^{\mbox{$\scriptscriptstyle{\rm HLO}$}} = 
      6944 \, (48)_{exp} (10)_{rad} \times 10^{-11}
$\cite{dTY04}.
Reference \cite{Hoecker04} already includes {\small KLOE}'s results. The
recent data of {\small CMD-2}~\cite{CMD2-05} and {\small
SND}~\cite{Achasov:2005rg,Achasov:2006} are not yet included.

The authors of \cite{ADH98} pioneered the idea of using vector spectral
functions derived from the study of hadronic $\tau$ decays~\cite{DHZ05} to
improve the evaluation of the dispersive integral.  However, the latest
analysis with {\small ALEPH}, {\small CLEO}, and {\small OPAL} data yields
$a_{\mu}^{\mbox{$\scriptscriptstyle{\rm HLO}$}} = 7110 \,
(50)_{exp} (8)_{rad} (28)_{SU(2)} \times 10^{-11}
$\cite{DEHZ03}, 
a value significantly higher than those obtained with $e^+e^-$ data
(see~\cite{BELLE-TAU} for recent preliminary results from Belle).
Isospin-breaking corrections were applied~\cite{MS88CEN}. Indeed, although
the precise {\small CMD-2} $e^+e^-\rightarrow \pi^+\pi^-$
data~\cite{Akhmetshin:2003zn} are consistent with the corresponding $\tau$
ones for energies below $\sim0.85$ GeV, they are significantly lower for
larger energies. {\small KLOE}'s $\pi^+\pi^-$ spectral function confirms
this discrepancy with the $\tau$ data. {\small SND}'s 2005
results~\cite{Achasov:2005rg} were compatible with the $\tau$ ones, but the
very recent preliminary reanalysis presented in~\cite{Achasov:2006} seems to
indicate that this is no longer the case.
This puzzling discrepancy between the $\pi^+\pi^-$ spectral functions from
$e^+e^-$ and isospin-breaking-corrected $\tau$ data could be caused by
inconsistencies in the $e^+e^-$ or $\tau$ data, or in the isospin-breaking
corrections which must be applied to the latter~\cite{eetau}.

The hadronic higher-order $(\alpha^3)$ contribution
$a_{\mu}^{\mbox{$\scriptscriptstyle{\rm HHO}$}}$ can be divided into two
parts:
$
     a_{\mu}^{\mbox{$\scriptscriptstyle{\rm HHO}$}}=
     a_{\mu}^{\mbox{$\scriptscriptstyle{\rm HHO}$}}(\mbox{vp})+
     a_{\mu}^{\mbox{$\scriptscriptstyle{\rm HHO}$}}(\mbox{lbl}).
$
The first one is the $O(\alpha^3)$ contribution of diagrams containing
hadronic vacuum polarisation insertions~\cite{Krause96}. Its latest value is
$a_{\mu}^{\mbox{$\scriptscriptstyle{\rm HHO}$}}(\mbox{vp})= -97.9 \,
(0.9)_{exp} (0.3)_{rad} \times 10^{-11} $~\cite{Hagiwara:2003da}.  The
second term, also of $O(\alpha^3)$, is the hadronic light-by-light
contribution. As it cannot be directly determined via a dispersion relation
approach using data (unlike the hadronic vacuum polarisation contribution),
its evaluation relies on specific models of low-energy hadronic interactions
with electromagnetic currents. Three major components of
$a_{\mu}^{\mbox{$\scriptscriptstyle{\rm HHO}$}}(\mbox{lbl})$ can be
identified: charged-pion loops, quark loops, and pseudoscalar ($\pi^0$,
$\eta$, and $\eta'$) pole diagrams. The latter ones dominate the final
result and require information on the electromagnetic form factors of the
pseudoscalars (see Secs.\ 2.4.1 and 2.5.5). In 2001 the authors
of~\cite{lbl1} uncovered a sign error in earlier evaluations of the
dominating pion-pole part. Their estimate of
$a_{\mu}^{\mbox{$\scriptscriptstyle{\rm HHO}$}}(\mbox{lbl})$, based also on
previous results for the quark and charged-pions loop parts~\cite{lbl2}, is
$
      a_{\mu}^{\mbox{$\scriptscriptstyle{\rm HHO}$}}(\mbox{lbl}) =  
      80\,(40)\times 10^{-11}
$.
A higher value was obtained in 2003 including short-distance
{\small QCD} constraints:
$
      a_{\mu}^{\mbox{$\scriptscriptstyle{\rm HHO}$}}(\mbox{lbl}) =
      136\,(25)\times 10^{-11}
$~\cite{MV03}.
Further independent calculations would provide an important check of this
contribution.

The {\small SM} prediction of the muon $g$$-$$2$ is given by the sum
$
    a_{\mu}^{\mysmall \rm SM} = 
         a_{\mu}^{\mysmall \rm QED} +
         a_{\mu}^{\mysmall \rm EW}  +
         a_{\mu}^{\mbox{$\scriptscriptstyle{\rm HLO}$}} +
         a_{\mu}^{\mbox{$\scriptscriptstyle{\rm HHO}$}}.
$
The discrepancies between recent {\small SM} predictions and the current
experimental value vary in a very wide range, from roughly 1 to 3 $\sigma$,
according to the values chosen for the hadronic contributions. If only
$e^+e^-$ data are employed, $a_{\mu}^{\mysmall \rm SM}$ deviates from
$a_{\mu}^{\mbox{$\scriptscriptstyle{\rm EXP}$}}$ by 2--3~$\sigma$.  The
analysis of this section shows that while the {\small QED} and {\small EW}
contributions appear to be ready to rival the forecasted precisions of
future experiments (like {\small E969}), much effort will be needed to
reduce the hadronic uncertainty.  This effort is challenging but possible,
and certainly well motivated by the excellent opportunity the muon $g$$-$$2$
is providing us to unveil (or constrain) ``new physics'' effects.  Once
again, a long-term program of hadronic cross-section measurements is clearly
warranted.

%---------------------------------------------------------------------------%
\subsubsection{Status of R at low energy}
\label{subsubsec:RSTATUS}

\begin{figure}[h]
\begin{center}
\epsfig{file=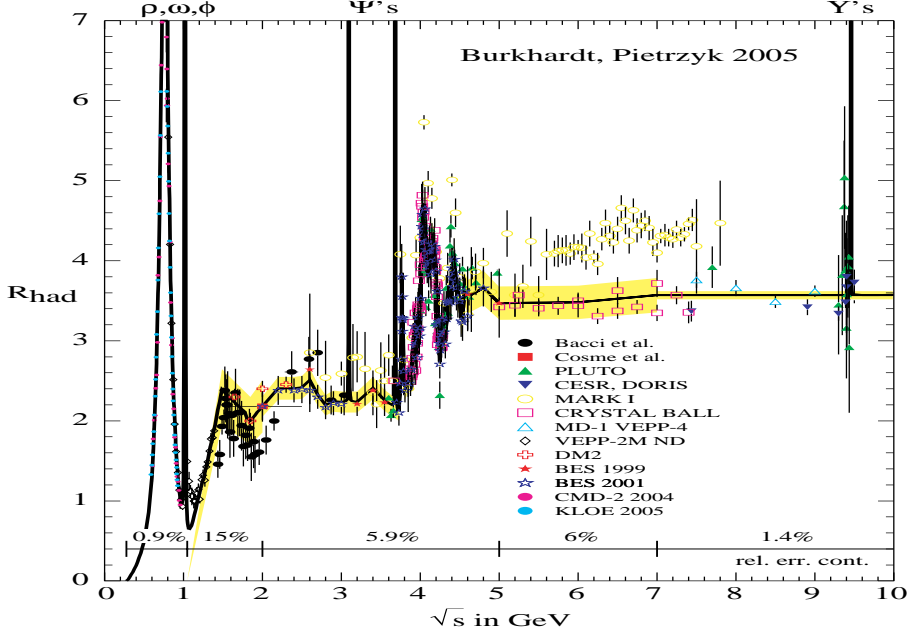,width=12.cm,height=8.5cm} 
\caption{An updated compilation of $R$ measurements from
ref.~\cite{BP05}. In the bottom line the overall uncertainties of the different
regions are reported.} 
\label{fig1}
\end{center}
\end{figure}

During the last thirty years the ratio $R$ has been measured by several
experiments. Usually, for energies below 2 GeV the cross section is measured
for individual channels, while above that value the hadronic final states
are treated inclusively.  Figure~\ref{fig1} shows an up-to-date compilation
of these data by Burkhardt and Pietrzyk~\cite{BP05}.  The main improvements
are in the region below 5 GeV (where the data are now closer to the
prediction of perturbative {\small QCD}): between 2 and 5 GeV, the {\small
BESII} collaboration reduced the error to $\sim 7\%$~\cite{Bai:2001ct}
(before it was $\sim 15\%$); below 1 GeV, the {\small CMD-2} and {\small
SND} collaborations at Novosibirsk, and {\small KLOE} at Frascati, measured
the pion form factor in the energy range around the $\rho$ peak with a
systematic error of $0.6\%$, $1.3\%$, and $1.3\%$ respectively.  In
Fig.~\ref{fig1}, the recent published results from the {\small BABAR} 
collaboration \cite{Aubert:2004kj}
on the cross sections $e^+ e^-$ to 3 and 4 hadrons are not yet included. The
uncertainty in the 1--2 GeV region is still 15\%~\cite{BP05}.

The measurement of the hadronic cross section has been usually performed by
varying the $e^+e^-$ beam energies.  An alternative approach,
recently~\cite{Binner:1999bt} proposed, consists of extracting
$\sigma_{had}$ from Initial State Radiation (ISR) events at flavour
factories, where the high luminosity of the machine compensates for the
reduced cross-section.  This method, successfully applied by {\small KLOE}
and {\small BABAR}, has the advantage of the same normalisation for each
energy point, even if it requires a very solid theoretical understanding of
radiative corrections, a precise determination of the angle and energy of
the emitted photon, and the full control of background events, especially
for events with the photon emitted in the final state {\small (FSR)}.  The
Karlsruhe-Katowice group computed the radiative corrections up to {\small
NLO} for different exclusive channels, implementing them in the event
generator {\small
PHOKHARA}~\cite{Rodrigo:2001jr,Kuhn:2002xg,Rodrigo:2001kf,Czyz:2002np,Czyz:PH03}.
The current precision for the $\pi^+\pi^-\gamma$ final state is $0.5\%$.

In the following we will consider the impact of {\small DAFNE}-2 on the 
hadronic cross-section measurements in the full accessible 
region [2$m_\pi$--2.5 GeV], by
considering three main energy regions.

\subsubsection*{$\pi^+\pi^-$ threshold region.}
\begin{figure}[h]
\begin{center}
\epsfig{figure=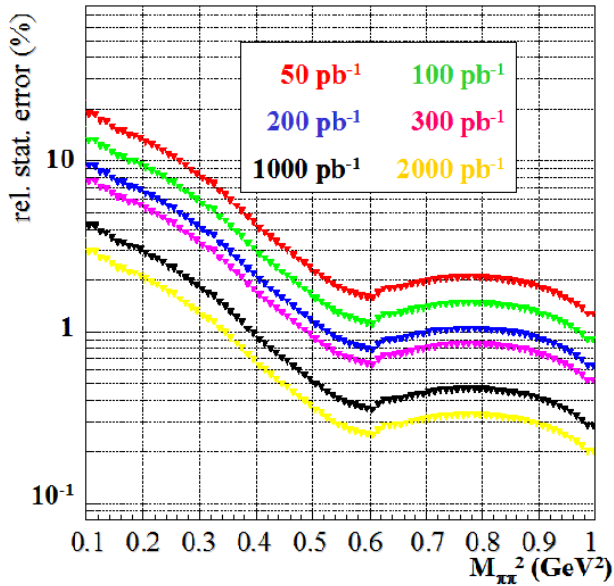,width=8.5cm}
\caption{\label{fig:threshimpact} The relative statistical
error on the cross section ${\rm d}\sigma/{\rm d}M^2_{\pi\pi}$ as a function
of the $\pi\pi$ invariant mass 
squared, $M_{\pi\pi}^2$, (bin width = 0.01 GeV$^2$),
according to different integrated luminosities. }
\end{center}
\end{figure}

The threshold region, [2$m_\pi$--0.5 GeV], provides 13\% of the total
$\pi^+\pi^-$ contribution to the muon anomaly: $a_\mu^{\rm HLO}$[2$m_\pi$--0.5
GeV] = $(58.0\pm2.1)\times 10^{-10}$~\cite{Hoecker04} To overcome the lack of
precise data at threshold energies, the pion form factor is extracted from a
parametrisation based on Chiral Perturbation Theory, constrained from
space-like data~\cite{Amendolia:1986wj}.
The most effective way to measure the threshold in the time-like region is
provided by ISR events, where the emission of an energetic photon allows to
study the two pions at rest.  However, at {\small DAFNE}, the process
$\phi\to\pi^+\pi^-\pi^0$, where one photon gets lost, is hundreds of
times more frequent than the signal, and therefore a precise measurement
requires an accurate evaluation of the background. Furthermore, irreducible
backgrounds due to 
$\phi\to\pi^+\pi^-\gamma$ are also present when running at the $\phi$
resonance peak.  The background issue can be largely overcome by running at
$\sqrt{s}<M_\phi$: such a possibility has been already explored by the KLOE
experiment, which is taking more than 200 pb$^{-1}$ of data at 1 GeV.
Figure \ref{fig:threshimpact} shows the statistical precision that can be
reached in the region below 1 GeV for different integrated luminosities,
with a bin width of 0.01 GeV$^2$. A statistics of 2 fb$^{-1}$ at 1 GeV will
allow to achieve a statistical error on $a_\mu^{\rm HLO}$ at threshold below
1\%. Notice that 
in order to maintain the systematic uncertainty at the same level, it is
important to take data below the $\phi$ peak, to reduce the backgrounds.

\subsubsection*{The $\rho$ peak region}

The $\pi^+\pi^-$ region between 0.5 and
1 GeV has been studied by different experiments.
{\small CMD-2}~\cite{Akhmetshin:2003zn} and {\small
SND}~\cite{Achasov:2005rg} performed an energy
scan at the $e^+e^-$ collider {\small VEPP--2M} ($\sqrt{s}\in$ [0.4--1.4]
GeV) with $\sim 10^6$ and $\sim 4.5\times 10^6$ events respectively, with
systematic errors ranging from 0.6\% to 4\% in the relative cross-section,
depending on the $2\pi$ energy region. The pion form factor has also been 
measured by {\small KLOE} using ISR, and results are also expected soon by 
{\small BABAR}. 
{\small KLOE} published a result~\cite{Aloisio:2004bu} based
on an integrated luminosity of 140 pb$^{-1}$, that led to a relative error
of 1.3\% in the energy region [0.6--0.97] GeV, dominated by systematics.
At the moment it has already
collected more than 2 fb$^{-1}$ at the $\phi$ meson peak, which represents,
around the $\rho$ peak, a statistics of $\sim 2 \times 10^{7}$
$\pi^+\pi^-\gamma$ events. 
{\small BABAR}~\cite{Aubert:2004kj} has already collected more than 300
fb$^{-1}$ at the $\Upsilon$ peak, and is going to collect about 1 ab$^{-1}$
by the end of data taking.
%photon tagging
%However, due to higher c.m.\ energy and
%photon tagging, this latter number corresponds to an equivalent luminosity
%of about 2 fb$^{-1}$ around the $\rho$--$\omega$ peak. 
The results of these
four experiments in the next few years will probably allow to know the
$\pi^+\pi^-$ cross-section for most of the $\rho$ shape with a relative
accuracy better than 1\% (even considering both statistical and systematic
errors). The discrepancies now present in the shape could be then washed
out. In this case,
%due to the highs statistics already collected by the 
%currents experiments, 
a significant improvement from
 DAFNE-2 is not envisaged on this region.

\subsubsection*{The 1--2.5 GeV energy region.}
The region [1--2.5 GeV], with an uncertainty of roughly 15\%, 
is the most poorly known, and contributes about 40\% to the 
uncertainty of the total dispersion integral for 
$\Delta^{(5)}_{had}(m_Z^2)$~\cite{BP05}.  It also provides
most of the contribution to
$a_{\mu}^{\mbox{$\scriptscriptstyle{\rm HLO}$}}$ above 1 GeV.
%The region [1--2.5 GeV] 
%is the most poorly known
%($\delta\Delta^{(5)}_{had}(m_Z^2)/\Delta^{(5)}_{had}(m_Z^2)\sim
%15\%$~\cite{deltaalpha}), contributing with $\sim 40\%$~\cite{deltaalpha} to the
%uncertainty on the dispersion integral at $m_Z^2$~\cite{deltaalpha}.  In
%addition, it gives the second relevant contribution to
%$a_{\mu}^{\mbox{$\scriptscriptstyle{\rm HLO}$}}$.

We  will now consider the impact of DAFNE-2 for 
inclusive and exclusive measurements separately:
\begin{itemize}
\item[-] {\bf Inclusive measurements:}

There is a systematic difference between the sum of exclusive channels and the
inclusive measurements~\cite{Hagiwara:2003da}, where most of the 
recent inclusive
data are from the early 80's 
(obtained with a total integrated luminosity of $200 \,nb^{-1}$).
\begin{figure}[h]
\begin{center}
\mbox{\epsfig{figure=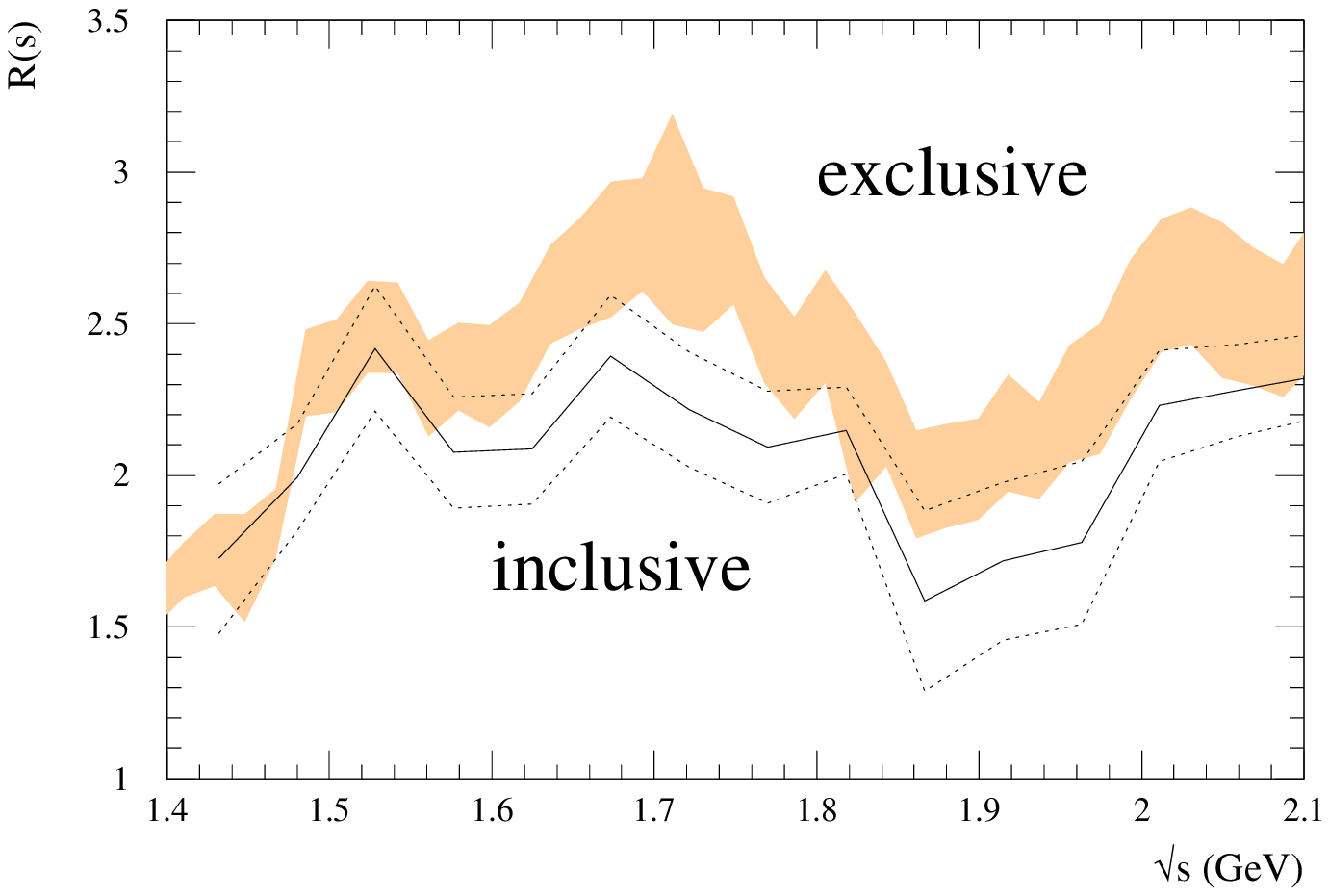,width=7.5cm}
\epsfig{figure=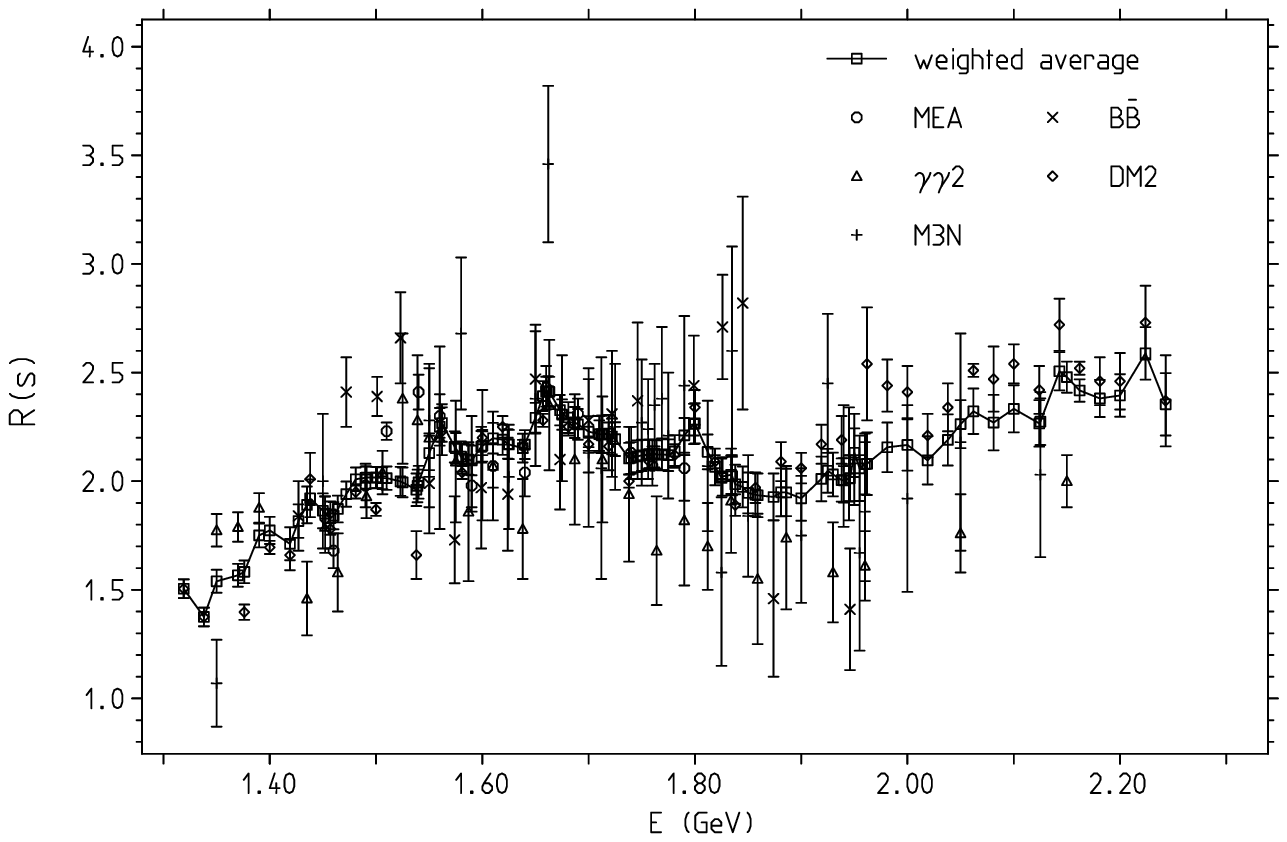,width=7.5cm}}
\caption{\label{fig:inclusive} Left: comparison between the
inclusive $R$ measurements and the sum of exclusive
channels~\cite{Hagiwara:2003da}.  Right: a
compilation~\cite{Jegerlehner:2003rx} of the most recent R inclusive
measurements in the same energy range. Notice that three out of the five
experiments whose data are presented in this figure, come from Adone at
Frascati in the seventies (namely MEA, $\gamma\gamma$2 and $B\overline{B}$). }
\end{center}
\end{figure}

Figure~\ref{fig:inclusive} ({\it left}) shows the comparison between the
inclusive and the sum of exclusive $R$ measurements in the energy range
[1.4--2.1] GeV, while the plot on the right collects the most recent
inclusive data in the same range~\cite{Jegerlehner:2003rx}.
%
%A new $e^+e^-$ collider, {\small VEPP--2000}, is expected to start data
%taking around 2007, with $L\simeq10^{32}~cm^{-2}~s^{-1}$.  It will cover the
%region $\sqrt{s}\in$ [0.4--2] GeV, with the aim of refining the physics
%program of {\small VEPP--2M} that was limited to $\sqrt{s}<1.4$ GeV.

%Most likely the current data from different experiments ({\small CMD-2},
%{\small SND}, {\small KLOE} and {\small BABAR}) will not be able to bring
%the error below $1\%$ in the energy region below $600$ MeV. This remains
%the main limitations on the theoretical evaluation of $a_{\mu}$.

With a specific luminosity of $10^{32} cm^{-2} sec^{-1}$, {\small
DAFNE-2} can perform a scan in the region from 1 to 2.5 GeV,
collecting an integrated luminosity of 20 pb$^{-1}$ per point (corresponding
to few days of data taking). By assuming an energy step of 25 MeV, the whole
region would be scanned in one year of data taking.
%
%assuming 1 week for point, which would include accelerator and
% detector inefficiencies.
%
A detector \`a la {\small KLOE}, plus some minor improvements such as a
finer calorimeter readout and an inner tracker (see \ref{DetAcc}), 
will be capable to perform
an inclusive $R$ measurement at the percent level. This would represent a
major improvement on this issue.

%
%It is mostly likely  that such an accuracy it's difficult to reach by 
%$ISR$, and therefore  can be reached 
%an inclusive measurment, can be reache by a scan, and therefore 
%If in the meantime no other inclusive result comes out,
%the improvement with respect to the old experiments is
%guaranteed. SHOULD BE QUANTIFIED?  
%

\item[-] {\bf Exclusive channels:}

A different issue concerns the exclusive measurements.
In this case, {\small BABAR} has published results on 
$e^+ e^-$ into 3 and 4 hadrons, obtained with an integrated luminosity 
of 89 fb$^{-1}$~\cite{Aubert:2004kj},
and it is expected to reach 1 ab$^{-1}$ by the end of the data taking.
%The same can be expected by Belle.
However, due to the ISR photon emission at the $\Upsilon(4s)$ resonance, 
the effective luminosity for tagged photon ($\theta_{\gamma}>20^o$)
in the energies below 2.5 GeV, will be of the order of few pb$^{-1}$ 
at full statistics. This is shown in Figure~\ref{fig:w}, where a bin width 
of 25 MeV and an overall efficiency of 10\% are assumed~\cite{Aubert:2004kj}.
\begin{figure}[h]
\begin{center}
\epsfig{figure=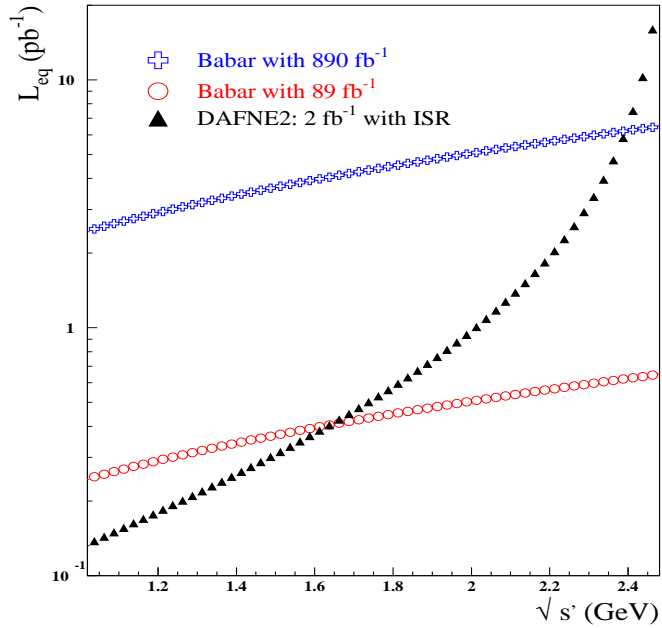,width=10.cm,height=10.cm}
\vspace{-1.cm}
\caption{\label{fig:w} Equivalent luminosity for: {\small BABAR} with 
890 fb$^{-1}$ 
(cross); {\small BABAR} with 89 fb$^{-1}$ (circle); 
DAFNE-2 with 2 fb$^{-1}$, using ISR 
 at 2.5 GeV (triangle). A bin width of 25 MeV is assumed. 
A polar angle of the photon larger than $20^o$ 
and an overall efficiency of 10\% are assumed~\cite{Aubert:2004kj}}
\end{center}
\end{figure}

%
%For the channels ....,we compare the relative statistical error of published
%BaBar (90 fb$^{-1}$), $extrapolated $BaBar by a factor ten in statistics 
%(O($1ab{-1}$)), and  DAFNE-II 
%with an integrated luminosity of 20 pb$^{-1}$ per point.
%
%
\begin{figure}[h]
\begin{center}
\epsfig{figure=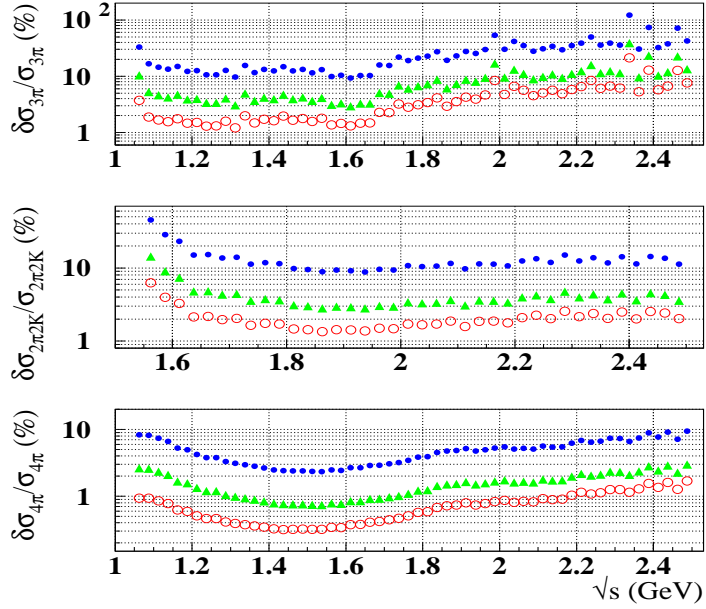,width=10cm,height=10cm}
\vspace{-1.cm}
\caption{\label{fig:impactscan} Comparison of the statistical accuracy in
the cross-section among DAFNE-2 with an energy scan with 20 pb$^{-1}$ per 
point ($\circ$); published {\small BABAR} results ($\bullet$), 
{\small BABAR} with full statistics ($\blacktriangle$)  for $\pi^+\pi^-\pi^0$
(top), $\pi^+\pi^-K^+K^-$ (middle) and $2\pi^+ 2\pi^-$ (down) channels. 
An energy step of 25 MeV is assumed.}
\end{center}
\end{figure}
Figure \ref{fig:impactscan}  shows the statistical error for the channels
$\pi^+\pi^-\pi^0$, $2\pi^+ 2\pi^-$ and $\pi^+\pi^-K^+K^-$, which can be
 achieved by  an energy scan at DAFNE-2 with 20 pb$^{-1}$ per point,
compared  
with {\small BABAR} with published (89 fb$^{-1}$), and 
full  (890 fb$^{-1}$) statistics.
As it can be seen, an energy scan allows to reach a statistical
 accuracy of the order of  1\% for most of the energy points.
% which 
% In this case
%an accuracy below 1\% can be expected on the dispersion integral, 
%provided that also the systematic error 
%can be pinned down to the same value.
   
We finally estimate the statistical accuracy which can be reached 
by DAFNE-2 using ISR at
 $\sqrt{s}=2.5$ GeV. 
\begin{figure}[h]
\begin{center}
\epsfig{figure=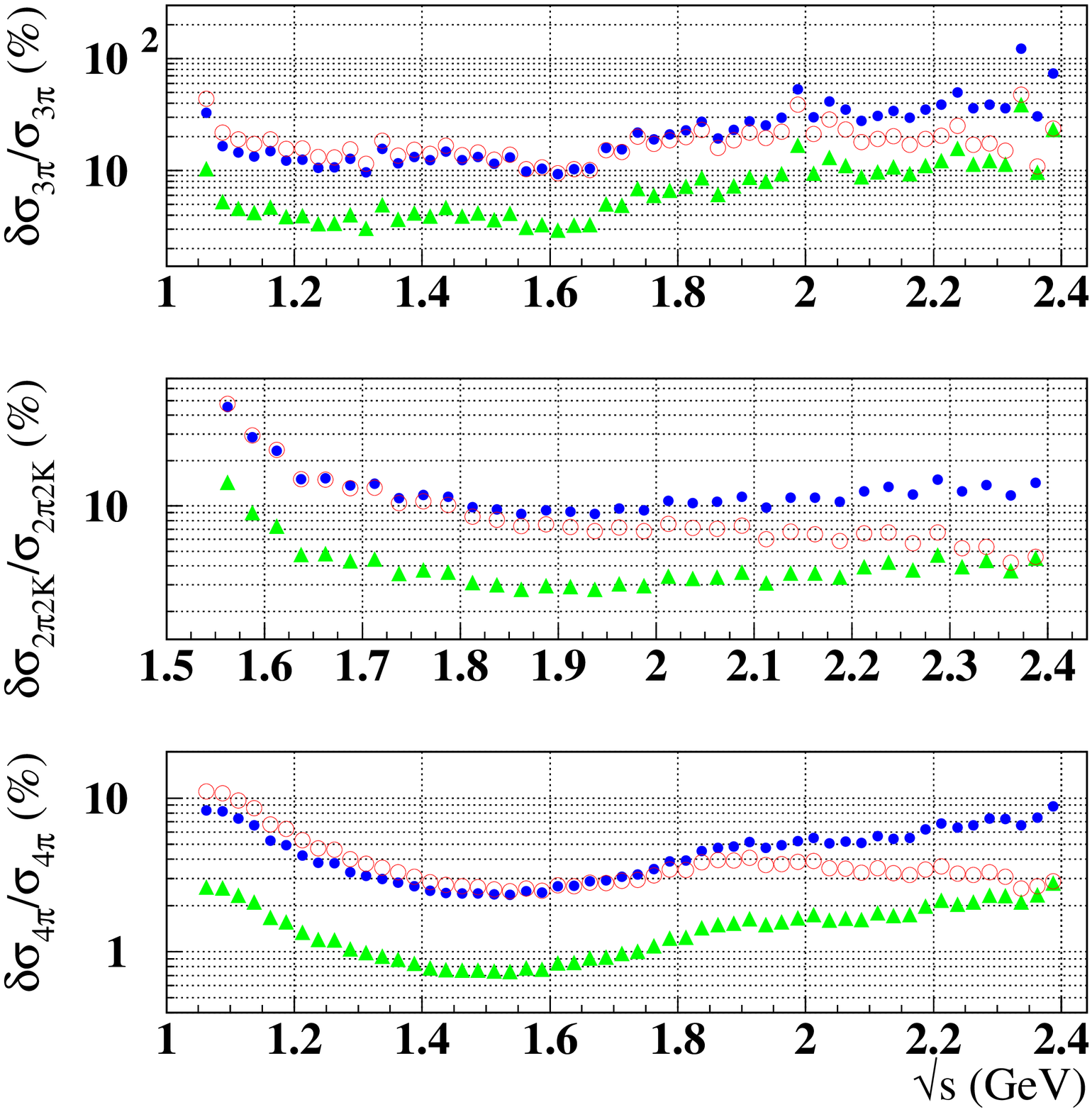,width=10cm,height=10cm}
\vspace{-1.cm}
\caption{\label{fig:impactisr} Comparison of the statistical accuracy in the
cross-section among DAFNE-2 with ISR at 2.5 GeV, 2 fb$^{-1}$ ($\circ$); 
published {\small BABAR} results ($\bullet$), 
{\small BABAR} with full statistics ($\blacktriangle$)  for $\pi^+\pi^-\pi^0$
(top), $\pi^+\pi^-K^+K^-$ (middle) and $2\pi^+ 2\pi^-$ (down) channels. 
A bin width of 25 MeV is assumed.}
\end{center}
\end{figure}
Figure \ref{fig:impactisr} shows the statistical accuracy for the same 
exclusive channels achieved by DAFNE-2 with 2 fb$^{-1}$ at 2.5 GeV, compared  
with {\small BABAR} with published (89 fb$^{-1}$), and full  (890 fb$^{-1}$) statistics.
In this case improvements from DAFNE-2 are not so significant.
\end{itemize}
%\subsubsection*{Considerations on the beam energy measurement.}
%\item[-] {\bf Considerations on the beam energy measurement:}

Finally we notice that an 
issue for this kind of measurement is the accuracy in the
determination of the center of mass energy. Based on the {\small KLOE} experience,
without resonant depolarization it's reasonable to obtain an accuracy on the
c.m.\ energy of O(${10^{-4}}$), i.e.\  100--200 keV. For a better precision
other methods, like
resonant depolarization, are needed.
\subsubsection{Conclusions}
In summary, the possibility to make precision tests of the Standard Model
in future experiments, requires a more accurate knowledge of the hadronic
cross-section in all the energy range between the  2$m_{\pi}$ threshold
 and 2.5 GeV. The region between 1 and 2.5 GeV is at
present the most poorly known and is crucial for the computation
of the hadronic corrections to the effective 
fine structure constant at the scale $m_Z$. In order to improve
%At threshold an accurate measurement is also 
the theoretical accuracy on $a_{\mu}$, a very accurate measurement at lower
energy would also be required. 
In both regions, {\small DAFNE}-2 can give important contributions.

\subsection{\bf Vector Mesons spectroscopy}
\label{Vecto}
Apart from allowing precision tests of the Standard Model, the measurement
of the $\sqrt{s}$ dependence of the cross sections of exclusive channels, 
represents the primary  source of information for the vector meson 
spectroscopy in the low energy region. 
This study of the vector meson spectroscopy is interesting to test and
to provide experimental inputs to the models of the strong interactions at
low energies.
Moreover the existence of glueballs and hybrid mesons, predicted by QCD in
this energy range and never observed in a clean way, can be investigated.

\subsubsection{Vector mesons below 2.5 GeV}

An $e^+e^-$ machine with 1.0$\leq\sqrt{s}\leq$ 2.5 GeV can give an
important contribution to the study of the vector mesons.
A high statistics scan of the energy region quoted above can: $(i)$ improve the
knowledge on the established vector mesons, $(ii)$ well measure the parameters
of other vector states, whose interpretation is still not clear, $(iii)$
search for possible new vector states.
Moreover, since there are discrepancies between some recent measurement of
exclusive cross sections of the {\small BABAR} Collaboration
\cite{Aubert:2004kj,babar6pi}
with the ISR method and the older energy scan measurements\cite{dm2}, a
test of the ISR method versus the energy scan one can be performed, by
running at the maximum $\sqrt{s}$ and comparing the results with the energy
scan with the same detector.

\subsubsection*{Established mesons}

In Tab.\ref{tab:vecmes} the generally accepted vector mesons
below 2.5 GeV are reported.
The agreement of the observed masses with the prediction of the quark
model\cite{godfrey} suggests the interpretation of these mesons as the
fundamental states and the first radial and orbital excitations of the
q\=q system. 

\begin{figure}[htbp]
\begin{tabular}{cc}
  \includegraphics[width=0.45\textwidth]{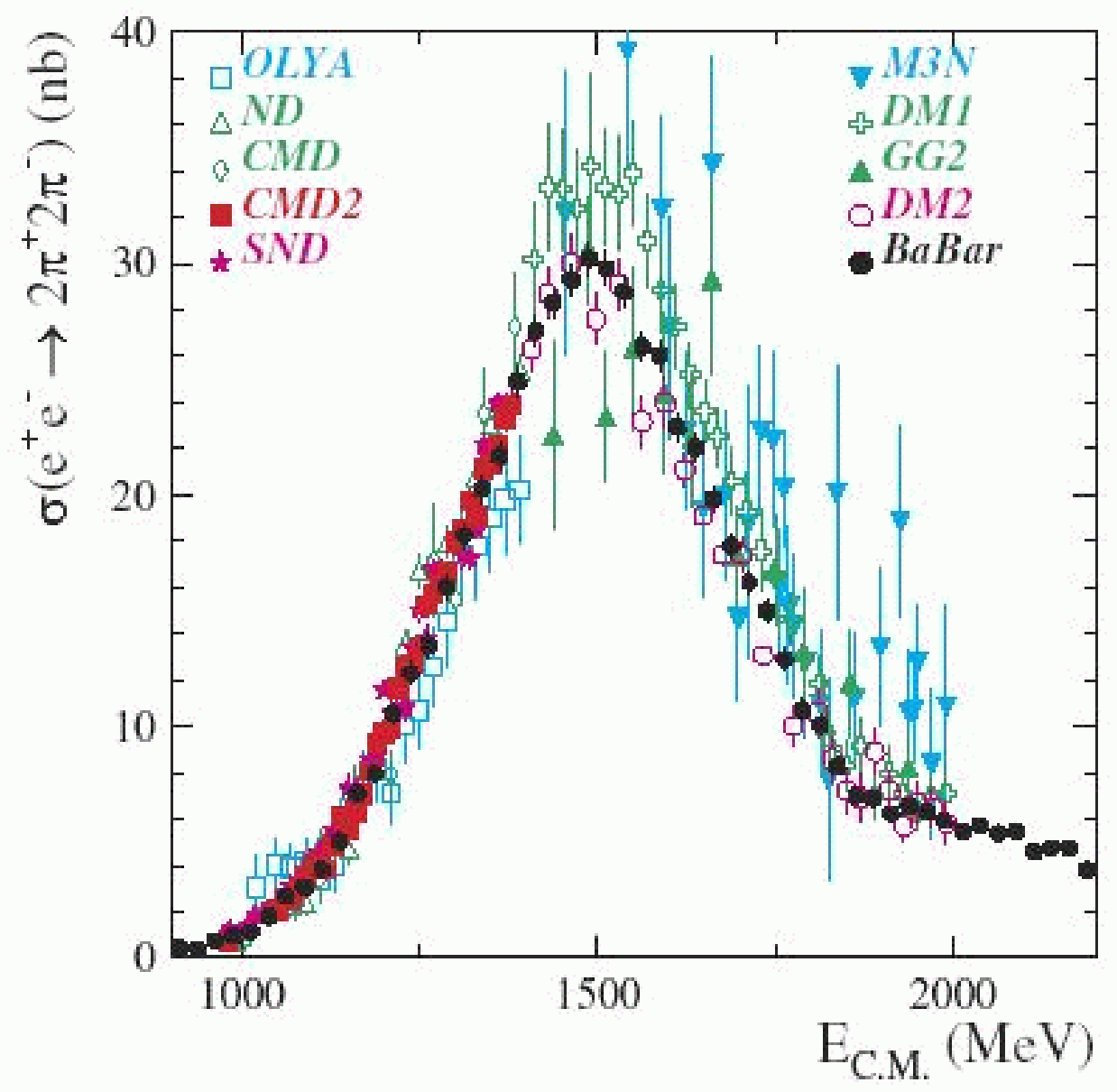} &
  \includegraphics[width=0.45\textwidth]{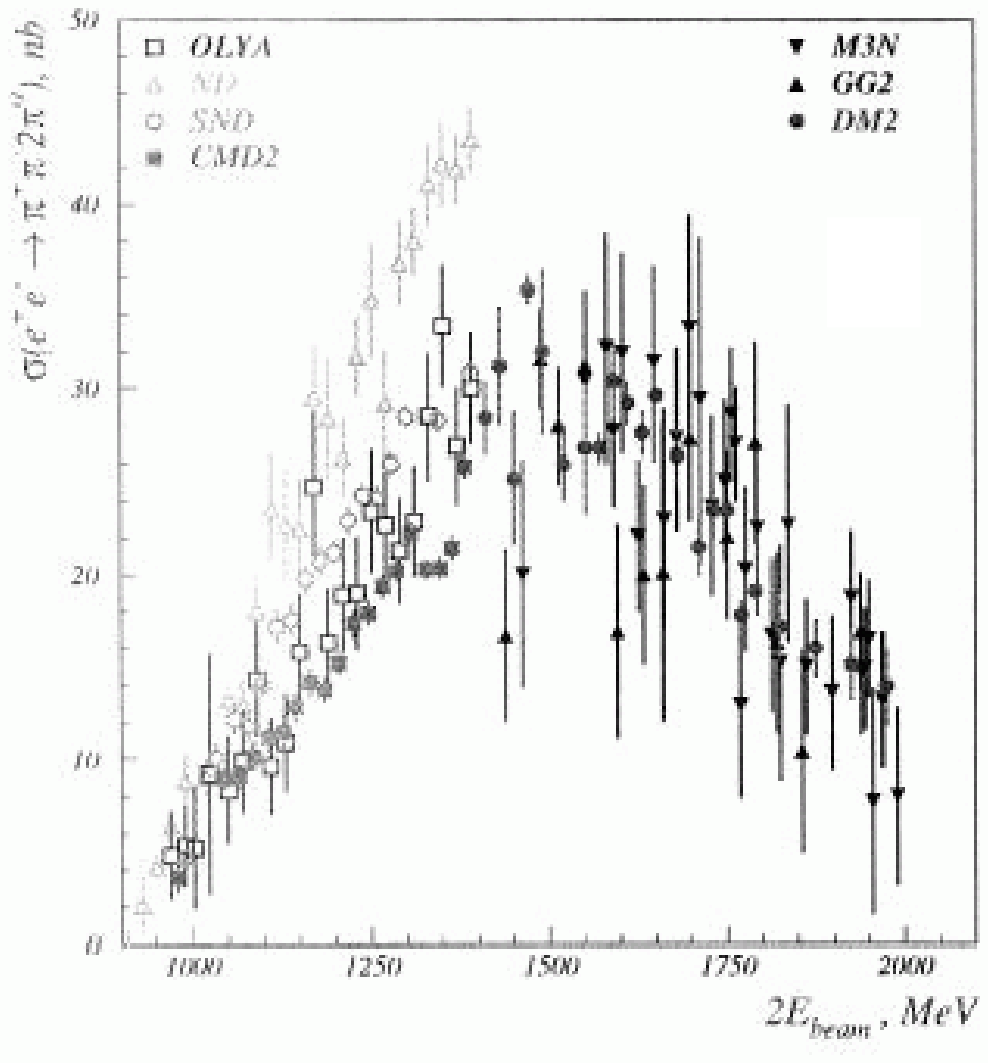} \\  
\end{tabular}
\caption{$e^+e^-\to 4\pi$ cross
  sections\cite{babar4pi,4picharg}.} 
\label{fig:4pi}
\end{figure}

\begin{table}[ht]
  \centering
  \begin{tabular}{| c | c | c | c |}
    \hline
    & (u\=u-d\=d)/$\sqrt{2}$ &  (u\=u+d\=d)/$\sqrt{2}$ & s\=s \\
    \hline
    1$^3S_1$ & $\rho(770)$  & $\omega(782)$  & $\phi(1020)$ \\
    2$^3S_1$ & $\rho(1450)$ & $\omega(1420)$ & $\phi(1680)$ \\
    1$^3D_1$ & $\rho(1700)$ & $\omega(1650)$ &  -           \\
    \hline
  \end{tabular}
  \caption{Classification of vector mesons.}
  \protect\label{tab:vecmes}
\end{table}
However this interpretation is not universally accepted\cite{donnachie},
since there are some inconsistencies with the predictions of the quark
model (in its $^3P_0$ version \cite{3p0}). 
In Fig.\ref{fig:4pi} are reported the cross sections of $e^+e^-\to 4\pi$;
according to the $^3P_0$ model, the $\rho_{2S}$ contribution to this final
state is negligible, while the $\rho_{1D}$ one is large and is dominated by
the ${a}_1(1260)\pi$ and ${h}_1(1170)\pi$ intermediate states, with
similar partial widths.
As ${h}_1\pi$ only contributes to the $\pi^+\pi^-\pi^0\pi^0$ final
state, while ${a}_1\pi$ contributes to both $\pi^+\pi^-\pi^0\pi^0$ and
$\pi^+\pi^-\pi^+\pi^-$, one would expect
$\sigma(e^+e^-\to\pi^+\pi^-\pi^0\pi^0) >
\sigma(e^+e^-\to\pi^+\pi^-\pi^+\pi^-)$ after the subtraction of the
$\omega\pi^0$ cross section from $\sigma(e^+e^-\to\pi^+\pi^-\pi^0\pi^0)$.
But experimentally one finds $\sigma(e^+e^-\to\pi^+\pi^-\pi^+\pi^-)\simeq
2\sigma(e^+e^-\to\pi^+\pi^-\pi^0\pi^0)$.
A possible explanation is a mixing of a vector hybrid
$\rho_H$ with the $\rho_{2S}$ and a small contribution of $\rho_{1D}$.
The observed 4$\pi$ cross sections will be explained by the fact that 
the dominant $\rho_H$ decay channel is ${a}_1(1260)\pi$\cite{fluxtube}.

A similar pattern can be envisaged for the isoscalar sector, ${i.e.}$
possible mixing of $\omega(1420)$ and $\omega(1650)$ with a hybrid $\omega_H$.

Concerning the s\=s mesons, if the $\phi(1680)$ is the 2$^3S_1$ state, the
$\phi_{1D}$ is still missing. 

Also the study of the radiative decays\cite{close} could
help in testing the possible mixing of these mesons with hybrids (see next
section).  

\subsubsection*{Gluonic mesons}

Hybrid mesons, $i.e.$ mesons with excited gluonic degrees of freedom are predicted
by QCD, with different, also exotic, quantum numbers. In the cases of $u$
and $d$ constituent quarks, the masses are predicted to be in the
region 1.3 -- 1.9 GeV.
There is general agreement on the mass ordering of such mesons:
$0^{-+}<1^{-+}<1^{--}<2^{-+}$.
There is also experimental evidence of exotic resonances,
$\pi_1(1400)$\cite{pi1} and $\pi_1(1600)$\cite{pi1bis}, both with
J$^{PC}$=1$^{-+}$.  
If $\pi_1(1400)$ is the lowest hybrid, the lightest vector hybrid could be
around 1.65 GeV, allowing the mixing pattern described above.
If, on the other hand, the lowest hybrid is $\pi_1(1600)$, one could expect
the lightest vector state at 1.9 -- 2.0 GeV, excluding the mixing with the
other vector mesons, but well inside the energy region covered by the machine
under consideration.
The I=1 vector hybrid should decay essentially into ${a}_1(1260)\pi$ or
$\rho\pi\pi$, then should be observable in the 4$\pi$ final state.
The I=0 vector hybrid should decay into $\rho\pi$ and $\rho(1450)\pi$,
allowing 3 and 5 $\pi$ final states.
Models predicts also strange hybrids (s\=sg) around 2.0 GeV, that should be
observable in $K\bar K\pi$ and $K\bar K\pi\pi$ final states.

Concerning glueballs, according to the lattice calculations\cite{glueball}, only the
scalar J$^{PC}$=0$^{++}$ is accessible at these energies via the radiative
decays of the vector mesons (see next section).
 
\subsubsection*{Other vector mesons}

Other vector mesons are present in the mass region under consideration.

The $\rho(1900)$ (J$^{PC}$=1$^{--}$ and I=1) is well established, measured
by various experiments, but with different values of mass and width as
reported in Tab.\ref{tab:rho1900}   
\begin{table}[htb]
  \centering
  \begin{tabular}{| c | c | c | c |}
    \hline
    Experiment & Process & Mass (GeV) & Width (MeV) \\
    \hline
    DM2       & $e^+e^-\to 6\pi$    & $\sim$ 1.93 & $\sim$ 35 \\
    FENICE\cite{fenicemh} & $e^+e^-\to hadrons$ & $\sim$ 1.87 & $\sim$ 10 \\
    E687\cite{e687}     & $3\pi^+ 3\pi^-$ photoproduction  & 1.91$\pm$ 0.01 & 33$\pm$ 13 \\ 
    {\small BABAR}\cite{babar6pi}   & $e^+e^-\to 3\pi^+ 3\pi^-$ & 1.88$\pm$ 0.03 & 130$\pm$ 30 \\
    {\small BABAR}   & $e^+e^-\to 2\pi^+ 2\pi^- 2\pi^0$
    & 1.86$\pm$ 0.02 & 160$\pm$ 20 \\
    {\small BABAR}\cite{solodov} & $e^+e^-\to 2\pi^+2\pi^-$& 1.88 $\pm$ 0.01 & 180 $\pm$ 20 \\
    {\small BABAR} & $e^+e^-\to\pi^+\pi^-2\pi^0$
    & 1.89$\pm$ 0.02 & 190$\pm$ 20 \\
    \hline
  \end{tabular}
  \caption{$\rho(1900)$ parameters. {\small BABAR} results are obtained using the
    radiative return method.}
  \protect\label{tab:rho1900}
\end{table}

The open questions than can be answered by a high statistics measurement
are: $(i)$ is the mass above or below the nucleon-antinucleon
threshold ?, and $(ii)$ the $\rho(1900)$ is large or narrow ?

These questions are connected to the interpretation of this particle, two
seem to be favoured: $(a)$ baryonium state, $(b)$ hybrid meson.
Baryonium can be either a diquark-antidiquark state with angular momentum
L=1\cite{jaffe}, or a N\=N quasi-nuclear bound state\cite{baryonium}. 
In both cases it should be strongly coupled to N\=N and produce some
visible signal, like threshold enhancements in the $e^+e^-\to N\bar N$
cross section.
A threshold enhancement, compatible with a resonance below p\=p
threshold\cite{fenicemh}, has been observed in the proton time-like form
factor by PS170\cite{apple} experiment at LEAR, and recently confirmed by
{\small BABAR}\cite{babarff}. 
Moreover BES-II Collaboration \cite{bbes} has observed an enhancement at the p\=p
threshold in J/$\psi\to p\bar p\gamma$, interpreted as the effect of a
resonance of mass 1859 MeV and total width smaller than 30 MeV, but with
different quantum numbers: J$^{PC}$=0$^{-+}$ and I=0.
More recently it has been identified with the X(1835) observed in
J/$\psi\to\eta^{\prime}\pi^+\pi^-\gamma$\cite{beshadron05}.
Other threshold enhancements have been reported by BES-II Collaboration
at the p$\bar\Lambda$ threshold, observed both in J/$\psi\to p\bar\Lambda
K^-$ and $\psi^{\prime}\to p\bar\Lambda K^-$, and by {\small BELLE}
Collaboration\cite{belle} at the p\=p, p$\bar\Lambda$ and
$\Lambda\bar\Lambda$ thresholds.

The hypothesis $(b)$ is supported by the fact that in some model, as
stated in the previous subsection, vector hybrids with 1.9 -- 2.0 GeV mass
and $\sim$ 100 MeV decay width are predicted, and by the fact that OBELIX
experiment did not observe evidence of baryonium type signal in \=np$\to
3\pi^+ 2\pi^-\pi^0$\cite{obelix}.

The $\rho(2150)$ has been observed by GAMS Collaboration\cite{gams} in
$\pi^- p\to\omega\pi^0 n$ and recently by BESII in
$\psi^{\prime}\to\pi^+\pi^-\pi^0$.

Two other vector states, $\omega(1250)$ and $\rho(1250)$ have been
reported in a recent reanalysis of the SND and CMD2 data on
$e^+e^-\to\pi^+\pi^-\pi^0$ and $e^+e^-\to\omega\pi^0$
respectively\cite{komada}. 

Finally the a vector X(1750), with 1753 MeV mass and 122 MeV total
width, observed by FOCUS\cite{focus} in diffractive photoproduction of
K$^+$K$^-$ deserves a clear interpretation.  

\subsubsection{Exclusive channels}

We give here a list of some interesting multihadronic channels, that can be
measured at {\small DAFNE}-2.

\subsubsection*{$e^+e^-\to\pi^+\pi^-\pi^0$}

\begin{figure}[htbp]
\begin{center}
  \includegraphics[width=0.75\textwidth]{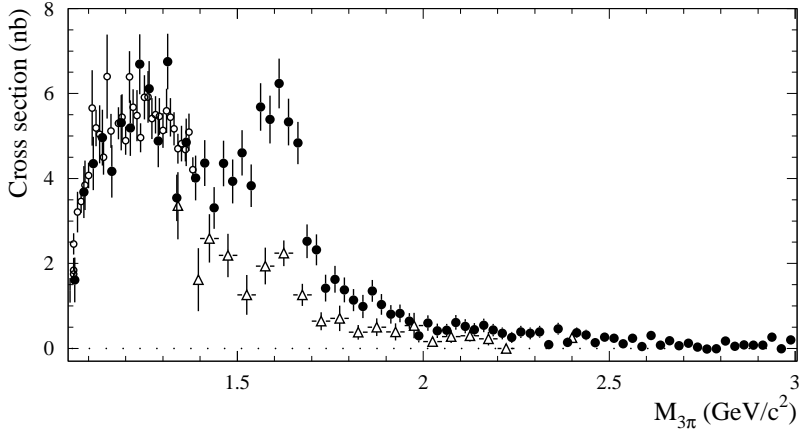}
\vspace{-0.75cm}  
\caption{ Invariant mass spectrum of $\pi^+\pi^-\pi^0$ from
    {\small BABAR} radiative return (full circles) compared with $e^+e^-$ scan
    data from DM2 (triangles) and CMD-2 (open circles).} 
\label{Matteo}
\end{center}
\end{figure}

The recent cross section measurement done by {\small BABAR} is in disagreement with the previous result of
the DM2 experiment at $\sqrt{s}\geq$ 1.3 GeV as shown in Fig.\ref{Matteo}, and this reflects in the
parameters of the two $\omega$ excitations, see Tab.\ref{tab:3pi}.

\begin{table}[htb]
\centering
\begin{tabular}{| c | c | c | c | c|}
\hline
 & \multicolumn{2}{|c|}{SND + DM2} & \multicolumn{2}{|c|}{{\small BABAR}} \\
\hline
 & Mass (MeV) & Width (MeV) & Mass (MeV) & Width (MeV) \\
\hline
$\omega(1420)$ & 1400 $\pm$ 140 & 870 $\pm$ 670 & 1350 $\pm$ 30 & 450 $\pm$
 100 \\
$\omega(1650)$ & 1770 $\pm$ 80  & 490 $\pm$ 240 & 1660 $\pm$
 10 & 230 $\pm$ 35 \\ 
\hline
\end{tabular}
\caption{$\omega(1420)$ and $\omega(1650)$ parameters from the recent
  measurements of $e^+e^-\to\pi^+\pi^-\pi^0$}
\protect\label{tab:3pi}
\end{table}

A new measurement of this final state is needed, also to test the ISR
method versus the energy scan one.

Furthermore, the interest of this final state is increased by the fact that
$\rho\pi$ is one of the preferred decay channels for an isoscalar vector
hybrid.  

\subsubsection*{$e^+e^-\to 4\pi$}

The $e^+e^-\to\pi^+\pi^-\pi^+\pi^-$ has been recently measured by {\small BABAR}
with the ISR method and is in good agreement with the previous
measurements, while the most recent high statistics results on
$e^+e^-\to\pi^+\pi^-\pi^0\pi^0$ are those of SND and CMD2 Collaboration,
but limited to the region $\sqrt{s}\leq$ 1.4 GeV.
Both cross section are described as dominated by the $a_1(1260)\gamma$
intermediate state, however a new measurement with higher statistics, also of
the angular distributions of the decay products\cite{4picharg}, could be
interesting. 

Being $a_1(1260)\gamma$, together with $\rho\pi\pi$ the main decay channels
expected for isovector vector hybrids, the 4$\pi$ channel is the more
promising for the search of such mesons.

\subsubsection*{$e^+e^-\to\pi^+\pi^-\pi^+\pi^-\pi^0$}

Two processes mainly contribute to this final state:
$e^+e^-\to\omega\pi^+\pi^-$, sensitive to the $\omega(1420)$ and
$\omega(1650)$ parameters, and $e^+e^-\to\eta\pi^+\pi^-$, which instead is
sensitive to $\rho(1450)$ and $\rho(1700)$.

Also this final state can be exploited to search for isoscalar vector hybrids
that could decay into $\rho(1450)\pi$. 

\subsubsection*{$e^+e^-\to 6\pi$}

This cross sections has recently been measured by {\small BABAR}\cite{babar6pi}; there
is good agreement with the previous measurements in the $e^+e^-\to
3\pi^+ 3\pi^-$ channel, while in the $e^+e^-\to 2\pi^+ 2\pi^- 2\pi^0$
there is some discrepancy with the DM2 data.

Furthermore these are the ``golden'' channels for the study of the
properties of the $\rho(1900)$. 

\subsubsection*{$e^+e^-\to K^+ K^-$, $K_S K_L$}

These final states can be exploited to extract the $\phi(1680)$ parameters.
The charged one is the final state in which FOCUS has observed the
vector state X(1750) in diffractive photoproduction.

\subsubsection*{$e^+e^-\to K\bar K\pi$, $K\bar K\pi\pi$}

These final states are interesting for the study of the $\phi(1680)$ and
for the search of strange vector hybrids, through the decay chains
$\phi_H\to K^\star K\to K\bar K\pi$, and $\phi_H\to K_1(1400) K\to
K^\star\pi K\to K\bar K\pi\pi$. 

\subsubsection*{$e^+e^-\to \phi f_0(980), \phi\eta, \phi\eta'$}
A combined study of the processes $e^+e^-\to \phi$f$_0(980)$ 
and $e^+e^-\to \phi\eta (\eta')$ with center of mass energies up to $\sim 3$ GeV, 
should help to shed light on the still controversial nature of the $f_0(980)$ scalar 
meson. \\
In fact as shown in Ref.~\cite{simone} it is possible to construct an 
analytic parameterization 
defined in the whole $q^2$-complex plane for a generic 
$\phi M$ transition form factor $F_{\phi M}(q^2)$ 
(where $M$ is any pseudoscalar or scalar 
light meson).\\
The main ingredients of this procedure are:
\begin{enumerate}
\renewcommand{\labelenumi}{\bf \alph{enumi}.}
\item the perturbative QCD counting and helicity rule~\cite{power-law} to describe
      the asymptotic behaviour;
\item data on the annihilation cross section $\sigma(e^+e^-\to\phi M)$
      and a Breit-Wigner parameterization in the resonance region that is from
      the theoretical threshold $(2M_\pi)^2$ up to $\sim (3$ GeV$)^2$;
\item the dispersion relations for the logarithm~\cite{dr-log} to perform
      the analytic continuation, below the threshold $(2M_\pi)^2$, down to $q^2=0$.
\end{enumerate}
The steps {\bf a} and {\bf b} of the procedure outlined above 
are strongly dependent on the 
nature of the meson $M$ under consideration. In fact the power
law asymptotic behaviour counts the hadronic fields in the final
state~\cite{power-law} (step {\bf a}) and, 
by invoking the quark-hadron duality~\cite{duality}, such a behaviour 
is restored also in the resonance region, which
is covered by the data and by the Breit-Wigner parameterization 
(step {\bf b}). Hence the value at 
$q^2=0$ of the transition form factor provided by step {\bf c}, 
is unambiguously linked to the assumed 
quark structure of $M$.\\
It follows that the radiative decay rate $\Gamma(\phi\to M\gamma)$, which is 
proportional to $F_{\phi M}(0)^2$, may be predicted, under different 
hypotheses about the nature of the meson $M$, and then compared with data. 
On this respect, a comparison among the $\phi M$ transition form factors, with 
$M=\eta$, $\eta'$, and f$_0$(980), is sensitive to the relative quark composition 
of these mesons.

\subsubsection{Threshold enhancements}

As stated above the experimental signatures of a baryonium are a
structure in some multihadronic final state and a threshold enhancement in the
baryon-antibaryon cross section.
Several thresholds should be accessible to a machine running up to 2.5 GeV:
$N\bar N$ and $\Lambda\bar\Lambda$ in which such enhancements have been
already observed ~\cite{apple,fenicemh,bbes,belle}, plus $\bar\Sigma^0\Lambda$
and its charge conjugate, and also $\Sigma\bar\Sigma$.

\subsubsection{Statistical considerations}

An energy scan of the region between 1.0 and 2.5 GeV with 15 MeV step, 
corresponding to 100 energy points, can be envisaged.
The less common among the exclusive processes listed above have cross
sections of the order of 1 nb.
Assuming a luminosity of 10$^{32}$ cm$^{-2}$ s$^{-1}$, in one $10^7$ year
data taking, it will be possible, for those final states, to collect about 10000
events per point reaching a statistical accuracy of about 1\%. 

\begin{figure}[htbp]
\begin{center}
  \includegraphics[width=0.5\textwidth]{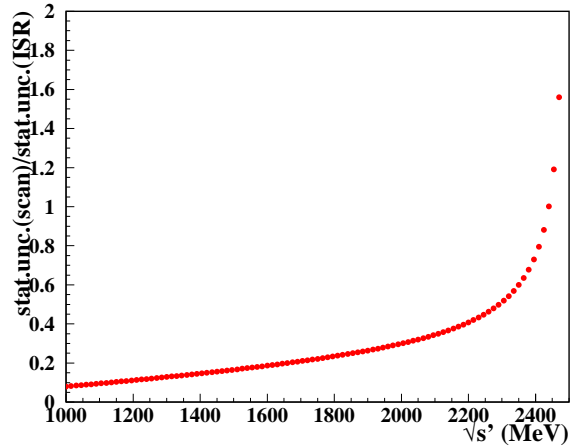} 
  \caption{Ratio of the statistical uncertainties of the
    energy scan method to the ISR
    one ($\sqrt{s^{\prime}}=\sqrt{s}-E_{\gamma}$). }
\label{fig:statratio}
\end{center}
\end{figure}

An alternative method to measure the multihadronic cross sections is the
ISR based one, by running the machine at the maximum $\sqrt{s}=$2.5 GeV.
The ratio of the statistical uncertainty achievable with the energy scan
method (15 MeV step, one year at L=10$^{32}$ cm$^{-2}$ s$^{-1}$), to the
ISR one in similar conditions (15 MeV step and an integrated luminosity of
1 fb$^{-1}$) is reported in Fig.\ref{fig:statratio}. Notice that for most
of the spectrum, the scan method is statistically more convenient.

The main competitors for these measurements are the B-factories that can
cover all the relevant energy region, for many exclusive channels. In any
case as clearly shown in Fig.\ref{fig:impactscan} of Sect.\ref{Hadro},
DAFNE-2 will collect larger statistics than the full {\small BABAR} data sample.

\subsection{\bf Radiative decays}
\label{Radia}

Radiative decays represent another important tool for studying the
structure of the hadrons. {\small DAFNE}-2 can contribute in two respects: first by
continuing the $\phi$ radiative decays program started by {\small KLOE} but also
profiting of the higher center of mass energy available by
looking for radiative decays of excited vector mesons.

The high production rate of $\eta$ and $\eta'$ mesons from the $\phi$ 
expected at DAFNE-2 will
allow the measurement of rare decays and the precise determination 
of kinematical distributions for processes with larger rates, thus 
providing an invaluable test-bed for QCD at low energies. We mention 
in particular the several tests of Chiral Perturbation Theory (ChPT in the
following) \cite{ChPT} that will be discussed in the following.
$\phi$ radiative decays are also an important source of light scalar
mesons: the well established $f_0(980)$, $a_0(980)$ and the questioned
$\sigma(600)$. 
An accurate measurement of the production branching ratio and of the 
mass spectra for the $\phi\to f_0(980)\gamma/a_0(980)\gamma$ decays 
can clarify the controversial exotic nature of the involved scalars.
The high luminosity of DAFNE-2 will provide an unprecedent statistics 
of the $\pi\pi/\eta\pi$ decay channels, already studied at KLOE, and
will open the possibility to search for the $K\overline{K}$ final state.
The existence of the $\sigma(600)$ meson can also be clarified by fitting
the low mass region for the $\pi\pi$ channel.
The search for exotic states can be also performed in the high energy 
option, where the radiative decays of excited vector mesons could
provide an evidence for the existence of hybrids or glueballs.

%==============================================================================
% Radiative decays from vector mesons higher than phi
%==============================================================================

%==============================================================================
% Phi radiative decays
%==============================================================================

\subsubsection{$\phi$ radiative decays}

In the high luminosity option {\small DAFNE}-2 can fulfil the physics program
involving $\phi$ radiative decays already performed at {\small DAFNE} with the 
{\small KLOE} detector. Tab.\ref{BRs} shows the rates of the main $\phi$ radiative
decays. The larger sample of $f_0(980)$ and $a_0(980)$ will
allow to access the $K\overline{K}$ decay channels while the high
intensity beams of tagged $\eta$ and $\eta'$ mesons can be used to
improve the search of rare $\eta$ decays and to provide a real $\eta'$ 
factory.
\begin{table}[ht]
  \centering
  \begin{tabular}{| c | c | c |}
    \hline
    final state & branching ratio & rate (evts/fb$^{-1}$) \\
    \hline
    $\eta\gamma$ & 1.3$\%$ & 3.9$\times 10^7$ \\
    $\pi^0\gamma$ & 1.25$\times 10^{-3}$ &  3.7$\times 10^6$\\
    $\eta^{\prime}\gamma$ & 6.2$\times 10^{-5}$ & 1.9$\times 10^5$ \\
    $\pi\pi\gamma$ & 1.1$\times 10^{-4}$ &  3.0$\times 10^5$\\
    $\eta\pi\gamma$ & 8.3$\times 10^{-5}$&  2.5$\times 10^5$\\
    \hline
  \end{tabular}
  \caption{Rates of the main $\phi$ radiative decays. f$_0$(980)$\gamma$ and
    a$_0$(980)$\gamma$ are the main contributions to the $\pi\pi\gamma$ and 
    $\eta\pi\gamma$ final states.}
  \protect\label{BRs}
\end{table}

%------------------------------------------------------------------------------
% Scalars
%------------------------------------------------------------------------------

\subsubsection*{Light scalar mesons: $f_0(980)$ and $a_0(980)$}

The $f_0(980)$ and $a_0(980)$ mesons are, respectively, the isospin 
singlet and the neutral element of the isospin triplet of the lowest 
mass scalars. Although their experimental evidence dates the beginning 
of the seventies \cite{Obs_f0,Obs_a0} and a lot of effort was spent since then 
to understand their controversial 
nature, the situation is still unclear.
Indeed, there are several theoretical models proposed to explain their 
composition, as ordinary $q\overline{q}$ mesons, 4-quark states or 
$K\overline{K}$ molecules \cite{Tornqvist,Jaffe,Molecule}.

In this context, one of the open questions is the $s$ quark content
of $f_0(980)$ and $a_0(980)$. Indeed, due to their quantum numbers
and mass degeneracy, a common large $s\overline{s}$ contribution is 
an evidence for an exotic nature of these particles. 
This is revealed by an higher coupling of the scalar mesons ($S$) to 
the $K\overline{K}$ final state with respect to $\pi\pi/\eta\pi$.
Using $\phi$ radiative decays, it is possible to extract the 
$K\overline{K}$ couplings also using the most copious decay chains
$\phi\to S\gamma\to\pi\pi\gamma/\eta\pi\gamma$ as already made by 
the VEPP-2M experiments \cite{SNDf0n,CMD2_f0a0,SNDa0n} and, with 
higher precision, by {\small KLOE} \cite{KLOEa0n,KLOEf0n,KLOEf0c}. However, 
this requires a modelling of the process.
With the higher luminosity expected at {\small DAFNE}-2, it is possible to
directly detect the $\phi\to[f_0(980)+a_0(980)]\gamma\to K\overline{K}\gamma$ 
decay chain, thus allowing a direct measurement of the couplings.
Having 50 fb$^{-1}$, the number of expected $K^0\overline{K^0}\gamma$ 
final state is in the range $2\div 8\times 10^3$ while two orders of 
magnitude more are expected for $K^+K^-\gamma$ \cite{AchasovKK}. 
Despite the higher statistics, the last decay channel is expected to 
be overwhelmed by an irreducible background due to 
$\phi\to K^+K^-$ events with final state radiation which is a factor 10
larger than the signal.

In the discussion of the $\gamma\gamma$ physics program (see Sect.\ref{Gamma})
we will show other complementary measurements on scalar meson physics at
{\small DAFNE}-2.  
%------------------------------------------------------------------------------
% Eta
%------------------------------------------------------------------------------

\subsubsection*{$\eta$ physics}

{\small DAFNE} has shown that a $\phi$ factory is actually one of the best 
places to study $\eta$ physics. Indeed precision results for the dynamics 
of $\eta\to 3\pi$ decays as well as upper limits on rare C and CP violating 
decays have been published by {\small KLOE} \cite{KLOEHAD05,KLOEeta3g,KLOEetapipi},
taking benefit from the high statistics available and the clean experimental 
signature characterised by a highly energetic, monochromatic, recoil photon 
and the possibility to maintain background well below the percent level. 
{\small DAFNE}-2 will open the opportunity to study rare and medium rare $\eta$ 
decays with great precision, as far as the integrated luminosity at the 
$\phi$ peak will reach the tens of fb$^{-1}$ domain:
$\sim 2\times 10^9$ $\eta$'s are produced with 50 fb$^{-1}$.
Let us briefly enumerate the channels of higher interest for {\small DAFNE}-2:
%Let us briefly enumerate the channels which can be of larger interest 
%for DAFNE-2:

\begin{enumerate}

\item{$\eta\to\pi^0\gamma\gamma$}\\
This decay's BR has been a puzzle for experimentalists over last 40 years 
or so, with its estimated value ranging from 25\% down to the recent
{\small KLOE} 
preliminary result of $(\,8.4\pm 3.0\,)\times 10^{-5}$ obtained with a 
sample of $68\pm 23$ candidate events in 450 pb$^{-1}$. 
The theoretical interest in this decay resides in offering a unique window 
on pure $p^6$ terms of the Chiral Lagrangian.
%It is of large interest from the theoretical point of view, since it
%offers a unique window on pure $p^6$ terms of the Chiral Lagrangian. 
The amount of events which can be collected 
at {\small DAFNE}-2, using a realistic efficiency extrapolated from the {\small KLOE} result, 
is about 200/fb$^{-1}$ allowing for both a precision measurement of the BR 
and for the first study of the $M_{\gamma\gamma}$ spectrum. Since it is
essentially a measurement based on photon counting (the main background
source being the $\eta\rightarrow\pi^0\pi^0\pi^0$ decay), it can profit from
an improvement of the calorimeter granularity (see sect.\ref{DetAcc}).

\item{$\eta\to\pi\pi$}\\
As for $K_L\to\pi\pi$ the two pion mode for the $\eta$ is CP violating. 
In the Standard Model it is further dynamically suppressed to the 
$O(10^{-27})$ level; possible contributions from the $\theta$ term of 
the QCD Lagrangian may well increase it, but constraints from neutron 
EDM show that this contribution cannot exceed $O(10^{-17})$. In some 
extensions of the SM it can be slightly increased up to $O(10^{-15})$ 
\cite{ShabalinEta}. 
{\small KLOE} has improved the limits on the $\pi^+\pi^-$ mode by an order of 
magnitude w.r.t. previous measurements, settling the upper limit at 
the $10^{-5}$ level. {\small DAFNE}-2 could explore the region down to 
$\approx 10^{-6}$ and could surely also improve the upper limit on
the $\pi^0\pi^0$ final state, which is currently only $3.3\times 10^{-4}$,
although {\small KLOE} has already good handles to refine it.

\item{$\eta\to\mu^+\mu^- (e^+e^-)$ and LF violating modes}\\
While the branching fraction of the $\eta\to\mu^+\mu^-$ decay has been 
measured, even if with large errors, the Standard Model expectations 
for the $e^+e^-$ mode are only at the $10^{-9}$ level, preventing its
observation even at {\small DAFNE}-2. It must be also stressed that QED background 
can be a relevant issue for both these modes. Anyhow, {\small DAFNE}-2 can 
improve the $\mu^+\mu^-$ BR determination, checking the unitarity bound 
($=4.3\times 10^{-6}$). It is obvious that a natural byproduct of the 
$\mu^+\mu^-$ and $e^+e^-$ searches will also result in a reduction on
the upper limit on the lepton flavour violating modes $\eta\to\mu^\pm e^\mp$.
Improvements of the detector particle ID capability w.r.t. {\small KLOE} will be
a good handle for this kind of searches.

\item{Dalitz and double Dalitz decays}\\
The
electromagnetic form factor of pseudoscalar mesons is an important
ingredient in the evaluation of the pseudoscalar pole part of the
light-by-light contribution to the muon anomalous magnetic moment (see
Sec. 2.2.3). Precise information can be extracted by studying Dalitz 
and, mainly, double Dalitz decays of the $\eta$, the latter being not yet 
observed so far. At {\small DAFNE}-2 a number as high as 3000/fb$^{-1}$ are expected 
to be produced, and even with a detection efficiency of few \% one could  
accurately measure the BR and spectrum for these decays. 

\item{$\eta\to\pi^+\pi^-e^+e^-$}\\
The study of this final state is very interesting, because it provides 
a test for possible CP violating mechanisms beyond the Standard Model. 
This can be achieved by studying the asymmetry in the angle between the 
$\pi^+\pi^-$ and $e^+e^-$ planes in the $\eta$ rest frame, which arises 
from the interference between CP-conserving and CP-violating amplitudes. 
The measured BR for this final state implies that 17000 events/fb$^{-1}$ 
of this kind would be produced at {\small DAFNE}-2.

\end{enumerate}

%------------------------------------------------------------------------------
% Eta'
%------------------------------------------------------------------------------

\subsubsection*{$\eta'$ physics}

%1x10 7 etap

{\small DAFNE}-2 would provide a real $\eta'$ factory via $\phi$ radiative decays,
with a production rate of $2\times 10^5$ $\eta'$/fb$^{-1}$. Notice that a
similar production rate can be obtained via
$\gamma\gamma\rightarrow\eta^{\prime}$ at $\sqrt{s}=2.5$ GeV. The
possibility to use both methods to obtain samples of $\eta^{\prime}$ in
completely different background and tagging configurations is to be
considered as very important.
%At the $\phi$ peak the number of $\eta'$ meson which are produced via the 
%magnetic dipole transition $\phi\to\eta'\gamma$ is as high as $2\times 10^5$
%per fb$^-1$ of integrated luminosity. 
Most of $\eta'$ branching fractions can be well measured 
and brought to the same accuracy of the best measured one, namely 
$\eta'\to\eta\pi^+\pi^-$ which is currently known with an error of 3\%. 
Since the error on the $\eta-\eta'$ mixing angle at {\small KLOE} is dominated by 
the knowledge of this BR, one could also try to improve it by measuring
simultaneously {\em all} the main $\eta'$ modes using the tagged recoil 
photon. Anyhow, it must be stressed that for some of the $\eta'$ decays 
the background from $\eta$ and/or kaon decays with same or similar final 
state could be a relevant issue. Apart from the mixing angle determination, 
many of the $\eta'$ final states are of interest in themselves, and can 
provide inputs to the phenomenology of low energy QCD. We will now briefly 
review these final states and their importance.

\begin{enumerate}

\item{$\eta'\to\pi\pi\eta$}\\
The $\pi^+\pi^-\eta$/$\pi^0\pi^0\eta$ modes account for about 44\%/21\% 
of all $\eta'$ decays. Their main interest is in studying the dynamics 
of the three bodies via the Dalitz plot technique. Since there is no 
tree contribution from VMD, the scalar mesons $\sigma(600)$ and $a_0(980)$ 
are believed to be dominant in the imaginary and real part of the matrix 
element amplitude respectively \cite{Fariborz99}. A recent full 
$p^4$ ChPT calculation with higher order resummation via Bethe-Salpeter 
equations has been performed to precisely predict the dynamics of these 
decays \cite{Borasoy}.  
The best experimental results currently available come from GAMS 
\cite{GAMS86} for the $\pi^0\pi^0\eta$ mode (about 6000 events) and from 
the VES collaboration for the $\pi^+\pi^-\eta$ mode \cite{VES_H05}. The 
latter have been obtained with about 20 thousands events in hadronic 
production (diffractive + charge exchange). 
The very abundant and clean sample which can be collected at {\small DAFNE}-2 could
bring the study of these Dalitz plots into a precision era, similarly to
the precise measurement of the $\eta\to 3\pi$ Dalitz plot parameters done
by {\small KLOE}.

\item{$\eta'\to\pi^+\pi^-\gamma$ (including $\rho\gamma$)}\\
As for the corresponding $\eta$ decay mode, this channel is sensitive to
the box anomaly contribution
%which appears in the Wess-Zumino-Witten term 
of the chiral Lagrangian. This term should manifest itself as a deviation 
from simple $\rho$ dominance in the observed dipion invariant mass
spectrum. Since this final state accounts for about 30\% of $\eta'$ decays
the production rate is quite high (60.000 events/fb$^{-1}$), allowing for 
precise fit to the spectrum. The existing measurements 
\cite{GAMS91,CryB97,L398} are based on few thousand events, 
and give sometimes opposite conclusions on the presence of the box anomaly 
term. 
As for the corresponding $\eta$ decay chain, C-parity violation of this
process is very interesting and can be tested by means of the charge 
asymmetry. 
%C-parity violation is of course another very
%interesting issue in this decay (as for the corresponding $\eta$ one) and
%can be tested by studying the charge asymmetry. 

\item{$\eta'\to\omega\gamma$}\\
This decay rate is quite small (about 3\%) and poorly known (10\% accuracy). 
However it is quite interesting since it can be related, together with the 
BR's of $\eta'\to\rho\gamma$, $\phi\to\eta'\gamma$ and with the $\eta'$ two 
photon width, to the gluonic content of the $\eta'$ \cite{Kou}. 
The search in the chain $\phi\to\eta'\gamma$ with $\eta'\to\omega\gamma$
and $\omega\to\pi^+\pi^-\pi^0$ will provide a clean signature due to the 
two almost monochromatic photons and the sharp $\omega$ mass peak. A
measurement at the level of $\leq 3\%$ can be reached, thus over constraining
the determination of the $\eta'$ gluonic content.
%A knowledge of this BR at say,
%3\% (well within the statistical capabilities of DAFNE-2) could be helpful
%in over constraining the gluonic content, whose value is essentially 
%determined by the measurements of $\eta'\to\rho\gamma$ and 
%$\phi\to\eta'\gamma$.  
 
\item{$\eta'\to\pi^+\pi^-\pi^0$}\\
This mode, as the corresponding $\eta$ decay, is due to the isospin
violating part of the strong Lagrangian and is in principle a source of
precise information about quark mass differences. Moreover the ratio of
this BR to the corresponding isospin conserving $\eta'\to\eta\pi\pi$ can
be related to the $\pi^0-\eta$ mixing \cite{GrossTreimanWilczek79}. From
the experimental point of view only a very weak upper limit exists ($<5\%$),
while theoretical expectations range in the $10^{-3}$ domain. 

\item{Dalitz decays}\\
As already mentioned for the $\eta$, Dalitz and double Dalitz decays 
can give precious information about the pseudoscalar e.m. form factors, 
which are key ingredients in evaluating the light-by-light scattering 
part of the muon anomalous magnetic moment. While double Dalitz decays 
seem to be outside the capabilities of {\small DAFNE}-2, with only few events 
produced per 10 fb$^{-1}$ of integrated luminosity, the single Dalitz modes, 
whose expected rates are two order of magnitudes larger and for which today 
only an upper limit exists, could be measured for the first time.

\end{enumerate}

%------------------------------------------------------------------------------
% Competitors
%------------------------------------------------------------------------------

\subsubsection*{Competing facilities}

In the panorama of the experimental programs foreseen in the next 
years, there are no strong competitors in the study of light scalar 
meson produced through $\phi$ radiative decays.
The primary goal of the experiments at VEPP-2000 ~\cite{VEPPtot} is 
the coverage of the $1.4\div2.0$ GeV region in the hadronic cross section 
measurement.

The $f_0(980)$ and $a_0(980)$ scalar mesons are currently studied by 
many experiments and for this certainly will continue in the following years.
Since the characteristics of the production mechanism is also 
sensitive to their 
nature, it will be still interesting in the future their study 
through $\phi$ radiative decay.

Concerning the $\eta/\eta'$ physics, the strongest competitors are the
experiments of the MAMI \cite{MAMI} and COSY \cite{COSY} facilities. 
After the end of the running period at the BNL laboratory, the Crystal
Ball \cite{CBALLatMAMI} detector was moved to Mainz in order to study 
$\eta(\eta')$ mesons produced in the reaction at threshold 
$\gamma p\to p \eta(\eta')$. The just concluded MAMI-B run provided a 
sample of $3\times 10^7$ $\eta$s and the foreseen upgrade to MAMI-C 
will open up the possibility of producing $\eta'$ mesons too. While the 
detector is well suited for the study of fully neutral decays, the simple 
tracking system and the absence of the magnetic field do not allow accurate 
measurements for decays involving charged particles.
The start of the data taking for the WASA experiment at COSY \cite{WASAatCOSY} 
is foreseen in January 2007. Here the $\eta$ and $\eta'$ mesons are 
produced in the $pp\to pp\eta(\eta')$ reactions and are precisely 
tagged by a forward spectrometer through the measurement of the $pp$ 
missing mass ($4\div 8$ MeV/FWHM). The central detector, made by a CsI 
electromagnetic calorimeter and a straw tubes drift chamber in a magnetic 
field, is optimised for events involving electrons and photons, with small 
particle ID capabilities.
Both Crystal Ball and WASA have a very high $\eta/\eta'$ production flux
and a precise meson tagging but they need a selective trigger and
have a large - and not perfectly known - level of multihadronic background.
Moreover, the running time which the experiments dedicates to these 
kind of measurements is limited and their reduced particle ID detector 
capabilities limit the number of possible measurements.
As already demonstrated by {\small DAFNE}, a $\phi$ factory offers a much cleaner
environment, allowing the usage of an unbiased trigger.

\subsubsection{Radiative decays in the high energy option}

Radiative decays of excited vector mesons can provide a tool to
separate $q\bar q$ states from hybrids according to the model of
Ref.~\cite{Close}. 
In the framework of the quark model, large partial widths are predicted for
some decay channels, that could be measured in an $e^+e^-$ machine running
at 1.0$\leq\sqrt{s}\leq$2.5 GeV.  
~
In particular the decays of $\rho(1450)\to f_2(1270)\gamma$, $\rho(1700)\to
f_1(1285)\gamma$, $\omega(1420)\to a_2(1320)\gamma$ and $\omega(1650)\to
a_1(1260)\gamma$ can be exploited, since their decay widths are predicted of
the order of 500 -- 1000 keV in the hypothesis that the mesons are $q\bar
q$, while the decay of the $\rho_H$ and $\omega_H$ hybrids to the same
particles are strongly suppressed.
In a similar way the structure of the $\phi(1680)$ can be tested through 
the decay $\phi(1680)\to f^{\prime}_2(1525)\gamma$ ($\Gamma\sim$ 200 keV),
while $\phi_H\to f^{\prime}_2(1525)\gamma$ is suppressed.
Also the angular distributions of these decays are calculated in
Ref.~\cite{Close} and can be measured, thus providing a further test of 
the structure of the vector mesons. 
Furthermore the decays $\omega(1650)\to a_0(1450)\gamma$ and
$\rho(1700)\to a_0(1450)\gamma$ can provide information on the properties
of the $a_0(1450)$, whose existence is questioned.

Radiative decays of the $\rho(1700)$ can also shed some light in the sector 
of the $f_0(1370)$, $f_0(1500)$, and $f_0(1710)$.
Two isoscalar scalar are expected in that mass region, so that the excess
can be explained with the presence of a scalar glueball that mixes with a
$(u\bar u+d\bar d)$ and a $s\bar s$ scalar meson.
In particular the ratio $\Gamma(\rho(1700)\to
f_0(1370)\gamma)$/$\Gamma(\rho(1700)\to f_0(1500)\gamma)$ is very sensitive 
to the mixing scheme, $i.e.$ the glueball is the lightest, the middle or
the heaviest of the three states.

The final states in which the above decays  can be studied are then
$\pi^+\pi^-\gamma$, $4 \pi\gamma$, and $\eta\pi\pi\gamma$ for the $\rho$
mesons, $\pi^+\pi^-\pi^0\gamma$ for the $\omega$ recurrences, in addition
to $\eta\pi^0\gamma$ and $K\bar K\gamma$ for $a_0(1450)\gamma$, and $K\bar
K\gamma$ for the $\phi(1680)$ decays. 
The corresponding cross sections should range from $\sim 10$ to few
hundreds pb, then in one year scan with 15 MeV energy step at 10$^{32}$
cm$^{-2}$ s$^{-1}$ luminosity, from $\sim$ 50 to 1000 events per energy
point are expected.

\subsection{\bf $\gamma\gamma$ physics}
\label{Gamma}

\def\g{\gamma}
\def\epiu{e^+}
\def\emeno{e^-}
\def\pipiu{\pi^+}
\def\pimeno{\pi^-}
\def\pizero{\pi^0}
\def\KL{K_L}
\def\KS{K_S}
\def\gg{\gamma\gamma}
\def\ee{\epiu\emeno}
\def\pipi{\pi\pi}
\def\pippim{\pipiu\pimeno}
\def\pizz{\pizero\pizero}

\subsubsection{Introduction}
\begin{figure}[htb]  
\begin{center}  
\epsfig{file=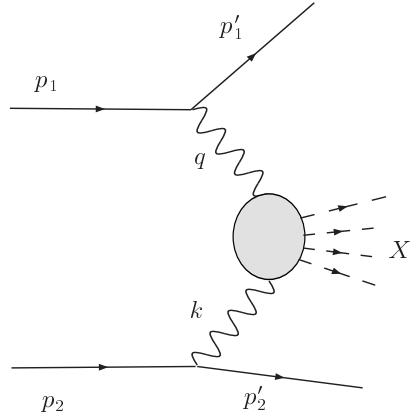,width=6.cm,height=5.6cm}
\caption{
Two-photon particle production in a $\ee$ collider.
}
\label{f:kine}  
\end{center} 
\end{figure}

The term ``$\gg$ physics" (or `two-photon physics") stands for the 
study of the reaction (see Fig. \ref{f:kine})
$$
\ee\,\to\,\ee \,\g^*\g^*\,\,\to\,\ee \,+\, X
$$
where $X$ is some arbitrary final state allowed by conservations laws. 
These processes, even though of ${\cal O} (\alpha^4)$, show a logarithmic 
dependence from the energy $E$ of the colliding beams that reflects in a 
not negligible cross section. It turns out that for $E$ greater than a 
few GeV the $\gg$ processes dominate with respect to the corresponding 
annihilation processes.

For quasi-real photons the number of produced events can be estimated from 
the expression:
\begin{equation}
N \,=\, L_{ee}\,\int\,{\rm d}W_{\gg}\,\frac{{\rm d}L}{{\rm d}W_{\gg}}\,
\sigma(\gg \to X)
\end{equation}
where $L_{ee}$ is the integrated luminosity, $W_{\gamma\gamma}$ is the photon-photon 
center of mass energy ($W_{\gg} = M_X$), ${\rm d}L$/${\rm d}W_{\gg}$ the 
photon-photon flux (in MeV$^{-1}$) and $\sigma$ is the cross section into a given final 
state. By knowing the fluxes of virtual photons emitted by the two 
colliding leptons, from the study of $\ee \to \ee \,+\, X$ one can really 
extract information on the process $\gg \,\to\, X$.

From the point of view of hadronic physics, photon-photon scattering \cite{penn}
complements the investigations of all the states which are directly coupled to one 
photon, i.e. states for which $J^{PC} = 1^{- -}$ and which proceed through the usual 
annihilation process. Indeed since the two-photon state is a $C = + 1$ state 
and the value $J = 1$ is excluded (Landau-Yang theorem), photon-photon scattering 
gives direct access to the study of states with $J^{PC} = 0^{\pm +},\,2^{\pm +}$.

The cross section $\sigma (\gg \to X)$ was studied over the past decades in 
$\ee$ colliders operated at c.m. energies of about 10 GeV or more. Concerning
the low-energy region $m_\pi \leq W_{\gg} \leq m_{f_0}$ 
the existing measurements are affected by two clear deficiencies:
\begin{itemize}
\item the large statistical and systematic uncertainties due to the 
relatively small data samples and relatively large background 
contributions;
\item the very small detection efficiency and particle identification 
ambiguities for low-mass hadronic systems.
\end{itemize}
Due to the combination of high luminosity and favourable kinematical 
conditions, {\small DAFNE}, equipped with the large multi-particle detector 
{\small KLOE}, offers the opportunity for new precision measurements of low-mass 
hadronic systems with high statistics and considerably smaller 
systematic errors.

\begin{figure}[htb]
\begin{center}  
\includegraphics[height=9cm]{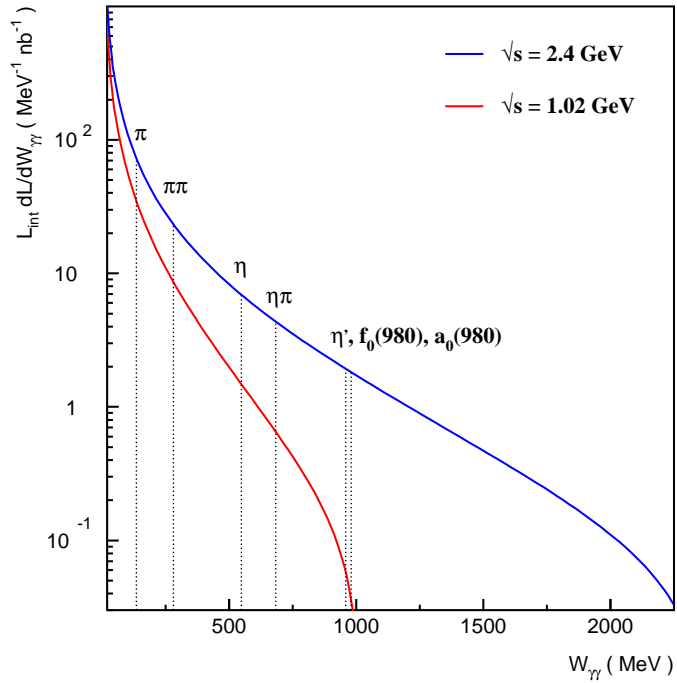}
\caption{
Photon-photon flux at DA$\phi$NE as function of $W_{\gg}$ for 
two values of $\sqrt{s}$ and an integrated luminosity machine 
$L_{int} = 1$ fb$^{-1}$.
}
\label{f:lum}
\end{center}
\end{figure}

This can be visualised by looking at the luminosity function given 
in Fig. \ref{f:lum}, showing some of the processes that can be 
investigated at $\sqrt{s} = 1.02$ GeV and the processes that will 
be available increasing the energy of the machine up to 
$\sqrt{s} = 2.5$ GeV.
%\begin{figure}[htb]
%%\begin{center}  
%\centering
%\epsfig{file=lumgg.ps,width=9cm}
%%\includegraphics{gg.eps}
%%\vspace{-1.cm}
%\caption{\footnotesize 
%Photon-photon flux at {\small DAFNE} as function of $W_{\gg}$ for 
%two values of $\sqrt{s}$. Two different cases for luminosity 
%of the machine are considered: $L_{ee} = 8 \times 10^{32}$ 
%cm$^{-2}$ s$^{-1}$ at $\sqrt{s} = 1.02$ GeV; $L_{ee} = 10^{32}$ 
%cm$^{-2}$ s$^{-1}$  at $\sqrt{s} = 2.4$ GeV.
%}
%\label{f:lum}  
%%\end{center} 
%\end{figure}

In order to isolate experimentally these processes and suppress systematic 
errors arising from non $\gg$-interactions, it is necessary to 
equip {\small KLOE} with (at least one) tagging systems to detect the scattered 
electrons.

A feasibility study for high-precision measurements of $\gg$-reactions 
leading to hadrons at {\small DAFNE} was carried out more than ten years ago 
\cite{alex}. The physics program and the characteristics of the tagging 
systems were investigated in detail. Although the results of this study 
are still valid, in the following we will re-consider some of the physics 
topics in light of the developments occurred since then.

\subsubsection{The process $\gg \to \pizz$: the $\sigma$ case}
One of the first attempts to describe nucleon-pion interactions within a
spontaneously broken $SU(2)_L\otimes SU(2)_R$ theory was the linear sigma 
model by Gell-Mann and L\'{e}vy~\cite{gell}. 

In this theory an `artificial' $\sigma$ field with the quantum numbers of vacuum is required, by chiral invariance, to couple to pions and nucleons, suggesting the existence of a $0^{++}$ particle to be looked for.
The natural process where a $\sigma$ contribution is expected to be important 
is the $\pi\pi\to\pi\pi$ scattering.
Experimental studies have never provided 
over the years a clear signal for it and the assessment of $\sigma$ in this channel
has become more and more controversial. 

Extending the linear sigma model from $SU(2)_L\otimes SU(2)_R$
to $SU(3)_L\otimes SU(3)_R$, to include the strange sector, 
a generalized sigma model 
with 9 scalars and 9 pseudoscalars can be built, see e.g.~\cite{schec}. 
Diverse solutions of this kind have been explored in the literature but
none of them has proved to be really effective at explaining data.

The only successful
approach to build a theory of pions at low energies is that proposed by
Callan-Coleman-Wess-Zumino (CCWZ) 
where the chiral symmetry is realized non-linearly and the
$\sigma$ field is removed from the spectrum. This construction 
is at the basis of modern Chiral Perturbation Theory (ChPT)~\cite{ecker}, 
the standard effective approach
to describe the interactions of the QCD pseudo-Goldstones at low energies.

Anyway there are persistent experimental indications of some structure in 
low energy $\pi\pi$ collisions. Many explanations of
such isoscalar enhancement have been provided during the years.
One of the most interesting results has been proposed recently in~\cite{leut}. 
It has been shown that the $\pi\pi$ scattering amplitude 
contains indeed a pole with the quantum numbers of vacuum, 
which we will call the $\sigma$  by analogy with the 
old linear $\sigma$ field, with a mass
of $M_\sigma=441^{+16}_{-8}$~MeV and a 
width $\Gamma_\sigma=544^{+25}_{-18}$~MeV.
This is also in reasonably good agreement, as for the mass predicted, 
with the observations made by the
E791 Collaboration at Fermilab~\cite{e791}:
in the $D\to 3\pi$ Dalitz plot analysis, E791 finds that almost
$46\%$ of the decay width proceeds through $D\to\sigma\pi$ with a 
$M_\sigma=478\pm 23 \pm 17$~MeV and
$\Gamma_\sigma=324\pm 40\pm 21$~MeV. 
BES~\cite{bessigma} has looked for $\sigma$ in $J/\psi\to\omega\pi^+\pi^-$ 
giving a mass value of $M_\sigma=541\pm 39$~MeV and a width of
$\Gamma_\sigma=252\pm 42$~MeV. For a summary of the experimental situation 
see~\cite{PDG04}.
%Above all, is this a real particle associated 
%to some field in some effective Lagrangian (an exotic states Lagrangian 
%for example) or is it just a pion rescattering effect?

The problem of assessing the existence and the nature of this state is not
confined to low energy spectroscopy. Just to mention a possible relevant 
physical scenario in which $\sigma$ could play a role,
consider the contamination of 
$B\to\sigma\pi$ in $B\to\rho\pi$ decays (possible because of the large 
$\sigma$ width). This could sensibly affect the isospin analysis for 
the CKM-$\alpha$ angle extraction~\cite{bsig}, as it could be tested if
the experimental 
precision on this measurement would grow. Recent studies of the
$\gamma$ angle through a Dalitz analysis of neutral $D$ decays, 
need the presence of a $\sigma$ resonance in the fits~\cite{babsig}.

Here we want to highlight the possibility that a $\sigma$ {\it resonance}
could be found (or disproved) in $e^+e^-$ collisions at DAFNE and DAFNE-2. We
consider 2 experimental options: 
a run at a center of mass energy of 1~GeV, a region where the $\phi$ 
backgrounds are considerably diminished, and better, a run at a center of mass
energy of 2.5 GeV in the high energy option of DAFNE-2. In the second option
the photon-photon center of mass energy $W_{\gamma\gamma}$ range can be
considerably extended as discussed in the following (see Sect.2.5.6). 
We consider in particular
the $e^+e^-\to e^+ e^-\pi^0\pi^0$, $\gamma$-fusion channel.
Consider that the $\gamma\gamma\to\pi^+\pi^-$ reaction
in this energy region is dominated by the Born term,
and is also characterized
by a large background given by $\gamma\gamma\to\mu^+\mu^-$.
From this point of view the $\pi^0\pi^0$ final state
provides the cleanest environment
where to look for a signal from the $\sigma$ meson.
Moreover if an isoscalar resonance is found in $\gamma\gamma\to\pi\pi$ data, 
this would further underscore the 4-quark 
hypothesis, since the sigma and the other sub-GeV scalar particles 
can hardly be explained as quark-antiquark states. 

\begin{figure}[htb]
\begin{center}  
\includegraphics[height=9.5cm]{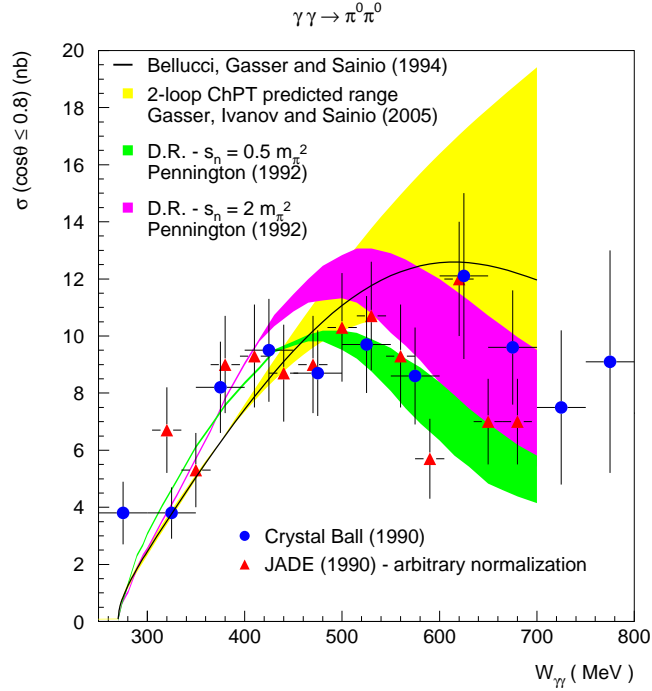}
\caption{Comparison of all the present data from Crystal Ball and JADE
  (arbitrarily normalized to the Crystal Ball data) to the 
  predictions based ChPT \cite{sainio} (solid line and yellow band)
  and on dispersion relation tecniques \cite{Penni1} (green and magenta bands). }
\label{Penni}
\end{center}
\end{figure}

On the theoretical side the process $e^+e^-\to e^+ e^-\pi^0\pi^0$ has been
considered in several papers.Some of these results are summarised in 
Fig.\ref{Penni}.
This includes the results of a 2-loop  ChPT calculation \cite{sainio,priv} for 
the $\gamma\gamma\to\pi^0\pi^0$ channel 
in the region of photon-photon c.o.m. energy from
about $2 m_\pi$ up to $700$~MeV, and the results of an approach based on 
dispersive techniques \cite{Penni1}.
These calculations are compared to the only available data, from Crystal Ball 
\cite{cball} and from JADE \cite{jade}, the 
latter having been rescaled to the 
normalization of the former. The large uncertainty in these data are such 
that no conclusion can be drawn on the agreement with either of the 
theoretical approaches, nor on the possible existence of a resonance-like 
structure in the region around 400-500 MeV as discussed in Ref.
\cite{workinprog} 
where a Breit-Wigner parametrization was used to model the departure from 
ChPT of a resonant isospin=0 contribution to the $\gamma\gamma\to\pi^0\pi^0$ 
amplitude. The possibility of a destructive interference between the isospin 0 
and 2 amplitudes, with a suppression of the $\sigma$ signal, was furhtermore 
discussed in Refs.\cite{Penni1,Penni2}.
 It is therefore clear that new data in the 
$\gamma\gamma\to\pi^0\pi^0$ channel are essential in order to see progress 
in this area. The size of the difference between the theoretical curves in 
Fig.\ref{Penni} and the size of the respective uncertainty bands, provide 
ambitious but useful benchmarks for the accuracy of new experimental 
measurements of these processes.
  
On the experimental side we notice first that an integrated luminosity of 
1 fb$^{-1}$ at $\sqrt{s}$=2.5 GeV allows in principle 
to reduce to about 2\% the uncertainty on the experimental
points in the 400 - 500 MeV region (50 MeV wide bins) 
shown in Fig.\ref{Penni} 
if the selection efficiency is 1 and 
no background contributes to the uncertainty. A slightly worse uncertainty 
can be obtained running at 1 GeV. However we notice that 
the $W_{\gamma\gamma}$ region 
where the effect of the $\sigma$ should be more evident
is affected by several backgrounds, such as
$e^+e^-\to\omega\pi^0\to\pi^0\pi^0\gamma$ with one lost photon,
that requires a crossed analysis 
of several distributions; the experimental 
resolution on $W_{\gamma\gamma}$ has to be considered also for the comparison 
between data and theoretical predictions. 
This certainly calls for a more selective analysis of data
making use of forward detectors to tag electrons; the details can be found 
in~\cite{workinprog} and are also discussed in Sect 2.5.6.

We believe that DAFNE-2 has the concrete opportunity to discriminate between the curves shown in Fig.\ref{Penni} and possibly find (or disprove) a 
resonant $\sigma$ in the cleanest possible channel. This can be done both in a
dedicated run at 1 GeV center of mass or, even better by running at the
maximum energy of 2.5 GeV to explore a larger $W_{\gamma\gamma}$ range.

\subsubsection{The two-photon widths of f$_0(980)$ and
  a$_0(980)$} 
Extending the measurement of $\gamma\gamma\rightarrow\pi\pi$ and
$\gamma\gamma\rightarrow\eta\pi$ to the $W_{\gamma\gamma}$ region around 1
GeV, the f$_0(980)$ and
a$_0(980)$
$\gamma\gamma$ widths can also be measured. 
This measurement is possible by running at the
maximum attainable centre of mass energy of 2.5 GeV, in order to maximise
the effective $\gamma\gamma$ luminosity in the 1 GeV region (see Fig.\ref{f:lum}).
In both cases a peak in the $W_{\gamma\gamma}$ dependence of 
the $\gamma\gamma\rightarrow\pi\pi (\eta\pi)$ cross section around the meson
mass allows to extract the $\gamma\gamma$ width.
\par
The $\gamma\gamma$ widths of f$_0(980)$ and a$_0(980)$ are rather poorly
known (relative uncertainties about 30\% see Ref.\cite{PDG04}). On
the other hand, due to the dependence on the fourth power of the constituent
charges their values are strongly related to the inner quark
structure. For a complete discussion of this issue see Ref.\cite{BP}. 

\subsubsection{The two-photon widths of the pseudoscalar mesons}
The topic of the mixing of the pseudoscalar (PS) mesons holds a 
central role in hadronic physics. In particular, $\eta$-$\eta'$ 
mixing has been actively investigated both from the theoretical and 
phenomenological side (for a review on the subject see Ref. 
\cite{feld02}).

Mixing can be described in two different basis:
the octet-singlet basis
with mixing angle $\theta$, and the quark-flavour basis
with mixing angle $\phi = \theta - \tan^{-1}(\sqrt{2})$.

In addition to the state mixing, the phenomenological analysis of decay or 
scattering processes involves also the weak decay constants defined by 
($P \equiv \eta, \eta'$)
$$
< 0 | A_{\mu}^k | P (q) > \,=\, i f_P^k q_{\mu} \qquad \qquad (k = 8,0; q,s)
$$
where $A_{\mu}^k$ are the neutral axial-vector currents. In the past it has 
frequently been assumed that the constants in the $\{\eta_8,\eta_0\}$ basis follow 
the same pattern of state mixing and depend on two parameters $f_8$ and $f_0$.
Recently, a theoretical investigation in the framework of ChPT \cite{kais} and a 
phenomenological analysis \cite{feld98a} have clearly shown that a correct 
treatment of the $\eta$-$\eta'$ system requires two mixing angles $\theta_8$ and 
$\theta_0$, which, as a consequence of flavour symmetry breaking, differ 
considerably. In principle, this more general mixing scheme should also apply 
to decay constants in the $\{\eta_q,\eta_s\}$ basis (where the constants
$f_q$ and $f_s$ are introduced),
but analysis \cite{feld98a} yields, practically, the same value for 
$\phi_q$ and $\phi_s$. This result gives support to the assumption
according to which
one mixing angle $\phi$ only is required 
to describe the decay constants mixing in the quark-flavour 
basis.

The value of the angle $\phi$ can be inferred from the analysis of many processes 
involving the $\eta$ and $\eta'$ mesons. This analysis has been performed in Ref. 
\cite{feld98b} and yields a weighted average 
$\bar{\phi} \,=\, (39.3 \pm\ 1.0)^\circ $
to be compared with the theoretical value (to first order in flavour symmetry breaking) 
$\phi_{\rm th} = 42.4^\circ$. Another, more recent analysis~\cite{rafel} has derived values
for the two angles: $\phi_q=(39.3\pm1.3)^{\circ}$ and $\phi_s=(41.4\pm1.4)^{\circ}$ 

The decay constants 
can be separately extracted from the two-photon decays of the $\eta$ and $\eta'$. 
By using a phenomenological estimate 
for $\phi$ and the experimental values \cite{PDG04}
$$
\Gamma (\eta \to \gg) \,=\, 0.510 \,\pm\, 0.026 \;{\rm keV} \qquad 
\Gamma (\eta' \to \gg) \,=\, 4.29 \,\pm\, 0.15 \;{\rm keV}
$$
one obtains \cite{feld98a}:
\begin{equation}
\frac{f_q}{f_\pi} \,=\, 1.07 \,\pm\, 0.04 \qquad 
\frac{f_s}{f_\pi} \,=\, 1.41 \,\pm\, 0.11
\end{equation}
where $f_\pi$ is the pion decay constants ($f_\pi$ = 131 MeV). For these decay 
constants the theoretical estimates to first order in flavour symmetry breaking 
are:
$$
f_q \,=\, f_\pi \qquad \qquad f_s \,=\, \sqrt{f_K^2 \,-\, f_\pi^2} 
\,=\, 1.41\,f_\pi\;.
$$
We see that $f_q/f_\pi$ is more than one standard deviation away from its 
theoretical estimate, while, even if its central value agrees perfectly with 
the theoretical prediction, the constants $f_s$ is not well determined. This 
situation is far from being satisfactory and calls for more precise measurements 
of the two-photon width of the $\eta$ and $\eta'$ mesons. Moreover notice
that even the $\pi^0$ two-photon width is poorly known (relative
uncertainty of $\sim 8\%$) and its determination can be
improved at DAFNE-2. Given the small value 
of these widths, the only way to pursue this experimental program is the study 
of meson formation in $\gg$ reactions. 
In Tab.\ref{ggtab} we report the estimates for the total production rate in
the process $e^+e^-\rightarrow e^+e^-P$ with $P$ a pseudoscalar meson. 
\begin{table}[ht]
  \centering
  \begin{tabular}{| c | c | c | c |}
    \hline
    $\sqrt{s}$ (GeV) & $\pi^0$ & $\eta$ & $\eta'$ \\
    \hline
    1.02 & 4.1$\times 10^5$ & 1.2$\times 10^5$ & 1.9$\times 10^4$ \\
    2.4 & 7.3$\times 10^5$ & 3.7$\times 10^5$ & 3.6$\times 10^5$ \\
    \hline
  \end{tabular}
  \caption{$e^+e^-\rightarrow e^+e^-P$ total rate for an integrated
    luminosity of 1 fb$^{-1}$ at two different center of mass energies. No
    tag efficiency is included in the rate calculation.}
  \protect\label{ggtab}
\end{table}

\subsubsection{Meson transition form factors}\label{sec:ffact} 
The study of the process $\ee \to\ee + PS$ when one of the final 
leptons is scattered at large angle gives access to the process 
$\gg^* \to PS$, i.e. with one off-shell photon. The amplitude 
of this process is given by (see Fig. \ref{f:kine})
\begin{equation}
T_{\mu\nu} \,=\, i \epsilon_{\mu\nu\rho\sigma}\,k^\rho\,q^\sigma\,
F_{P\gg^*} (Q^2)\,,
\label{definiz}
\end{equation}
where $F_{P\gg^*}$ is the photon-meson transition form factor\footnote{We remark that
the same pion form factor, but in the time-like region, intervenes 
in the so called Dalitz decay $\pi^0 \to \ee\g$ (see 
Ref. \cite {kampf}).}. Here 
we assumed $k^2 = 0$ and, by neglecting the electron mass, we defined
$$Q^2 = - q^2 =2E_1E'_1(1-\cos\theta_1) \neq 0$$
where $E_1$, $E'_1$ are the energies of the initial and final lepton,
respectively, and $\theta_1$ is the scattering angle.
From equation\ref{definiz} one obtains:
\begin{equation}
\Gamma (P \to \gg^*) \,=\, \frac{\pi \alpha^2}{4}\,M_P^3\,
F_{P\gg^*}^2 (Q^2)\,.
\end{equation}

There has been a considerable effort to predict and measure these form 
factors. In the framework of pQCD the leading order prediction for the 
asymptotic behaviour is:
$$
\lim_{Q^2 \to \infty}\,Q^2\,F_{P\gg^*}(Q^2) \,=\, \sqrt{2} f_P\,.
$$
Instead, from the axial anomaly in the chiral limit of QCD it is possible 
to deduce the behaviour of these form factors in the limit $Q^2 \to 0$. 
For $\pi^0$ and $\eta$ one has:
$$
\lim_{Q^2 \to 0}\,F_{P\gg^*}(Q^2) \,=\, \frac{1}{2 \sqrt{2} \pi^2}\,
\frac{1}{f_P}\,,
$$
to leading order in $m_u^2/M_P^2$ and $m_d^2/M_P^2$, where $m_u$ and $m_d$ 
are the masses of the $u$ and $d$ quarks. Due to the large mass of the quark 
$s$ this result does not hold for the $\eta'$. Furthermore, the authors of 
Ref. \cite{brods} proposed a simple-pole formula connecting these two regimes:
$$
F_{P\gg^*}(Q^2) \,=\, \frac{1}{2 \sqrt{2} \pi^2 f_P}\,
\frac{1}{1 + Q^2/\Lambda_P^2}
$$
with $\Lambda_P^2 = 4 \pi^2 f_P^2$. 

\begin{figure}[htb]
\begin{center}  
\includegraphics[width=9.cm]{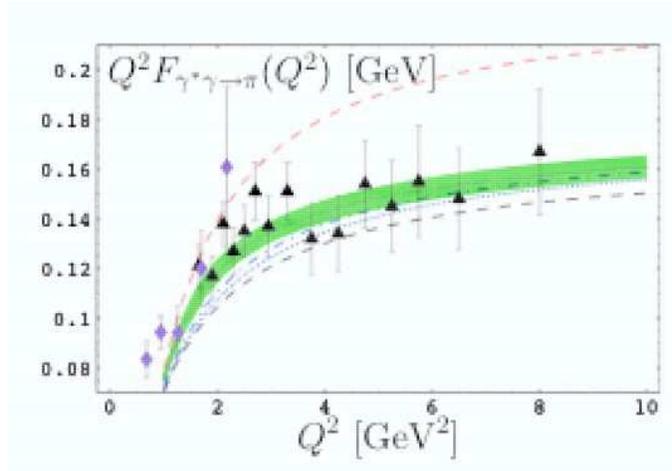}
\caption{Pion transition form factor in comparison with 
CELLO (diamonds) and CLEO data (triangles). The curves refer to 
different theoretical models (figure taken from Ref. \cite{baku}).
}
\label{f:ffact}
\end{center} 
\end{figure}

From the experimental side these functions can be obtained from the
measurements of the differential rates $d\sigma(e^+e^-\rightarrow
e^+e^-+P)/dQ^2$ when one of the virtual photons is emitted at small angle
(i.e. is nearly real), while the other is tagged by detecting one of the
final leptons emerging at a finite angle respect to its original flight direction.
This kind of study has been performed in the past by CELLO \cite{cello} 
and CLEO \cite{cleo}, and the $Q^2$ evolution of the form factors turned out to be consistent 
with theoretical expectations (see Fig. \ref{f:ffact}). 
The $Q^2$ region covered by the whole dataset extends from 0.5 to 8 GeV$^2$. 
These data were also used to extract the slope $a_{\pi}$ of the pion form factor 
$F_{P\gg^*}(Q^2)$. For example, for the $\pi^0$ it turns out \cite{kampf}
\begin{eqnarray}
a_\pi \!\!&=&\!\! 0.0326 \,\pm\, 0.0037 \qquad ({\rm CELLO}) \\
a_\pi \!\!&=&\!\! 0.0303 \,\pm\, 0.0017 \qquad ({\rm CLEO})\;.
\end{eqnarray}
However, these extrapolations are model dependent and a direct and accurate 
determination would be important also in light of the role played by this 
parameter in the determination of the hadronic light-by-light scattering 
contribution to the muon anomalous magnetic moment (see Sect.2.2.3). 

In principle, the measurement of the slope parameter can be performed at {\small DAFNE} by 
implementing the {\small KLOE} detector with a somewhat large angle tagging system (see Ref. 
\cite{alex}).

\subsubsection{Experimental considerations.}
Measurements related to $\gg$ physics have been performed in previous experiments 
with or without tagging the two-photon events by detection of the scattered electrons.
 
Tagging can be performed on one side or on both sides (single or double tag mode), 
allowing almost unambiguous identification of the events coming from the 
$\gg$-interactions. Unfortunately, that comes at the price of a significant yield reduction.
% estimated from previous experiences to be a factor 5 to 10 for each detected electron.
Moreover, the electrons are detected in specifics angular and energy ranges, 
producing a distortion of the invariant mass spectrum of the $\gg$ system that 
can be effectively reconstructed. Thus, the actual needs for a tagging system have 
to be carefully evaluated.

\subsubsection*{Why tagging is needed?} 
The measurement of the $\gg\to\pipi$ cross sections and of the pseudoscalar mesons 
radiative widths have to be regarded as second generation experiments. Lower systematic 
errors are therefore required together with high statistics, calling for a
strong background reduction.

The main source of background comes from annihilation processes, the worst situation 
represented by a machine working at a center of mass energy corresponding to the peak  
of the $\phi$ meson resonance. In this case, $\phi$ decays with one or more particles 
undetected can mimic the $\gg$ final states, with production rates three or four order 
of magnitudes larger.

%, as it is shown in table~\ref{table:phibkgd}. 
%\begin{table}[htb]
%\vspace{5cm}
%\caption{ Main background sources from $\phi$ decays.}
%\label{table:phibkgd}
%\end{table}
%As an example consider the decay chain
%\begin{equation}
%   \phi\to\KS\KL\to\pizz X(undetected), 
% \end{equation}
%where the $\KL$ escapes detection because it does not decay inside the tracking 
%system and does not leave a significant signal crossing the calorimeters, with 
%a probability $P(\KL ~undetected)$ which can be reasonably larger than 10\%.
% The cross section for this process is given by
%\begin{equation}
%    \sigma_{\ee\to\pizz X} ~=~ 
%    \sigma_{\ee\to\phi} ~\times ~BF(\phi\to\KS\KL) \times BF(\KS\to\pizz) 
%    \times P(\KL ~undet);
% \end{equation}
%it is of the order of 100~nb, to be compared to $\sigma(\ee\to\ee\pizz)\sim 
%10~{\mathrm{pb}}$.\\

\begin{figure}[htb]
\begin{center}  
\includegraphics[height=11cm]{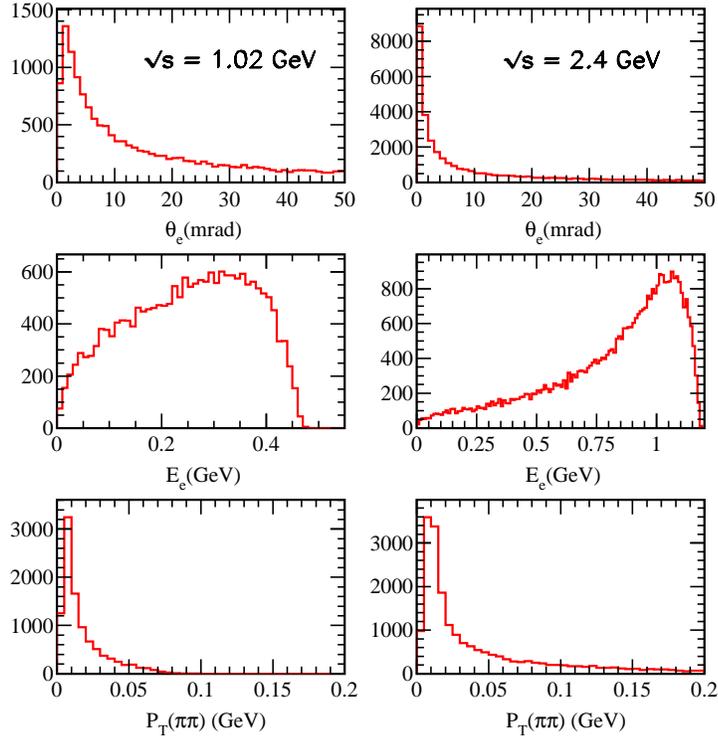}
\caption{
Energy and angle distribution of scattered electrons for the 
two machine energy options. The distribution of the total 
transverse momentum of the $\gg$ system is shown in the 
bottom plots.
}
\label{fig:kine}
\end{center}
\end{figure}

In order to suppress this background one can take advantage of the fact that the $\gg$ 
system has essentially zero transverse momentum, as shown in Fig. \ref{fig:kine},  
contrary to background coming from $\ee$ annihilation, where one or more particles are 
not detected.
%\begin{figure}[htb]
%\begin{center}  
%\includegraphics[height=9cm]{e_theta_510.ps}
%\caption{\footnotesize Energy and angle of scattered electrons in the case for a 
%machine energy $\sqrt{s}=1.02$ GeV.
%}
%\label{fig:kine}
%\end{center}
%\end{figure}

A cut on this variable gives a rejection factor of the order of few tens, 
depending on the background type. It should be noted that at low energy 
colliders 
the typical two-photon selection criterion $E_{\rm vis}/E_{\rm cm}$ (i.e. the ratio 
of the visible over the center of mass energy) does not help, because of the low 
particle multiplicity. We can therefore conclude that two-photon reactions cannot 
be studied at the $\phi$ peak without a suitable tagging system. 

For a collider running at higher energies, the level of background due to hadronic events
would be much lower, but not negligible and tagging would help to reach the necessary 
rejection factor. 

It is worthwhile to mention that tagging the two-photon events would also be useful 
to reject a possible background for particular measurements of one-photon processes, 
like the measurement of the $\ee\to\pippim$  cross section from Initial State Radiation 
events.

\subsubsection*{Requirements for a tagging system} 
In a low energy $\ee$ collider, as those considered here, tagging can be performed more 
easily than with machines that works at higher energies, like the old PEP and PETRA 
colliders and the new $B$-meson factories, as a consequence of the greater average 
scattering angle of the electrons, $\bar{\theta_e}\,\simeq\,m/E_{\rm beam} = 1$ mrad.

This advantage can be exploited only partially owing to the limits imposed by the 
low-$\beta$ insertion quadrupoles and by the minimum angle covered by the central detector, 
which is of the order of $200 \div 300$ mrad for a typical general purpose detector. 
Only a minor fraction  of the scattered electrons enter the central detector, while most 
of them follow a trajectory which departs from the main beam orbit after several meters.
A tagging system should therefore consist of one or more detectors located in specific
regions along the beam line, where the electron yields would be most effective. 

A design of the tagging system can be conceived only when a reasonable scheme of 
the machine layout is available. Independently from its final design, the desired 
features of the tagging system can be summarised as follows:
\begin{itemize}
\item it should be able to record the electrons from $\gg$ reactions over as  
      large as possible angular and energy ranges;
\item it should be able to identify the nature of the hitting particle 
      (i.e. separate electrons from muons and pions produced by the $\ee$ 
       annihilation); 
\item it should possibly give informations about the energy and the scattering 
      angle of the detected electrons;
\item a high electron flux in the same angular region due to machine background 
      (beam-gas bremsstrahlung) and radiative Bhabha events is expected. So fast 
      detectors with relatively good radiation hardness are required. 
      A photon detector could help vetoing the majority of these backgrounds. 
\end{itemize}

\begin{figure}[htb]  
\begin{center}  
\includegraphics[height=6.5cm]{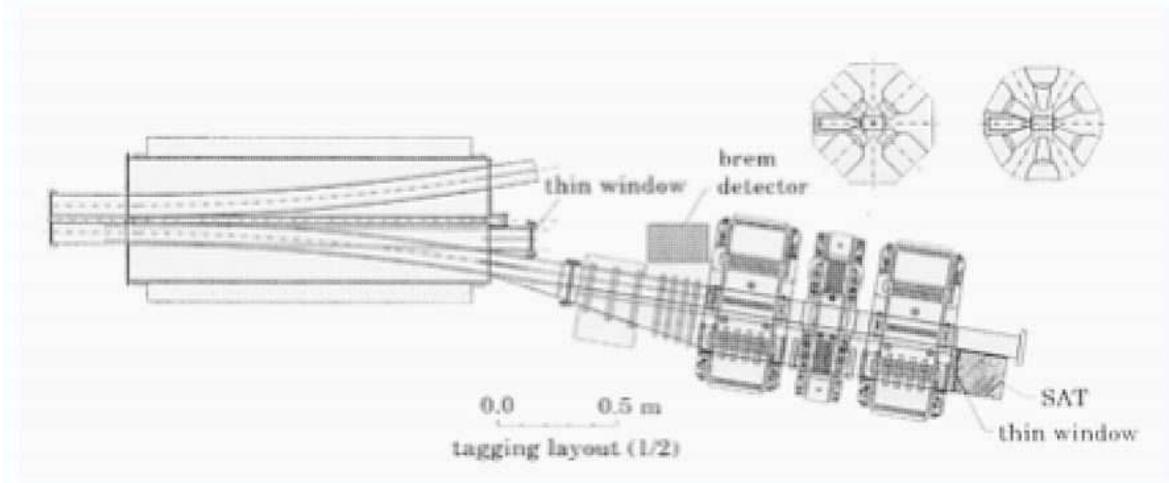}
\caption{ 
A possible location for the small-angle tagger (SAT) in the 
present layout of {\small DAFNE}.
}
\label{f:sat} 
\end{center} 
\end{figure}

To get an idea about the possibility to equip a low energy $\ee$ collider with a 
tagging system we can refer to the proposal submitted at the beginning of the 
{\small DAFNE} project \cite{alex}. The system is composed of both small 
(SAT) and wide (WAT) angle tagging detectors. The SAT accepts electrons emitted 
forward at an angle lower than 20 mrad. It is located at about 8.50 m from the 
Interaction Point (IP), following the Split Field Magnet (SFM), that is the weak 
dipole which separates horizontally the beams in two independent rings. The 
electrons produced in $\gg$ processes are affected by a larger bending inside the 
dipole with respect to the primary beam, because of their lower energies, so that 
they are sufficiently separated by the beam to be collected somewhere downward the SFM. 
Fig \ref{f:sat} shows a scheme of the SAT detector, taken from \cite{alex}. The 
detector is located externally to the beam pipe and extends horizontally from a 
distance of 4 to 30 cm with respect to the beam line. The beam pipe should be shaped 
in order do not absorb the electrons that would  cross it at small angles. 

\begin{figure}[htb]
\begin{minipage}[b]{0.48\textwidth}
\vspace{-0.6cm}
\includegraphics[height=.33\textheight]{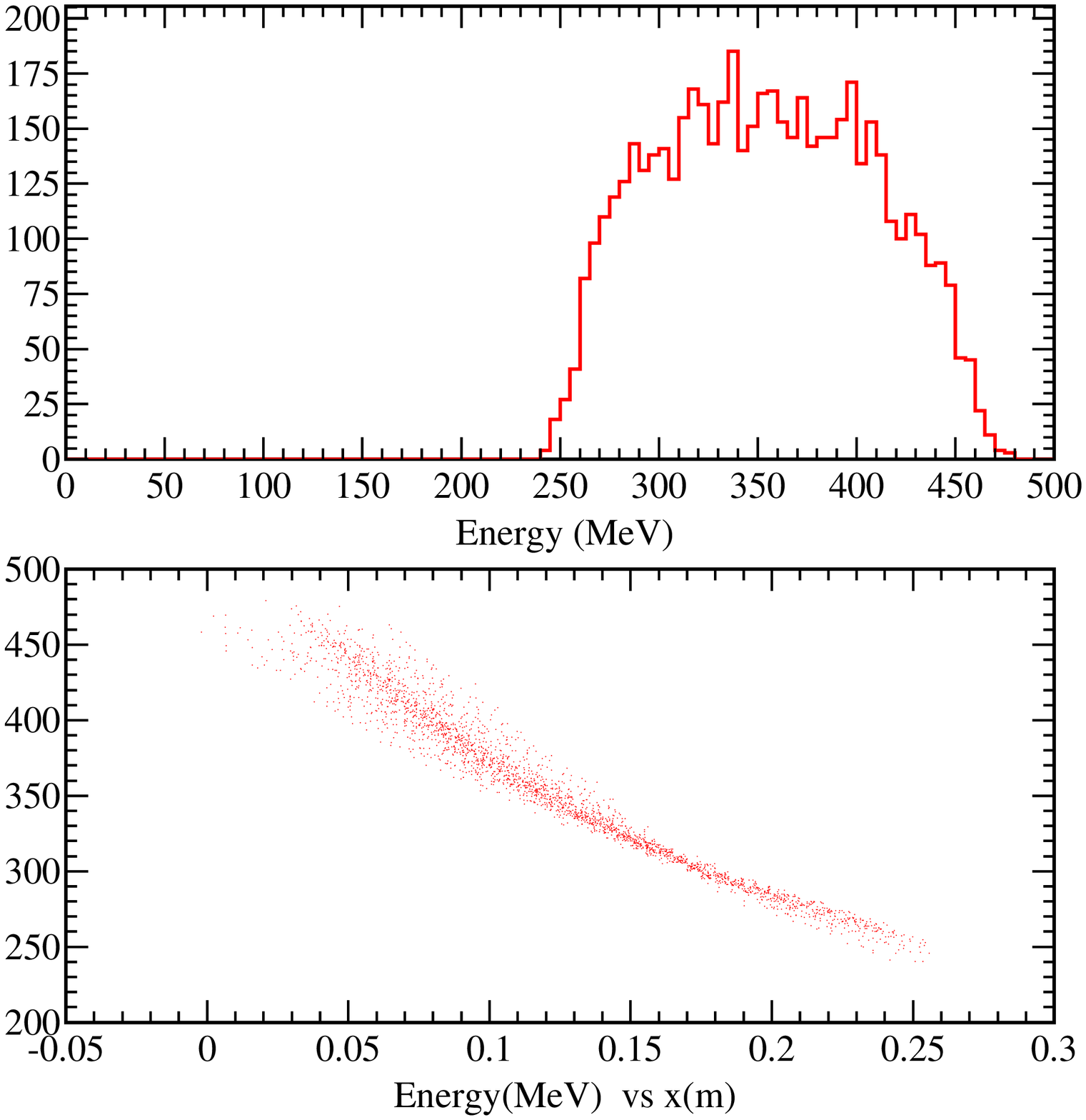}
\end{minipage}
\hfill
\begin{minipage}[b]{0.45\textwidth}
\includegraphics[height=.33\textheight]{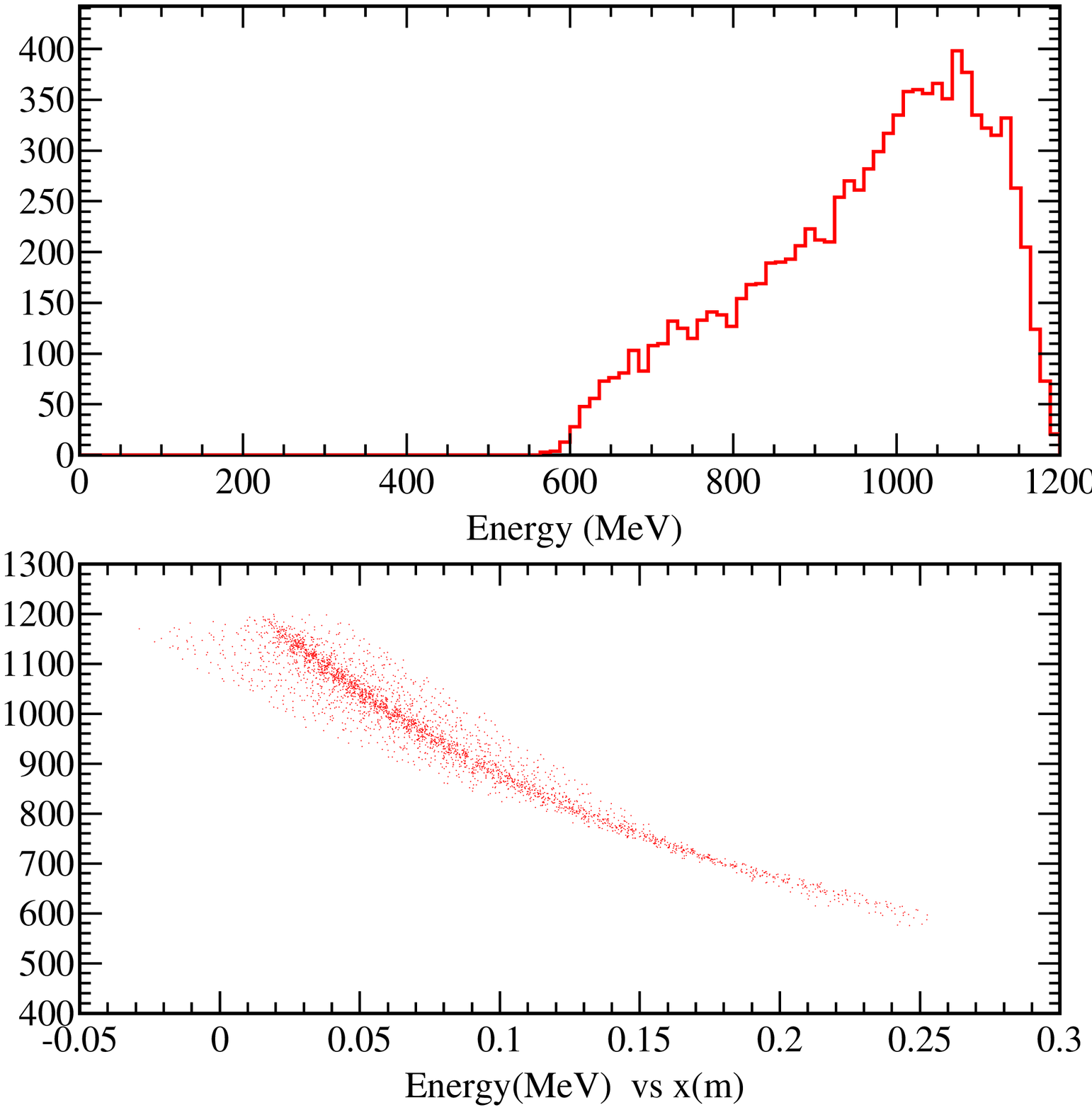}
\end{minipage}
\caption{Energy (top) and energy vs hit position 
on the SAT for scattered electrons. Electrons with an hit position 
$x < 0.04$ m are not yet escaped from the beam pipe and can not be 
detected. Left plots are for $\sqrt{s}=1.02$ GeV, while right plots 
are for $\sqrt{s}=2.4$ GeV.}
\label{fig:EvsX_SAT}
\end{figure}

Fig. \ref{fig:EvsX_SAT} shows the energy of the collected electrons and the horizontal 
coordinate of their impact point on the SAT. The energy-position correlation provides 
an energy measurement with a few percent accuracy. The most energetic electrons follow 
an orbit close to the primary beam and cannot be collected by the SAT. 
%\begin{figure}[htb]
%\begin{minipage}[b]{0.48\textwidth}
%\vspace{-0.6cm}
%\includegraphics[height=.3\textheight]{sat_510.ps}
%\end{minipage}
%\hfill
%\begin{minipage}[b]{0.45\textwidth}
%\includegraphics[height=.3\textheight]{sat_1200.ps}
%\end{minipage}
%\caption{\footnotesize Energy (top) and energy vs hit position on the SAT 
%for scattered electrons. Electrons with an hit position $x < 0.05$ m are 
%not yet escaped from the beam pipe and can not be detected. Left plots are 
%for $\sqrt{s}=1.02$ GeV,  while right plots are for $\sqrt{s}=2.4$ GeV.}
%\label{fig:EvsX_SAT}
%\end{figure}

\begin{figure}[htb]
\begin{minipage}[b]{0.45\textwidth}
\vspace{-0.6cm}
\includegraphics[height=.33\textheight]{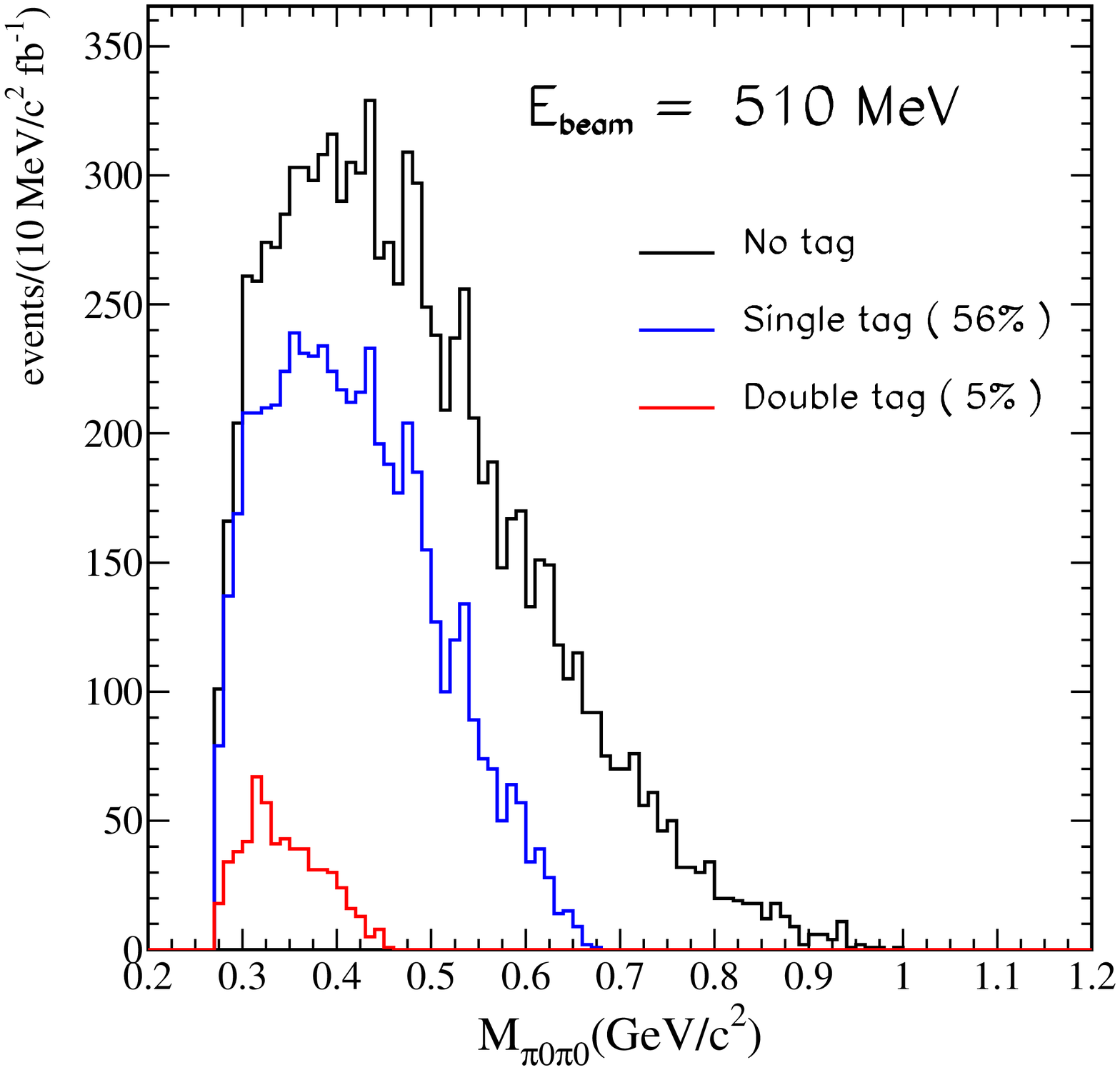}
\end{minipage}
\hfill
\begin{minipage}[b]{0.45\textwidth}
\includegraphics[height=.33\textheight]{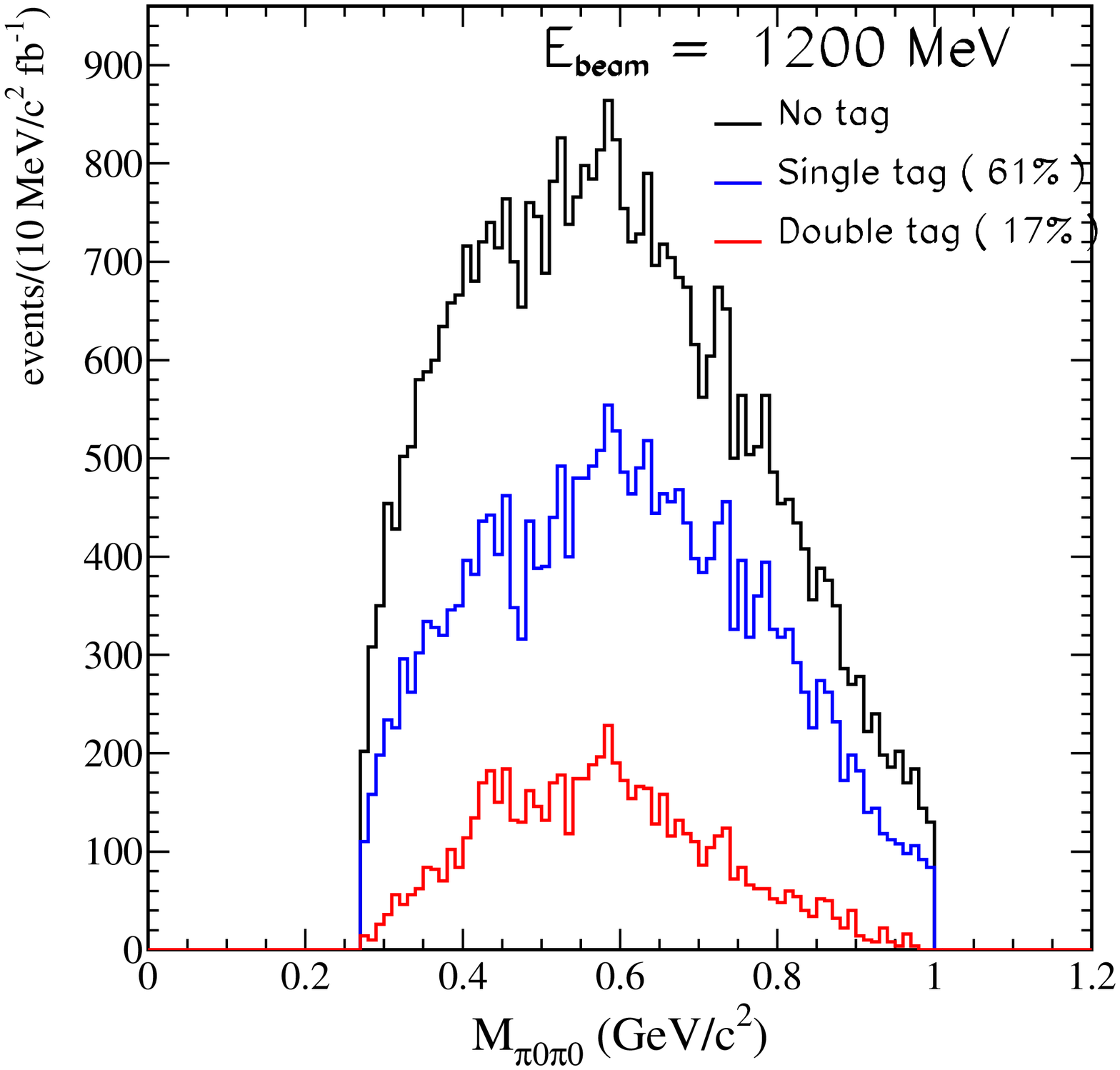}
\end{minipage}
\caption{Invariant mass of the $\gg\to\pizz$ system 
in no-tag (black line), single-tag (blue) and double-tag (red) mode. 
Number of entries corresponds to the number of events expected for a 
10 MeV/$c^2$ bin size and an integrated luminosity of 1 fb$^{-1}$. 
Left plots are for $\sqrt{s}=1.02$ GeV while right plots are for 
$\sqrt{s}=2.4$ GeV. The percentages in parenthesis are the overall reduction
factors due to the single and double tagging. }
\label{fig:Wgg_tag}
\end{figure}

A cut is also observed on the minimum electron energy, mainly due to the correlation 
between the energy and the angle of the scattered electrons. That results in a 
limitation of the maximum invariant mass of the $\gg\to\pizz$ system that can 
be tagged. As it is shown in Fig. \ref{fig:Wgg_tag}, for $E_{\rm beam} = 510$ MeV, 
even in single tag mode, $W_{\gg}$ is limited to $\sim$ 500 MeV/$c^2$ at most.
These results have been obtained with a Montecarlo generator based on the
Weizsacker - Williams
approximation \cite{cour} and a ChPT two-loop cross section for $\gg\to\pizz$ 
\cite{sainio}.
For $E_{\rm beam}=$ 1200 MeV, the $W_{\gg}$ has been limited to 1 GeV/$c^2$ 
at generation level because of the poor reliability of the ChPT approximation above 
these energies.

The transport of the scattered $e^\pm$ is simulated according to the {\small DAFNE} magnet 
optics, with the intensity of the magnetic fields adjusted for the two different beam 
energies.
%\begin{figure}[htb]
%\begin{minipage}[b]{0.5\textwidth}
%\vspace{-0.6cm}
%\includegraphics[height=.3\textheight]{minv_510.ps}
%\end{minipage}
%\hfill
%\begin{minipage}[b]{0.5\textwidth}
%\includegraphics[height=.3\textheight]{minv_1200.ps}
%\end{minipage}
%\caption{\footnotesize Invariant mass of the $\gg\to\pizz$ system in no-tag (black line), 
%single-tag (blue) e double-tag (red) mode. Left plots are for $\sqrt{s}=1.02$ GeV while 
%right plots are for $\sqrt{s}=2.4$ GeV.}
%\label{fig:Wgg_tag}
%\end{figure}

A WAT located in the interaction region, covering scattering angles larger than few degrees, 
would add about 10\% of events in single tag mode. It is reasonable to conceive a detector 
in this region able to measure both the track angle and energy. This provides the quantity 
$Q^2 = 2 E_{\rm beam} E' (1-\cos\theta_e)$, that is a measurement of the degree of virtuality 
of the photon. The $Q^2$ values expected in this region range between $10^{-4} \div 10^{-1}$ 
GeV$^2$ (see also Ref. \cite{song}). This is a very favourable situation to measure the 
$\pizero$ transition form factors (see sec. \ref{sec:ffact}) where the virtual photon is 
provided by the electron tagged at wide angle and the real photon is associated to an 
electron tagged in the SAT or even untagged, which can be assumed nearly on-shell.  

\subsubsection{Final Remarks}
In the following we summarise the main results coming out from our
studies. We considered two different working energies for the machine:
\begin{itemize}
\item{$\sqrt{s}=1.02$ GeV. The energy range accessible to $\gamma\gamma$
    reactions is effectively limited to $W_{\gamma\gamma}\sim 600$
    MeV. This energy range allows the measurement of the two-photon width
    and of the slope of the transition form factor for $\pi^0$ and
    $\eta$. As for the $\sigma$ meson (assuming the resonance parameters
    quoted above) the resonance shape cannot be measured over its whole
    extension. From the experimental side, given the huge background
    associated to the $\phi$-peak, this physics program cannot be exploited
    without a tagging system. As shown in the left panel of
    Fig.\ref{fig:Wgg_tag} tagging of electrons scattered at small angle
    results in a further limitation of the accessible  $W_{\gamma\gamma}$
    region.}
\item{$\sqrt{s}=2.5$ GeV. The $W_{\gamma\gamma}$
    region accessible in this case extends over 1 GeV. Therefore the
    physics program outlined in the previous sections can be fully
    exploited. Even in this case a tagging system is needed to have a
    complete control of the hadronic backgrounds. This appears to be in any case 
    a crucial condition in order to reach the required precision
    levels.}
\end{itemize}

\subsection{\bf Hadron Form Factors in the time-like region}
\label{Bario}

\subsubsection{Introduction: the physics case}
\label{sec:intro}

The form factors of hadrons, as obtained in electromagnetic processes, provide
fundamental information on their internal structure, i.e. on the dynamics
of quarks and gluons in the nonperturbative confined regime. A lot of data for
nucleons have been accumulated in the space-like region using elastic electron 
scattering (for a review, see Ref.~\cite{gao} and references therein). While the 
traditional Rosenbluth separation method suggests the well known scaling of the 
ratio $G_E/G_M$ between the electric and magnetic Sachs form factors, new 
measurements on the electron-to-proton polarization transfer in 
$\vec{e}^- p \to e^- \vec{p}$ scattering reveal strongly contradicting results, 
with a monotonically decreasing ratio for increasing momentum transfer 
$-q^2=Q^2$~\cite{jlab}. This in turn reflects in an approximate 
$1/Q$ trend of the ratio $F_2/F_1$ of the Pauli to Dirac form factors in the 
presently explored range $2\leq Q^2 \leq 5.6$ GeV$^2$~\cite{gao}, 
which is in contradiction with the $1/Q^2$ trend predicted by perturbative 
QCD and, more generally, by dimensional counting rules~\cite{brodsky}. This fact
has stimulated a lot of theoretical work in order to test the reliability of the
Born approximation underlying the Rosenbluth method (see Ref.~\cite{2gamma,2gamma-1} 
and references therein). 

\begin{figure}[h]
\begin{center} 
\psfig{file=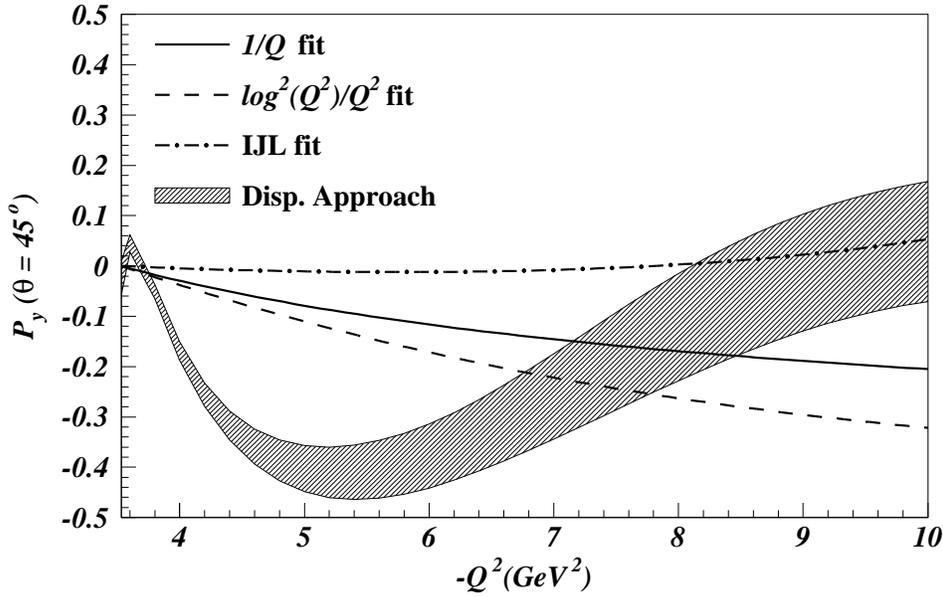, width=13cm}
\end{center} 
\caption{Predicted proton polarization $P_y$ for the $e^+ e^- \to \vec{p} \bar{p}$ 
process at scattering angle $\theta = 45^o$. Solid line for the analytic continuation 
of the $F_2/F_1 \propto 1/Q$ fit in the space-like region~\cite{1/q}; dashed
line for the $\log^2(Q^2)/Q^2$ fit~\cite{logq2}; dot-dashed for a fit from Iachello,
Jackson and Lande~\cite{iachello}. The lined band is obtained by means of
dispersive analysis of the ratio $G_E/G_M$, based on space-like and time-like
data~\cite{dr-app}.}
\label{fig:pacetti}
\end{figure}

In any case, the above scenario makes it critical to deepen our knowledge of $G_E$ and 
$G_M$ also in the time-like region by mapping the $Q^2$ dependence of their moduli and 
phases. In fact, while space-like form factors of stable hadrons are real because of the
hermiticity of the electromagnetic Hamiltonian, time-like form factors, as they can
be explored in $e^+ e^- \to H \bar{H}$ or $p \bar{p} \to \ell^+ 
\ell^-$ processes, are complex because of the residual interactions of the 
involved hadrons $H$ (protons $p$). Their absolute values can be extracted by 
combining the measurement of total cross sections and center-of-mass (c.m.) 
angular distributions of the final products. The phases are related to the 
polarization of the involved hadrons. For example, in $e^+ e^- \to \vec{B}\bar{B}$ 
reactions with spin-$\half$ baryons the normal polarization $\mathcal{P}_y$ to the 
scattering plane is proportional to the phase difference between $G_E$ and 
$G_M$~\cite{dubnick}. Such a polarization is present even if the electron and positron
beams are not polarized. It is extremely sensitive to the theoretical input, as it is
evident in Fig.~\ref{fig:pacetti}, and it can discriminate among analytic 
continuations to the time-like region of models that successfully reproduce the proton 
$G_E/G_M$ data in the space-like region~\cite{brodsky1,egle1}. 

%%%%%%%%%%%%%% Fig. Py (ex Fig.20)

%%%%%%%%%%%%%%%%%%%%%%%%%%%%%%%%%%%%%%%%%%%%%%%%%%%%%%%%%%%%%%%%%%%%%%%%%%

Experimental knowledge of form factors in the time-like region is poor and it
regards only pions and nucleons (for a review see Ref.~\cite{egle1}). As for 
the latter, there are no polarization measurements, hence the phases are unknown. 
The available unpolarized differential cross sections were integrated over a wide 
angular range, and data for $|G_M|$ were extracted under the hypothesis that 
either $G_E=0$ or $|G_E|=|G_M|$. While the first hypothesis is arbitrary, the 
second one is true only at the physical threshold $q^2\equiv s =4M^2$, with $M$ 
the nucleon mass; therefore, the relative weight of $|G_E|$ and $|G_M|$ in the cross
section is yet unknown. As for the neutron, only one measurement is 
available by the {\small FENICE} collaboration~\cite{fenice} for $s \leq 6$ 
GeV$^2$, which displays the same previous drawback. 

Nevertheless, these few data reveal very interesting (and puzzling) properties.
In fact, the form factors are analytic functions of $q^2$ in the whole domain. 
Therefore, the 
analytic properties and phases in the time-like region are connected to the 
space-like region by dispersion relations~\cite{dr}. In particular, $|G_M|$ should
asymptotically become real and scale as in the space-like region. However, a fit to 
the existing proton $|G_M|$ data for $s \leq 20$ GeV$^2$ is compatible 
with a size twice as larger as the space-like result~\cite{e685-2}. Moreover, 
the very recent data from the {\small BABAR} collaboration on 
$|G_E|/|G_M|$~\cite{babar} show that the ratio is surprisingly larger than 1, 
contradicting the space-like results with the polarization transfer method~\cite{jlab} 
and the previous time-like data from {\small LEAR}~\cite{lear}. Also the few neutron 
$|G_M|$ data are unexpectedly larger than the proton ones in the corresponding $s$ 
range~\cite{fenice}. Finally, all the available data show a steep rise of $|G_M|$ 
for $s \sim 4M^2$, suggesting the possibility of interesting (resonant) 
structures in the unphysical region (for more details, see Ref.~\cite{baldini}). 

The possible upgrade of the existing {\small DAFNE} facility~\cite{LoI} to enlarge the 
c.m. energy range from the $\phi$ mass to 2.5 GeV, would allow to explore the 
production of baryons from the nucleon up to the $\Delta$. 
Therefore, in the following we will review the formalism necessary to extract 
absolute values and phases of baryon form factors from cross section 
data (Sec.~\ref{sec:formulae}). We will also make numerical simulations of the
experimental observables (Sec.~\ref{sec:mc}), in order to explore under which
conditions {\small DAFNE}-2 could give leading contributions in this field 
(Sec.~\ref{sec:end}).

%%%%%%%%%%%%%%%%%%%%%%%%%%%%%%%%%%%%%%%%%%%%%%%%%%%%%%%%%%%%%%%%%%%%%%%%%%%%%%%%

\subsubsection{Survey of the formalism}
\label{sec:formulae}

The matrix element for the reaction $e^+ e^- \to B \bar{B}$, where an electron and
a positron with momenta $k_1, k_2,$ annihilate into a spin-$\half$ baryon and an 
antibaryon with momenta $p_1,p_2,$ can be obtained by crossing of the
corresponding matrix element for elastic $e^- B$ scattering. There are several
equivalent representations; here, we use the one involving the axial 
current~\cite{egle2}. The matrix element can be fully parametrized in terms of 
three complex form factors $G_E(s,t), G_M(s,t), A(s,t)$, which are 
functions of $s=(p_1+p_2)^2$ and $t=(k_2-p_1)^2$. In the Born approximation, 
$G_E, G_M,$ reduce to the usual Sachs form factors and depend on $s$ only, while $A=0$. 
We also define $\Delta G_E(s,t)$ and $\Delta G_M(s,t)$ the non-Born contributions to 
$G_E(s)$ and $G_M(s)$. 

By replacing the $t$ dependence in the form factors with $\cos\theta$, where $\theta$ 
is the scattering angle between the incoming positron and the produced baryon, charge 
conjugation invariance imposes general symmetry properties of the Born and non-Born 
amplitudes with respect to the $\cos\theta\rightarrow-\cos\theta$ 
transformation~\cite{2gamma-1}. In particular, $\Delta G_{E,M}$ are antisymmetric,
while $A$ is symmetric. If we neglect bilinear combinations of the non-Born terms
$\Delta G_{E,M}, A$, the unpolarized cross section contains the pure Born term and 
the interference between Born and non-Born contributions. Its general angular
dependence is given by~\cite{noi}
\begin{eqnarray}
\frac{d\sigma^o}{d\cos\theta} &= &\frac{d\sigma^{\mathrm{Born}}}{d\cos\theta} + 
\frac{d\sigma^{\mathrm{int}}}{d\cos\theta} \nn \\
&= &a_0(s) \ +\  a_1(s) \, \cos^2\theta \ +\ \nn \\  
&+ &\cos\theta \, [c_0(s)\ +\  
c_1(s)\, \cos^2\theta \ +\  c_2(s)\, \cos^4\theta\  +\  \dots ]\; ,
\label{eq:ang-dep}
\end{eqnarray}
where $a_0, a_1$, are real combinations of $|G_E|$ and $|G_M|$, while $c_i$ 
$(i=0,1,\dots)$  are coefficients incorporating effects from the $s$ dependence
of $\Delta G_E, \Delta G_M$, and $A$. In our analysis, we will take 
$\Delta G_{E,M}=0$ and $c_i(s)=0$ for $i=1,2,\dots$. The result should 
represent somewhat a lower bound to the actual absolute strength of non-Born 
contributions. 

In this framework, the unpolarized cross section in the c.m. frame of the annihilation 
becomes
\bea
\frac{d\sig^o}{d\cos\theta} &\approx 
& a(s) \, [ 1+ R(s) \, \cos^2 \theta ] - b(s) \, \mbox{\rm Re}
[G_M(s)\,A^\ast(s,t)]\, \cos \theta \; , \label{eq:unpolxsect} \\
a(s) &= &\frac{\alpha^2 \pi}{2s}\,\frac{1}{\tau}\,\sqrt{1-\frac{1}{\tau}}\,
\left( \tau |G_M|^2 + |G_E|^2\right) \; , \label{eq:prova} \\ 
b(s) &= &\frac{2\pi \alpha^2}{s}\,\frac{\tau -1}{\tau} \; , \label{eq:prova2} \\
R(s) &= &\frac{\tau |G_M|^2-|G_E|^2}{\tau |G_M|^2+|G_E|^2} \; ,
\label{eq:prova3}
\eea
%\bea
%\frac{d\sig^o}{d\cos\theta} &\approx 
%& a(s) \, [ 1+ R(s) \, \cos^2 \theta ] - b(s) \, \mbox{\rm Re}
%[G_M(s)\,A^\ast(s)]\, \cos \theta \; , \label{eq:unpolxsect} \\
%a(s) &= &\frac{\alpha^2 \pi}{2s}\,\frac{1}{\tau}\,\sqrt{1-\frac{1}{\tau}}\,
%\left( \tau |G_M|^2 + |G_E|^2\right) \; , \quad 
%b(s) = \frac{2\pi \alpha^2}{s}\,\frac{\tau -1}{\tau} \; , \quad 
%R(s) = \frac{\tau |G_M|^2-|G_E|^2}{\tau |G_M|^2+|G_E|^2} \; , \nn
%\eea
where $\alpha$ is the fine structure constant and $\tau = s/4M^2$. 
If we neglect for the moment the non-Born contribution, measurements of $d\sig^o$ 
at fixed $s$ for different $\theta$ allow to extract the angular asymmetry $R$, 
which can be combined with a measurement of the total cross section $\sig^o$ to 
separate $|G_E|$ from $|G_M|$. This procedure is the time-like equivalent of the 
Rosenbluth separation in the space-like region, but with the advantage that the 
time-like $s=q^2$ is automatically fixed: only the scattering angle needs to be 
changed, while keeping a space-like $Q^2=-q^2$ constant requires also to 
simultaneously vary the beam energy. In this framework, any deviation from the Born 
$(1+R\cos^2 \theta )$ behaviour can be attributed to non-Born contributions. 
In general, the latter can come from the $\cos\theta$ dependence of each one of 
$\Delta G_E, \Delta G_M,$ or $A$. Several independent observables are needed to 
better constrain and disentangle two-photon exchange mechanisms, including the 
polarization of the recoil proton and/or of the electron beam.

For spin-$\half$ baryons with polarization $\mathbf{S}_B$, the cross section is
linear in the spin variables, i.e. $d\sig = d\sig^o \, (1+ \mathcal{P} \, 
\mathcal{A} )$, with $d\sig^o$ from Eq.~(\ref{eq:unpolxsect}) and
$\mathcal{A}$ the analyzing power. In the c.m. frame, three 
polarization states are observable~\cite{dubnick,brodsky1}: the longitudinal 
$\mathcal{P}_z$, the sideways $\mathcal{P}_x$, and the normal $\mathcal{P}_y$. 
The first two ones lie in the scattering plane, while the normal points in the 
$\mathbf{p}_1 \times \mathbf{k}_2$ direction, the $x, y, z,$ forming a 
right-handed coordinate system with the longitudinal $z$ direction along the momentum of
the outgoing baryon. Here, we will concentrate on $\mathcal{P}_y$, 
because it is the only observable that does not require a polarization in the 
initial state~\cite{dubnick,brodsky1}. With the above approximations, it can be
deduced by the following spin asymmetry
\bea
\mathcal{P}_y &= &\frac{1}{\mathcal{A}_y}\, 
\frac{d\sig^\uparrow - d\sig^\downarrow}{d\sig^\uparrow + d\sig^\downarrow} 
\nn \\
&\approx &\frac{b(s)}{2\sqrt{\tau -1}\, d\sig^o}\, \sin \theta \,\nn \\ 
&\times &\left \{ \cos \theta \, \mathrm{Im}\left[ G_M(s) \, G_E^\ast(s) \right] - 
\sqrt{\frac{\tau -1}{\tau}}\, \mathrm{Im}\left[ G_E(s) \, A^\ast(s,t) 
\right] \right\} \; .
\label{eq:py}
\eea
%\bea
%\mathcal{P}_y &= &\frac{1}{\mathcal{A}_y}\, 
%\frac{d\sig^\uparrow - d\sig^\downarrow}{d\sig^\uparrow + d\sig^\downarrow} 
%\nn \\
%&\approx &\frac{b(s)}{2\sqrt{\tau -1}\, d\sig^o}\, \sin \theta \, 
%\left\{ \cos \theta \, \mathrm{Im}\left[ G_M(s) \, G_E^\ast(s) \right] - 
%\sqrt{\frac{\tau -1}{\tau}}\, \mathrm{Im}\left[ G_E(s) \, A^\ast(s) 
%\right] \right\} \; .
%\label{eq:py}
%\eea
The Final State Interactions (FSI) between the final baryons may produce the phase 
difference in the form factors which emerges through the imaginary part of their 
interference. The spin asymmetry~(\ref{eq:py}) can be nonvanishing even without 
polarized lepton beams, because it is produced by the mechanism 
$\mathbf{p}_1 \times \mathbf{k}_2\cdot \mathbf{S}_B$, a time-reversal odd combination 
which is forbidden in absence of FSI and, in general, in the Born approximation for the 
space-like elastic scattering.

The normal $\mathcal{P}_y$ vanishes at the end-points $\theta =0, \pi$ and at the 
physical threshold $\tau =1$. The Born contribution has a typical $\sin 2\theta$ 
behaviour, any deviation being due to non-Born terms. Interestingly, at 
$\theta = \pi/2$ the Born contribution vanishes, and $\mathcal{P}_y$ gives direct 
insight to the amplitude for multiple photon exchanges~\cite{egle2}. 
The measurement of $\mathcal{P}_y$ alone does not completely determine the phase 
difference of the complex form factors. By defining with $\delta_E$ and $\delta_M$ the
phases of the complex $G_E$ and $G_M$, respectively, the Born contribution is 
proportional to $\sin (\delta_M - \delta_E)$, leaving the ambiguity between 
$(\delta_M - \delta_E)$ and $\pi - (\delta_M - \delta_E)$. Only
the further measurement of $\mathcal{P}_x$ can solve the problem, because 
$\mathcal{P}_x \propto \mbox{\rm Re}(G_M^{}\,G_E^\ast) \propto \cos (\delta_M -
\delta_E)$~\cite{brodsky1}. But at the price of requiring a polarized electron 
beam.

%%%%%%%%%%%%%%%%%%%%%%%%%%%%%%%%%%%%%%%%%%%%%%%%%%%%%%%%%%%%%%%%%%%%%%%%%%%%%%%

\subsubsection{Numerical simulations}
\label{sec:mc}

A Monte Carlo simulation was performed for the unpolarized cross section 
$d\sig^o$ and the normal polarization $\mathcal{P}_y$ using the approximations 
described in the previous section. From the expression of $a(s)$ in 
Eq.~(\ref{eq:unpolxsect}) and the known $|G_M(s)|\sim 1/s^2$
scaling~\cite{brodsky}, events were randomly sorted in the $4<s <50$ 
GeV$^2$ range using the $1/s^5$ distribution. Then, only those ones in 
agreement with the Born term of $d\sig^o$ were accepted. An initial sample of 
$280\,000$ events has been considered with the cut $4<s<6$ GeV$^2$. Since the 
total cross section for $e^+ e^- \to p \bar{p}$ is approximately 1 nb, at the 
foreseen luminosity of $10^{32}$ cm$^{-2}$s$^{-1}$ this sample can be collected in 
one month with efficiency 1. The error bars in the following figures are purely 
statistical: they are obtained by making 10 independent repetitions of the 
simulation.

Several extensions to the time-like region of models for nucleon form factors in the
space-like region can be considered~\cite{brodsky1,egle1}. For practical reasons, here 
we have considered the parametrizations of Refs.~\cite{iachello,lomon}, because they 
have been updated in Ref.~\cite{egle1} by including all the available space-like and
time-like data in the fit. Moreover, these models release separate parametrizations for 
the real and imaginary parts of $G_E$ and $G_M$, as they are needed in 
Eqs.~(\ref{eq:unpolxsect}) and (\ref{eq:py}). Being both based on the Vector-Meson 
Dominance (VMD) hypothesis, nevertheless they give drastically different results for 
$\mathcal{P}_y$ as a function of $s$~\cite{egle1}. However, in the $s$ range here 
explored the first one produces very small $P_y$, which are statistically 
distinguishable from the second one but often consistent with zero. Therefore, in the 
following we will consider only observables produced by the updated parametrization of 
Ref.~\cite{lomon}. 

%%%%%%%%%%%%%%%%%%%%%%%%%%%%%%%%% ex Fig. 21 %%%%%%%%%%%%%%%%%%%%%%%%%%%%%%%%%%

\begin{figure}[h]
\begin{center} 
\psfig{file=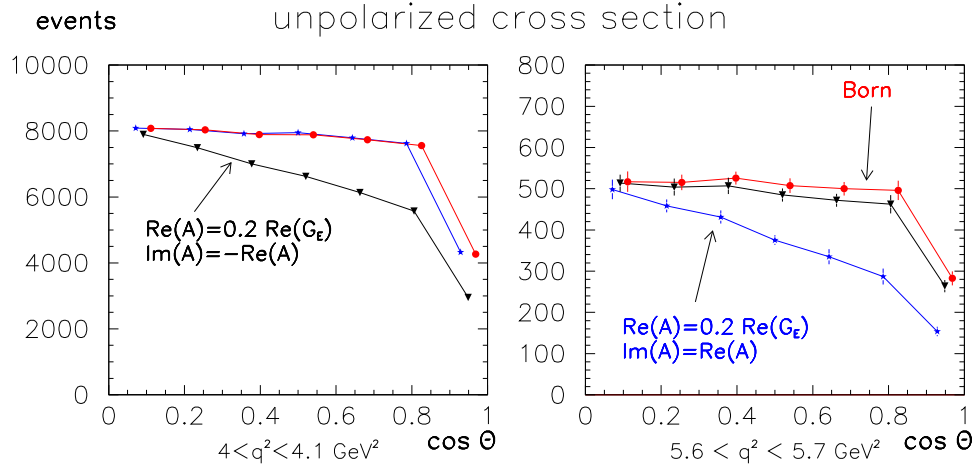, width=13cm}
\end{center} 
\caption{\label{fig:unpolxsect} Angular distribution of $e^+ e^- \to p \bar{p}$
events according to Eq.~(\protect{\ref{eq:unpolxsect}}). Left panel: $52\,000$
events at c.m. squared energy $4<s<4.1$ GeV$^2$. Right panel: $3\,500$ events at
$5.6<s<5.7$ GeV$^2$. Filled (red) circles for the Born contribution, 
downward (black) triangles and (blue) stars for two different choices of 
the non-Born axial form factor (see text). Statistical error bars only; lines are 
drawn to guide the eye.}
\end{figure} 

%%%%%%%%%%%%%%%%%%%%%%%%%%%%%%%%%%%%%%%%%%%%%%%%%%%%%%%%%%%%%%%%%%%%%%%%%%

In Fig.~\ref{fig:unpolxsect}, the angular distribution of $e^+ e^- \to p\bar{p}$
events is shown according to the unpolarized cross section~(\ref{eq:unpolxsect}).  
In the left panel, approximately $52\,000$ events are accumulated for the c.m.
squared energy $4<s<4.1$ GeV$^2$; in the right panel, $3\,500$ events for 
$5.6<s<5.7$ GeV$^2$. Filled (red) circles represent the Born 
contribution. For the non-Born correction, we have assumed 
$\mathrm{Re}[A(s)] \approx C \, \mathrm{Re}[G_E(s)]$, since 
asymptotically the dimensional counting rules give the same $1/s^n$ trend 
irrespective from the number of virtual photons exchanged. We choose $C=0.2$ as 
a sort of upper limit corresponding to 6\% of two-photon radiative corrections 
required to restore the agreement between space-like cross sections obtained 
with the Rosenbluth and the polarization transfer methods~\cite{2gamma}. 
Downward (black) 
triangles and (blue) stars correspond to take $\mathrm{Im}(A) = - \mathrm{Re}(A)$ 
and $\mathrm{Im}(A) = \mathrm{Re}(A)$, respectively, which reflects our ignorance about
the behaviour of the two-photon amplitude. The effect seems clearly detectable with 
one choice or the other, while for intermediate $s$ the result overlaps with the Born 
one. In Ref.~\cite{noi}, the angular distribution was fitted with
\begin{equation}
N(\cos\theta ) = n \, \left[ 1 - B\, \cos\theta + C\, \cos^2\theta \right] \; .
\label{eq:fit}
\end{equation}
The parameter $C$ allows for reconstructing the ratio $|G_E/G_M|$ within 5-10\%,
once model inputs are used from Refs.~\cite{iachello,lomon} or from a simple dipole form~\cite{noi}. 
The parameter $B$ introduces a left-right asymmetry in the $\cos\theta$ distribution, 
which is related to two-photon exchange according to Eq.~(\ref{eq:unpolxsect}). This
term can be identified and estimated provided that $|A|$ is larger than 5\% of 
$|G_M|$ and the relative phases of the form factors do not produce severe 
cancellations~\cite{noi}. We stress that the above statements depend on the
approximations discussed in the previous section, in particular on the truncation of
the expansion~(\ref{eq:ang-dep}). Finally, the error bars are negligible;
nevertheless, the angular coverage should be limited to $\cos\theta < 0.85$, i.e. 
$\theta > 30$ deg., because the number of counts drops fastly for smaller angles. 

%%%%%%%%%%%%%%%%%%%%%%%%%%%%%%%%% ex Fig. 22 %%%%%%%%%%%%%%%%%%%%%%%%%%%%%%%%%%

\begin{figure}[h]
\begin{center} 
\psfig{file=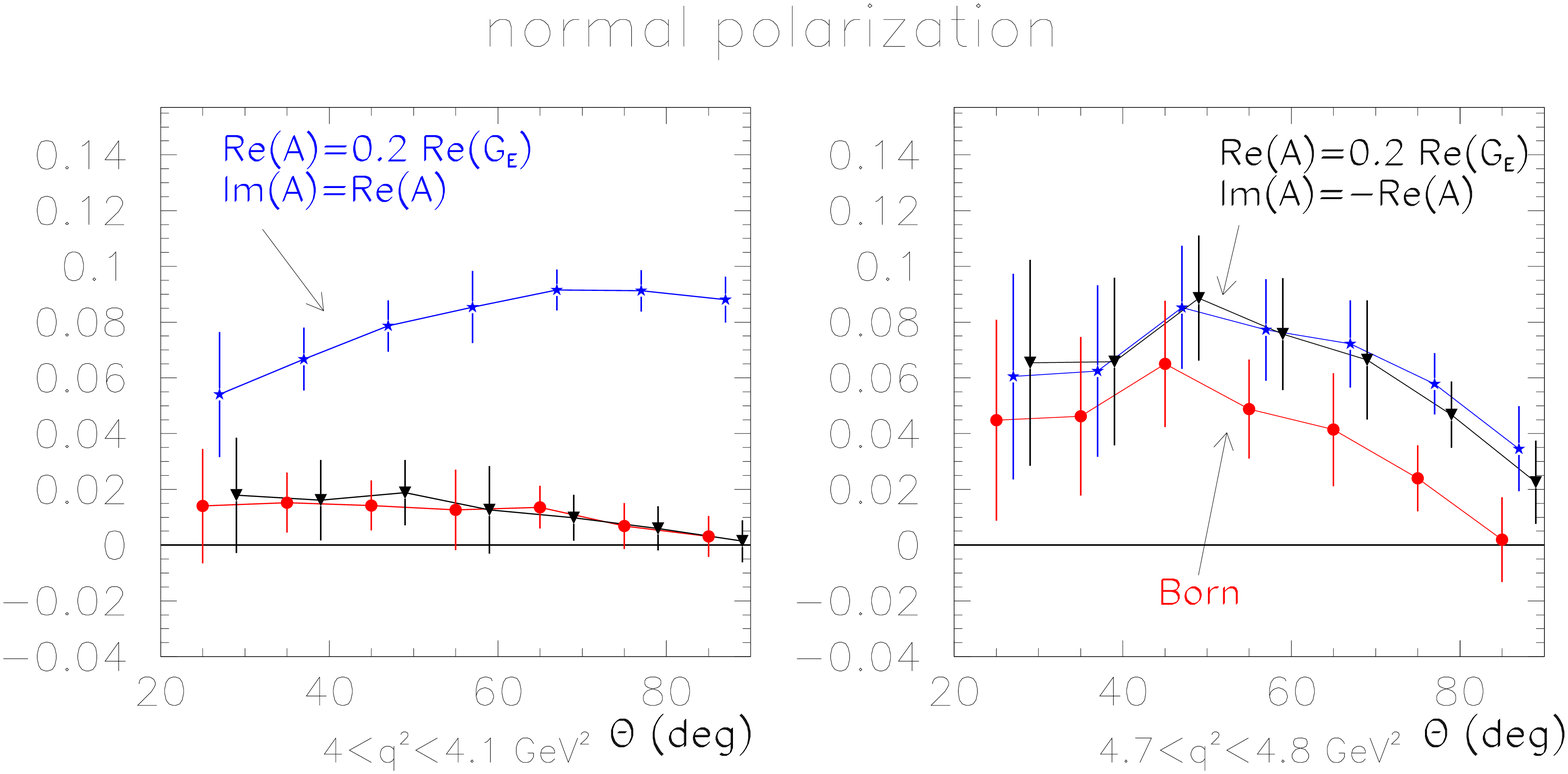, width=13cm}
\end{center} 
\caption{\label{fig:py} Spin asymmetry of $e^+ e^- \to \vec{p} \bar{p}$ normal to
the scattering plane, according to Eq.~(\protect{\ref{eq:py}}) with 
$\mathcal{A}_y=1$. Left panel: $52\,000$ events at c.m. squared energy 
$4<s<4.1$ GeV$^2$. Right panel: $9\,000$ events at $4.7<s<4.8$ GeV$^2$. Same 
notations as in previous figure.}
\end{figure} 

%%%%%%%%%%%%%%%%%%%%%%%%%%%%%%%%%%%%%%%%%%%%%%%%%%%%%%%%%%%%%%%%%%%%%%%%%%

In Fig.~\ref{fig:py}, the spin asymmetry for the polarized $e^+ e^- \to \vec{p}
\bar{p}$ process is shown in its component $\mathcal{P}_y$ normal to the scattering
plane, according to Eq.~(\ref{eq:py}) with $\mathcal{A}_y=1$. Notations are as 
in the previous figure. In the left panel, approximately $52\,000$ events are 
accumulated for the c.m. squared energy $4<s<4.1$ GeV$^2$; in the right panel, 
$9\,000$ events for $4.7<s<4.8$ GeV$^2$. The $\mathrm{Im}(A)=\mathrm{Re}(A)$ 
non-Born correction can be separated at the lowest $s$, reaching an asymmetry 
of almost 10\% at $\theta \sim 90$ deg. where the Born term vanishes, according 
to Eq.~(\ref{eq:py}). This is not possible at higher $s$ because of larger 
error bars. Nevertheless, the $\sin 2\theta$ tail of the Born term becomes 
clearly visible, reaching its (statistically nonvanishing) maximum at 
$\theta \sim 45$ deg., again according to Eq.~(\ref{eq:py}).

%%%%%%%%%%%%%%%%%%%%%%%%%%%%%%%%%%%%%%%%%%%%%%%%%%%%%%%%%%%%%%%%%%%%%%%%%%%%

\subsubsection{Final remarks}
\label{sec:end}

At the foreseen luminosity of $10^{32}$ cm$^{-2}$s$^{-1}$ {\small DAFNE}-2 can collect
$e^+ e^- \to B \bar{B}$ events at the considerable rate of 0.1 Hz. Moreover, 
extension of the available c.m. energy $\sqrt{s}$ up to $\sim 2.5$ GeV would allow 
to study the production of several baryons 
$B=p,n,\Lambda,\Sigma^0,\Sigma^\pm$. From the point of view of the
detector, the experimental program outlined here is characterised by two 
peculiar aspects typically not satisfied by conventional general purpose 
detectors (like {\small KLOE}, see Sect.\ref{DetAcc}): 
the first is related to the measurement of the baryon
polarisation, and the second to the detection of neutrons with kinetic
energies between few and few hundred MeV. In fact, while the polarization of
the $\Lambda$ can be easily studied by looking at the angular distribution of its
decay products, insertion of a polarimeter around the interaction region is
required to measure the polarisation of nucleons (proton and
neutron). Moreover, special care has to be devoted to the
problem of detecting the $n$ other than the $\bar{n}$ to have $n\bar{n}$ 
coincidences. This issue has been analysed, together with all the others
concerning the measurement, in the Letter of Intent expressed in 2005~\cite{LoI}. 
Before 2011, 
which can be a reasonable estimate of the time schedule for the {\small DAFNE}-2 
update, three competitors will be active: {\small VEPP}-2000~\cite{VEPPtot}, operating 
with the same luminosity but limited to nucleon detection only; 
{\small BABAR}~\cite{babar}, that will finish its data taking in 2008, exploring 
larger $\sqrt{s}$ but for unpolarized protons only; {\small BES-III}~\cite{Liu04}, 
starting in 2007, with a higher luminosity but detecting only unpolarized 
protons. Few years later, the {\small PANDA} collaboration~\cite{panda} should start 
taking data at the same luminosity for Drell-Yan events with unpolarized $p\bar{p}$ 
pairs at $2M<\sqrt{s}<5$ GeV. It should be followed, in few more years, by the 
{\small PAX}~\cite{pax} collaboration that will consider polarized collisions. 

In summary, we have shown that a sample of about $300\,000$ events, which is a factor 
$\sim 50 \, \veps$ larger than the present {\small BABAR} one~\cite{babar} (with 
$\veps$ the efficiency), can be collected at {\small DAFNE}-2 in approximately one 
month of dedicated run. Since the event distribution falls approximately like $1/s^5$, 
a good angular coverage of $p\bar{p}$ is needed. In particular, selection of a 
specific angle to separate the $s$ dependence of different theoretical 
contributions does not help. 

Moreover, unpolarized cross section $d\sig^o$ and spin asymmetry $\mathcal{P}_y$ 
normal to the scattering plane, have been simulated by introducing crude 
simplifications in the non-Born term, which reflect the present theoretical ignorance 
about this contribution. In this context, from the angular tail of the results it 
seems that the proton detector should be best positioned in the range 30-70 deg. 
with respect to the beam direction (or 110-150 deg. because of the symmetry of the
formulae). In fact, in this range Born and non-Born contributions to $d\sig^o$ can
be separated to extract information on the absolute values of the proton form
factors with a 5-10\% uncertainty (and, indirectly, also on the phase of $G_M$ via its 
interference with the non-Born term, provided the latter is 
at least larger than 5\% of
$|G_M|$) \cite{noi}. The separation is possible also for $\mathcal{P}_y$ at the lowest
$s$. For higher $s$, error bars are larger but in the same angular range the 
absolute size of $\mathcal{P}_y$ is maximum and the relative phases of the complex 
form factors might be extracted from the Born term, the non-Born correction being 
small.

Finally we mention that, as pointed out in Ref.~\cite{Tornll}, the
$B\bar{B}$ final state opens the possibility to study the spin
correlations predicted by Quantum Mechanics, or, assuming Quantum
Mechanics, to test the CP invariance in the $\Lambda$ decay. 
Particularly promising appears
the case of $\Lambda\bar{\Lambda}$. In fact the $\Lambda\rightarrow\
p\pi^-$ decay plane is related to the spin direction, and represents a
natural polarimeter. By calling $\kappa_{\Lambda}$ and
$\kappa_{\bar{\Lambda}}$ the unit vectors orthogonal to the decay
planes of $\Lambda$ and $\bar{\Lambda}$, the distribution of the
scalar product $\kappa_{\Lambda}\cdot \kappa_{\bar{\Lambda}}$ is
sensitive to deviations from Quantum Mechanics. An analysis based on
about 700 events $J/\psi\rightarrow\Lambda\bar{\Lambda}$ has been done
by {\small DM2}~\cite{DM2EPR}.

\subsection{\bf Low Energy Kaon-nucleus interactions}
\label{Kaons}

%\footline={\hfil}
%\footline={\hss\tenrm\folio\hss}

\vglue 1.0truecm

The interest in the field covered here is of a systematic rather than 
exploratory nature: information on low--energy $K N$ interactions is scarce 
and of a poor statistical quality when compared to the corresponding $\pi N$ 
ones\cite{PDG04,2.}. The low quality of these data reflects in turn on our knowledge 
of the parameters of the $K N$ interaction, remarkably worse than in the 
$SU(3)_f$--related $\pi N$ case\cite{3.}. On top of this situation, one must add 
the problem of understanding kaonic hydrogen (and deuterium) level--shifts and 
widths\cite{4.}, whose recent experimental determinations, despite having finally 
come out with the expected sign ~\cite{5.}, are still 
awaiting an adequate explanation for their magnitude~\cite{6.}.

Data at very low momenta and at rest are essential to clarify many of the 
above--mentioned problems\cite{2.}: the two interaction regions of {\small DAFNE} 
are small--sized sources of low--momentum, tagged $K^\pm$'s and 
$K_L$'s, with negligible contaminations (after suitable cuts on angles 
and momenta of the particles are applied event by event), in an 
environment of very low background radioactivity. This situation is simply 
unattainable with conventional technologies at fixed--target 
experiments. It is therefore of interest to consider the feasibility of 
low--energy, $K^\pm N$ and $K_L N$ experiments at {\small DAFNE}.

Rates to be expected in a simple apparatus at {\small DAFNE} (similar in 
geometry to {\small KLOE}), filled by an almost ideal gas at room temperature 
(such as $^4He$/$^3He$ or $H_2$/$D_2$), are (for a luminosity of $8 \times
10^{32}\ cm^{-2} s^{-1}$) of the order of about $10^7$
two--body events per year (in $H_2$ gas at atmospheric pressure, taken 
as yardstick because of the better known interactions), of which about 
$6.1 \times 10^6$ elastic scattering events, $3.8 \times 10^6$ $\pi^+ 
\Sigma^-$ and about $1.6 \times 10^6$ for each of the remaining four 
two--body channels $\pi^0 \Sigma^0$, $\pi^0 \Lambda$, $\bar K^0 n$, and 
$\pi^- \Sigma^+$. The 
above rates are enough to measure angular distributions in all channels, and 
also the polarisations for the self--analysing final--hyperon states, 
particularly for the decays $\Lambda \to \pi^- p$, $\pi^0 n$ and $\Sigma^+ 
\to \pi^0 p$. One could also expect a total of about $10^5$ three-body 
final--state and $10^4$ radiative--capture events, which should allow a good 
measurement on the absolute rates for these processes as well.

Such an apparatus will need: tracking for incoming and outgoing charged 
particles, time--of--flight measurements (for charged--particle 
identification), a moderate magnetic field (due to the low momenta involved) 
for momentum measurements, and a system of converters plus scintillators for 
photon detection and subsequent geometrical reconstruction of $\pi^0$ and 
$\Sigma^0$ decays.

The above figures for $K^\pm$ rates do not include particle losses in the 
beam--pipe wall and in the internal tracking system, which were assumed 
sufficiently thin (e. g. of a few hundred $\mu m$ of low--$Z$ material), 
nor rescattering effects in a nuclear target 
such as $^4He$. We have indeed checked that, due to the shape of the angular 
distribution of the kaons, particle losses are contained (mostly at small 
angles, where $K$--production is negligible and events would be hard 
to be reconstructed), and momentum losses flat around $\theta = \pi / 2$ 
(where most of the $K^\pm$'s are produced): even for a total thickness of the 
above--mentioned materials of 1 $mm$, kaon momenta do not decrease below 100 
$MeV/c$ and losses do not grow beyond a few percents. Rather, one could 
exploit such a thickness as a low--momentum, thin moderator, to span the 
interesting region just above the charge--exchange threshold at $p_L(K^-) 
\simeq$ 90 $MeV/c$, measurements which would add precious, additional 
constraints on low--energy amplitude analyses\cite{7.}.

Acceptable rates can thus be achieved, orders of magnitude above those of the 
existing data at about the same momentum, i.e. to the lowest--energy 
points of the British--Polish Track--Sensitive Target (TST) Collaboration, 
taken in the mid and late 70's at the NIMROD 
accelerator.

Since losses do not affect $K_L$'s, a detector of the kind sketched above, 
could be used without any problem to study low--energy $K_L \to K_S$ 
regeneration and charge--exchange in gaseous targets, providing essential 
information for this kind of phenomena. Also, a {\small DAFNE} detector dedicated 
to kaon experiments on 
gaseous $H_2$ and $D_2$ can continue its active life to measure $K^+$--, 
$K^-$--, and $K^0_L$--interactions on heavier 
gases as well ($He$, $N_2$, $O_2$, $Ne$, $Ar$, $Kr$, $Xe$), exploring not 
only the properly nuclear aspects of these interactions, such as nucleon 
swelling in nuclei\cite{8.}, but also producing $\pi \Sigma$, $\pi \Lambda$ and 
$\pi \pi \Lambda$ systems at invariant masses below the elastic $\bar K N$ 
threshold in the so--called unphysical region, with statistics substantially 
higher than those now available\cite{9.}, due to the $\simeq 4 \pi$ geometry 
allowed by a colliding--beam--machine detector.

For interactions in hydrogen, the c.m. energy of the pion--hyperon states 
is limited by momentum 
conservation to the initial one, equal (neglecting energy losses) to $w = 
(m_p^2 + m_K^2 + m_p M_\phi)^{1/2}$, or $1442.4\ MeV$ for incident 
$K^\pm$'s and $1443.8\ MeV$ for incident $K_L$'s. As already mentioned, 
energy losses for charged kaons can be exploited to explore $K^- p$ 
interactions down to the charge--exchange threshold at $w = 
1437.2\ MeV$, corresponding to a $K^-$ laboratory momentum of about $90\ 
MeV/c$. For interactions in nuclei, momentum can instead be carried away by 
spectator nucleons, and the inelastic channels explored down to threshold. The 
possibility of reaching energies below the $\bar K N$ threshold allows 
exploration of the unphysical region, containing two resonances, the $I = 0$, 
$S$--wave $\Lambda(1405)$ and the $I = 1$, $J^P = {3\over2}^+$ $P$--wave 
$\Sigma(1385)$, observed mostly in production experiments (and, in the first 
case, with limited statistics\cite{9.}): the information on their 
couplings to the $\bar K N$ channel relies entirely on extrapolations 
of the low--energy $\bar K N$ data. The coupling of the $\Sigma(1385)$ to the 
$\bar K N$ channel, for instance, determined via dispersion 
relations involving the total sum of data collected at $t \simeq 0$, still 
carries uncertainties which at their best are of the order of 50 \% 
of the $SU(3)_f$ prediction\cite{10.}. As for the 
$\Lambda(1405)$, even its spectroscopic classification is an open problem, 
given the paucity of the best available 
data. We could add that recently the presence of a second state 
has been claimed\cite{11.}, and to test such a claim would of course be 
important, for the role the 
state has both for kaonic atoms and the determination of the low--energy 
parameters of the $K N$ interactions.

A formation experiment on bound nucleons, in an (almost) $4 \pi$ 
apparatus with good efficiency and resolution for low--momentum $\gamma$'s 
(such as {\small KLOE}), can measure a channel such as $K^- p \to \pi^0 \Sigma^0$ 
(above threshold), or $K^- d \to \pi^0 \Sigma^0 n_s$ (both above and below 
threshold), which is pure $I = 0$: up to now all analyses on the 
$\Lambda(1405)$ have been limited to charged channels, and assumed the 
$I = 1$ contamination to be either negligible or smooth and non--interfering 
with the resonance signal. Since models for the $\Lambda(1405)$ 
differ mostly in the details of the resonance shape, 
and it is precisely the shape which could be changed even by a 
moderate interference with an $I = 1$ background, such measurements would be 
decisive. Having in the same apparatus and at almost the same energy
tagged $K^-$ and $K_L$ produced at the same point, one can further separate 
$I = 0$ and $I = 1$ channels with a minimum of systematic uncertainties, by 
measuring all channels $K_L p \to \pi^0 \Sigma^+$, $\pi^+ \Sigma^0$ and $K^- 
p \to \pi^- \Sigma^+$, $\pi^+ \Sigma^-$, besides, of course, the 
above--mentioned, pure $I = 0$, $K^- p \to \pi^0 \Sigma^0$ one. It must be 
noted that the recent claim for two states\cite{11.} is 
based on a low--statistics measurement ~\cite{12.} of the reaction $K^- 
p \to \pi^0 \pi^0 \Sigma^0$: an analysis of 
all $\pi \pi Y$ ($Y = \Lambda, \Sigma$) channels, possible with much higher 
statistics at {\small DAFNE}, would be therefore highly desirable.

Another class of interesting processes which are expected, 
at a much smaller rate, from {\small DAFNE}'s kaons are the radiative capture 
ones $K^- p \to \gamma \Lambda$, $\gamma \Sigma^0$ and $K_L p \to \gamma 
\Sigma^+$ (both in hydrogen and nuclei), and $K^- n \to \gamma \Sigma^-$ 
and $K_L n \to \gamma \Lambda$, $\gamma \Sigma^0$ (only in nuclei). Up to 
now only searches for photons emitted after stops of $K^-$'s in liquid 
hydrogen and deuterium have been performed with some success: spectra are 
dominated by photons from unreconstructed $\pi^0$ and $\Sigma^0$ 
decays, and separating signals from this background poses serious 
difficulties. Indeed these experiments were able to produce only an estimate of 
the respective branching ratios\cite{13.}. The $4 \pi$ geometry at 
{\small DAFNE}, combined with the 
``transparency'' of a {\small KLOE}--like apparatus, its high efficiency for 
photon detection and its good resolution for spatial reconstruction of the 
events, should make possible (in an $H_2$/$D_2$ experiment) the full 
identification of the final states and therefore the measurement of the 
absolute cross sections for these processes, although in flight and not at 
rest.

A first proposal would be the following: before building a 
dedicated apparatus for low--energy experiments on gaseous 
targets, one could equip {\small KLOE} with a less restrictive 
trigger, that could select the interactions of anti--kaons (tagged by the 
particles on the opposite side, be they either $K^+$'s or $K^0_S$'s) with 
the gas filling the chamber and reconstruct off-line the pion--hyperon, 
pion--pion--hyperon and single--$\gamma$--hyperon spectra for all 
charge combinations. Such data would contain both the $\Lambda$(1405) and the 
$\Sigma$(1385), including their interference, plus the effects of 
rescattering inside the remainder of the $^4He$ target. The latter 
will further feed -- via charge--exchange processes -- also such ``exotic'' 
combinations as $\Sigma^{\pm}\pi^{\pm}$, allowing a better understanding of 
the nuclear--medium distortions on the ``elementary'' processes $\bar K N \to 
\pi Y$, $\bar K N \to \pi \pi Y$ and $\bar K N \to \gamma Y$. {\small KLOE} (or a 
similar, scaled down apparatus) 
is unique for such a scope: the need for a good efficiency and high 
resolution for low--energy $\gamma$'s (motivated for {\small KLOE} by decays such 
as $\phi \to \gamma (a_0,f_0)$ and the reconstruction of very low--momentum 
$\pi^0$'s) allows also the identification and reconstruction of $\Sigma^0$'s 
through their decay to $\Lambda \gamma$, virtually impossible in any other 
detector with an almost 100 \% efficiency. On the other hand, the very high 
efficiency for $\gamma$ detection, combined with the high intensity of the 
source and the ease with which one can discriminate between kaons and pions 
(not to mention leptons) from the $\phi$ decays, allows an unprecedently 
clean determination of radiative capture events (even if in a slightly 
more complex target than hydrogen or deuterium).

As a closing remark one can add that contaminations due to the presence of 
a small admixture of other gases in helium, or to the tungsten wires running 
across the chamber, are not that important for the mass spectra (they amount 
to -- small -- distortions in the nucleon distribution functions, which the 
``elementary'' amplitudes have to be convoluted with, with respect to those 
for pure $^4$He), and even less for the ratio of $\gamma Y$ (or $\pi \pi Y$) 
to $\pi Y$ spectra.
\subsection{\bf Conclusions}
We have seen that {\small DAFNE}-2 offers the possibility to carry on a
wide physics program, mainly based on the possibility to increase the center
of mass energy up to 2.5 GeV but also by profiting of a high luminosity run
at the $\phi$ resonance energy. This program includes important
measurements that are essential for the most precise tests of the Standard 
Model, and a large
number of relevant measurements in the field of hadronic physics. Some of
these measurements can be done only in the {\small DAFNE}-2 environment.  
We remark that  
such a program considerably broadens the physics potential of the entire
project being complementary to the Kaon decays program. 
In Sec.\ref{DetAcc} we will discuss the main requirements of this program to
the foreseen detectors.

\section{A higher energy option: the $\tau$-charm factory}
\label{Tauc}

\newcommand\scatt{\mathrm{scatt}}
%%%%%%%%%%%%%%%%%%%%%%%%%%%%%%%%%%%%%%%%%%%
\def\beq{\begin{equation}}   \def\eeq{\end{equation}}
\def\bea{\begin{eqnarray}}   \def\eea{\end{eqnarray}}
\newcommand{\al}{\alpha}
\newcommand{\bet}{\beta}
\newcommand{\as}{\alpha_s}
\newcommand{\GeV}{\,\mbox{GeV}}
\newcommand{\MeV}{\,\mbox{MeV}}
\newcommand{\matel}[3]{\langle #1|#2|#3\rangle}
\newcommand{\state}[1]{|#1\rangle}
\newcommand{\ve}[1]{\vec{\bf #1}}
\newcommand{\vep}{\varepsilon}
\newcommand{\gsim}{\lower.7ex\hbox{$ \;\stackrel{\textstyle>}{\sim}\;$}}
\newcommand{\lsim}{\lower.7ex\hbox{$ \;\stackrel{\textstyle<}{\sim}\;$}}
%%%%%%%%%%%%%%%%
\def\rarr{ \rightarrow}
\def\epem{ e^+e^- }
\def\c2{CLEO~II.V}
\def\ccb{ c\bar{c} }
\def\d0d0{ D^0\bar{D}^0 }
\def\p0p0{ P^0\bar{P}^0 }
\def\ddb{ D\bar{D} }
\def\qp2{ \Bigl| \frac{q}{p} \Bigr|^2 }
\def\pq2{ \Bigl| \frac{p}{q} \Bigr|^2 }
\def\rarr{ \rightarrow }
\def\larr{ \leftarrow }
\def\jp{ J/\psi }
\def\Jp{ J/\psi }
\def\pspr{ \psi^\prime }
\def\ps2s{  \psi(2S) }
\def\q2{ $q^2$ }
\def\cm2s1{ $\,{\rm cm}^{-2} {\rm s}^{-1}$}
\def\d0{D_2^{*0}}
\def\d+{D_2^{*+}}
\def\dst{D_2^*}
\def\mev{ \,{\rm MeV}/c^2  }
\def\gev{ \,{\rm GeV}/c^2  }
\def\tc{~ $\tau$-charm ~}
%\newcommand{\note}[1]{\marginpar{\tiny #1}}
%%%%%%%%%%%%%%%%
% some macros which might be helpful: (used in my thesis)
\newcommand{\krz}{\ensuremath{{K}{}^*(892)^0}}
\newcommand{\krzb}{\ensuremath{\overline{K}{}^*(892)^0 }}
\newcommand{\krzmndk}{\ensuremath{D^+ \rightarrow \krzb \mu^+ \nu}}
\newcommand{\krzlndk}{\ensuremath{D^+ \rightarrow \krzb \ell^+ \nu_\ell}}
\newcommand{\philndk}{\ensuremath{D_s^+ \rightarrow \phi(1020)\; \ell^+ \nu_\ell}}
\newcommand{\phimndk}{\ensuremath{D_s^+ \rightarrow \phi(1020)\; \mu^+ \nu }}
\newcommand{\kkmndk}{\ensuremath{D_s^+ \rightarrow K^+ K^- \mu^+ \nu }}
\newcommand{\phiendk}{\ensuremath{D_s^+ \rightarrow \phi(1020)\; e^+ \nu_e }}
\newcommand{\kpimndk}{\ensuremath{D^+ \rightarrow K^- \pi^+ \mu^+ \nu }}
\newcommand{\clmunew}{\ensuremath{\textrm{CL}\mu_{(\textrm{new})}}}
\newcommand{\gevcsq}{\ensuremath{\textrm{GeV}/c^2}}
\newcommand{\ccbar}{\ensuremath{c\bar{c}}}
\newcommand{\thv}{\ensuremath{\theta_\textrm{v}}}
\newcommand{\thl}{\ensuremath{\theta_\ell}}
\newcommand{\costhv}{\ensuremath{\cos\thv}}
\newcommand{\sinthv}{\ensuremath{\sin\thv}}
\newcommand{\costhl}{\ensuremath{\cos\thl}}
\newcommand{\costhlsq}{\ensuremath{\cos^2\thl}}
\newcommand{\qsq}{\ensuremath{q^2}}
\newcommand{\yvar}{\ensuremath{\qsq{}/\qsq_{\rm max}}}
\newcommand{\bw}{\ensuremath{\textrm{B}_{\phi}}}
\newcommand{\mkpi}{\ensuremath{m_{K\pi}}}
\newcommand{\mkk}{\ensuremath{m_{K^+ K^-}}}
\newcommand{\amp}{\ensuremath{0.36~\exp(i \pi/4)} \ensuremath{(\textrm{GeV})^{-1}}}
\newcommand{\prd}[1]{Phys.~Rev.~D \textbf{#1}}
\newcommand{\plb}[1]{Phys.~Lett.~B \textbf{#1}}
\newcommand{\kpimnbr}{\ensuremath{{\Gamma(\kpimndk{}) / \Gamma(\kpipi{})} }}
\newcommand{\kpimnbrb}{\ensuremath{{\Gamma(\kpimndk{}) \over \Gamma(\kpipi{})} }}
\newcommand{\krzlnbrb}{\ensuremath{{\Gamma(\krzlndk{}) \over \Gamma(\kpipi{})} }}
\newcommand{\krzmnbrb}{\ensuremath{{\Gamma(\krzmndk{}) \over \Gamma(\kpipi{})} }}
\newcommand{\phimnbrb}{\ensuremath{{\Gamma(\phimndk{}) \over \Gamma(\phipi{})} }}
\newcommand{\philnbrb}{\ensuremath{{\Gamma(\philndk{}) \over \Gamma(\phipi{})} }}
\newcommand{\phimnbr}{\ensuremath{{\Gamma(\phimndk{}) / \Gamma(\phipi{})} }}
\newcommand{\kpipi}{\ensuremath{D^+ \rightarrow K^- \pi^+ \pi^+ }}
\newcommand{\gevc}{\ensuremath{\textrm{GeV}^2/c^2}}
\newcommand{\phipi}{\ensuremath{D_s^+ \rightarrow \phi(1020)\; \pi^+ }}
\newcommand{\krzresult}{\ensuremath{ 0.602 ~\pm~0.010~{\rm(stat)}~\pm~0.021~{\rm (sys)}}}
\newcommand{\krzresultstat}{\ensuremath{ 0.602 ~\pm~0.010~}}
\newcommand{\rvvalue}{\ensuremath{1.549 \pm 0.250 \pm 0.145}}
\newcommand{\rtwovalue}{\ensuremath{0.713 \pm 0.202 \pm 0.266}}
\newcommand{\rvresult}{\ensuremath{r_v = \rvvalue{}}}
\newcommand{\rtworesult}{\ensuremath{r_2 = \rtwovalue{}}}
\newcommand{\rtwo}{\ensuremath{r_2}}
\newcommand{\rthree}{\ensuremath{r_3}}
\newcommand{\rvee}{\ensuremath{r_v}}

%%%%
\renewcommand{\rmdefault}{ptm}
\newcommand{\Header}{
  \begin{tabular}{rl}
  \hspace{-.4cm}
%  \special{psfile=logo.eps
   %   voffset=-100  %-10
  %    hoffset=200  %-30
 %     hscale=200 vscale=200 angle=0}
% \includegraphics[width=2.0in]{logo.eps}
      &
    \renewcommand{\arraystretch}{0.5}
%    \begin{tabular}{r}
%      {\hspace{1cm}~\LARGE\sffamily LABORATORI~ NAZIONALI~ DI~ FRASCATI}\\
%      \\
%      {\Large\sffamily SIS-Pubblicazioni}\\
%    \end{tabular}
    \renewcommand{\arraystretch}{1}
  \end{tabular}
  \vskip 1cm
  \begin{flushright}
  \renewcommand{\arraystretch}{0.5}
    \begin{tabular}{r}
      {\underline{LNF-05/31 (P)}}\\    % insert here the preprint number
      {\small 22 dicembre 2005} \\      % insert here the preprint Date
     {\tt taucharm7.tex}\\
    \end{tabular}
  \end{flushright}
  \renewcommand{\arraystretch}{1}
  \vskip 1 cm
  }

%%%%%

\subsection{\bf The physics case for  a $\tau$-charm factory at LNF}
\subsubsection{Overview}
In this section we discuss the physics that could be addressed with 
a \tc  factory at Frascati, the impact  and  competitiveness of
such  a  
research program with respect to the existing and planned ones.
 When the LHC will possibly have directly probed the existence of new 
physics at the TeV scale, 
further progress may be achieved with low energy experiments in high 
precision  
flavour physics. A dedicated \tc factory with unprecedented 
luminosity 
and much better systematics, to be ready in 2013 (we assume at least 5
years of construction after the end of {\small DAFNE}) , could  be essential to 
constrain 
the new physics with intensive flavour studies. 
\par
Within flavour physics, the charm quark occupies a peculiar place. It is 
the lightest 
among the heavy quarks, i.e., the quarks with mass  larger than the 
QCD scale $\Lambda_{QCD}$. Besides, it is the only up-type heavy
quark that hadronises, 
 which favours 
searches for Beyond the Standard Model (BSM) physics, since due to 
GIM mechanism 
the box diagram 
processes are suppressed in the Standard 
Model (SM). 
Therefore, the charm quark is a good testing ground for 
problems still open
%pQCD models and 
%lattice gauge theories aimed to  unravel the mysteries still lurking 
in 
the B sector. The charm quark is also a probe for BSM physics in the
LHC era. For both statements, an obvious caveat has to be formulated,
that is, the charm quark mass sits right in the middle of a range
heavily populated of light quark resonances, the 2 GeV region. Final
state interactions are particularly important, and should be carefully
taken into account. Final state interactions, on the other hand, are useful
for finding CP violation effects: they are an essential ingredient to find 
direct CPV through partial width asymmetries and, while in T-odd moments
they can fake CP violation, nonetheless they can be disentangled at a \tc 
factory. Finally, the charm quark mass can be extracted
both from hidden charm and open charm decays, which provides a valuable
validation check.
\par
Charm physics has been the driving force for modern detector development, 
and it is now investigated by B-factories and  \tc factories  
\cite{Bianco:2003vb}\cite{burdmanshipsey}\cite{QWG}. 
A \tc factory is an $\epem$ collider running at center of mass energy from
$\ccb$ charmonium $(\jp)$, to the $\psi(2S)$ (the $\tau \bar \tau$
threshold), the $\psi(3770)$ (the $\ddb$ threshold) and
above to the $D_s\bar D_s$ threshold. A collider attaining a center of
mass energy of 3.8\,GeV investigates a very broad spectrum of
very important topics, such as charmonium spectroscopy, $\tau$ lepton
decays, and D meson decays.
\par
Studying charm decays produced at threshold in $e^+e^-$ annihilation 
offers several 
special advantages. Most of the charmonium spectroscopy results have come
from $\epem$ storage rings, where $J^{PC}=1^{--}$ states are formed
directly to lowest order. The non-vector states such as $^3P_J$ are
observed in two-step processes like $\epem \rightarrow \psi(2S)
\rightarrow \ccb + \gamma$.
 Threshold production of charmed hadrons leads to
very clean low  
multiplicity final states with very low backgrounds. Clean events also 
imply favourable  conditions for 
neutrino reconstruction via missing mass techniques.
One can employ tagged events to obtain the {\em absolute} values  
of charm hadron branching ratios in a model independent way.  
The widths for $D^+ \to \mu ^+ \nu$  
and $D^+_s \to \mu ^+ \nu$ can be measured with unrivalled control over  
systematics.  Finally, with the charm hadrons being produced basically at 
rest the time  
evolution of $D^0$ decays cannot be measured directly. Yet by comparing  
EPR correlations in $D$ decays produced in $e^+e^- \to D^0\bar 
D^0$,  
$e^+e^- \to D^0\bar D^0 \gamma$ and $e^+e^- \to D^0\bar D^0\pi ^0$  
one can deduce whether oscillations are taking place or not, as explained 
in
detail in \cite{Bianco:2003vb}.  
\par
 Several proposals have been discussed over the  
last 15 years for \tc factories with the ambitious goal  
of achieving luminosities up to the  
$10^{33} - 10^{34}$\cm2s1   range for the c.m. energy interval 
of 3 - 5 GeV. 
A \tc factory project was proposed at CERN\cite{CERN87} (1987), at
SLAC\cite{SLAC89} (1989), in
Spain\cite{SPAIN90} (1990), at JINR\cite{JINR92} (1992) and
Argonne\cite{ARGONNE1994} (1994). It became reality with the CESR-c/CLEO-c
and BEPC-II/BES-III
projects, described in Sect.\ref{SECT:COMPETITORS}.
\par
Exploring all the possibilities for the future of the LNF in the panorama 
of  high energy physics during
and after the LHC era includes also  \tc physics.
Expertise for both machine and detector does exist for (although
the infrastructures of the LNF are probably not  compatible with) a \tc
project, a collider with an energy 
in the center of mass of 3.8 GeV, and maximum luminosity of 
10$^{34}$\cm2s1, i.e. an order of magnitude above the
BEPC-II design value. Such a machine could be based on a double symmetric
ring collider, flat beams in multi-bunch
operation, normal conducting magnets, one interaction region, and 
on-energy
injection system\cite{LNFF5}\cite{Raimondi05}.  
In the following, we shall assume an integrated 100~fb$^{-1}$ integrated
Luminosity per $10^7$s year.
\par
 The detector is ambitious, but working examples already exist and 
operate. 
CLEO-III/CLEO-c \cite{ref4qgwg}\cite{ref5qgwg}, for instance, is a 
solenoidal $4\pi$ hermetic 
detector with tracker, RICH, ECAL, potent DAQ and open trigger. No vertex 
detector is 
necessary in principle if the symmetric beam option is used, since $\ddb$ 
pairs are produced 
at rest in the lab reference system.
\par
In this section we shall review the \tc relevant  physics topics in hidden 
and
open charm physics, both SM- and BSM-related. In particular we shall 
underline
the items functional to BSM characterisation in the LHC era, and discuss 
the relative merits of a \tc factory with respect to competition.

\subsubsection{Hidden charm physics}

Heavy quarkonium, being a multi-scale system, offers a precious window into
the transition region between high energy and low energy QCD and thus
a way to study the behaviour of the perturbative series and the
nontrivial vacuum structure. The existence of energy levels below,
close and above threshold, as well as the several production
mechanisms, allows one to test the population of the QCD Fock space in
different regimes and eventually to search for novel states with
nontrivial glue content (hybrids, glueballs). Besides, a 
study of the final state in charmonium decays will open a novel tool for studying low
energy  spectroscopy and hadronization.
\par
The diversity, quantity and accuracy of the data collected
at several accelerator experiments (BES, KEDR, CLEO-III, CDF, D0,
B-factories, Zeus, H1, RHIC) makes quarkonium an extremely relevant
and timely system to study. In the near future, even larger data
samples are expected from the CLEO-c and BES-III upgraded
experiments, and in perspective at the LHC and at Panda at GSI.
%BEPCII starting in 2008
\par
From the theory point of view, the recent progress 
in the formulation  of non-relativistic effective field
theories (EFT) for heavy quarkonium (NRQCD, pNRQCD) \cite{eft} and the related
lattice implementation, makes it possible  to go beyond phenomenological models and,
for the first time, provide a
unified  and systematic description of all aspects of
heavy-quarkonium physics.  This, together with the huge
flow of experimental data, allows to use quarkonium as a
benchmark for our understanding of QCD, for the precise determination
of relevant SM parameters and for search of physics BSM. 
\par
A  \tc machine with the characteristics discussed above would allow for
record samples of $J/\psi$, $\psi(2S)$, $\psi(3770)$ with unmatched
systematics and with very rich and interesting physics, complementary
to  
the LHC program, that includes the following topics.
 \subsubsection*{Precise extraction of SM parameters from quarkonium.}
  Ground state observables may be studied in the framework of
 perturbative QCD \cite{eft,pNRQCD,pert}.  These studies are relevant because they may allow,
 in principle, the precise extraction of some of the fundamental
 parameters of the SM, like the heavy quark masses and the
 strong coupling constant.
 %Heavy quarkonium is one of the most suitable system to extract a
 %precise determination of the mass of the heavy quarks $b$ or $c$.
 A recent analysis performed by the
 Quarkonium Working Group \cite{QWG}, and based on all the previous determinations
 existing in the literature, indicates that
 the quark mass extraction from heavy quarkonium involves an error of
 about 50 MeV both in the bottom ($1$\% error) and in the charm ($4$\% error) case.
 Such error is already very competitive with extractions coming from other physical systems
 and in the future it  should be reduced  further up to $ 40$\% (30 MeV). 
\par
 The present PDG \cite{PDG04} determination of $\alpha_s$ from heavy quarkonium pulls down
 the global $\alpha_s$ average quite noticeably, due to an error that has been largely
 underestimated.
 Using the latest development in the calculation of
 relativistic corrections and on the treatment of perturbative series \cite{eft},
 it is conceivable to use 
 electromagnetic and hadronic decay widths, whose experimental 
 accuracy is already sensitive to next-to-leading corrections, to provide a competitive 
 sources of $m_c$ and $\alpha_s(m_c)$.
 \subsubsection*{Charmonium decays and transitions}
The study of decay observables has witnessed in the last years a remarkable 
progress. New experimental measurements, mainly coming from {\small BELLE}, BES, CLEO 
and E835 have improved existing data on inclusive, 
electromagnetic and several exclusive decay channels 
as well as on several electromagnetic 
and hadronic  transition amplitudes \cite{QWG,PDG04}. 
In some cases the new data have not only led  
to a reduction of the uncertainties but also to significant shifts in the 
central values. The error analysis of several correlated measurements
has evolved and improved our determination of quarkonium branching fractions.
 New data have  led to the discovery of new states.
 From a theoretical point of view several heavy quarkonium decay
observables may be studied nowadays in the framework of
effective field theories of QCD \cite{nrqcd,eft,decay}.
\par
By collecting huge statistical data samples one
can open the era 
of precision measurements on several charmonium decays and transitions.
In particular:
 \begin{itemize}
\item{\it Electromagnetic transitions.}
 1-2\% precision measurements on many radiative transitions
 will be possible, allowing access to the suppressed (M1, M2 and E2) amplitudes, which are
 mostly dependent on higher-order corrections and better test
 different theoretical approaches \cite{QWG,m1,m1lattice}.
% This is particularly needed  
% due to the existence of recent results on dipole transitions obtained 
% directly in QCD \cite{m1,m1lattice}.
 The transition $J/\psi \to \gamma \eta_c$ 
 is a good example. Such transition is presently known with a 
 $30$\%  error (only one  direct transition measurement \cite{crb}, 
several measurements of the BRs
 $J/\psi \to \eta_c \gamma \to \phi \phi \gamma$
 combined with  one independent measurement \cite{bel} of $\eta_c \to \phi\phi$) and enters 
 several charmonium BRs. Its uncertainty sets therefore 
 the experimental error of several measurements.
 %recent  calculation of M1
 Runs at the $\psi(2S)$ energy 
 will  also provide a very large sample of tagged $J/\psi$ decays (as more
  than half of these mesons decay to $J/\psi$), but 
 are also an excellent source of $\chi_c$'s, and, as recently shown, of $h_c$'s.
  \item{\it Radiative and hadronic decays.}
 Radiative and hadronic charmonium decays  involve several open puzzles.
 For example in exclusive hadronic decays the $\rho \pi$  puzzle 
 in $J/\psi$ and $\psi(2S)$ decays  is related to the anomalously small
 $\psi(2S)\to\rho\pi$ decay with respect to the $J/\psi \to\rho\pi$
 decay, which is the largest two body hadronic decay mode of the $J/\psi$ 
\cite{QWG,Bianco:2003vb}.
 New states  can be discovered in charmonium decays. An example is  the enhancement 
 in the radiative $J/\psi$ decay in proton-antiproton pair recently observed by BES
\cite{bbes}  and identified also with a candidate for baryonium.
 %$J/\psi$ to $X(1835)$
 \item{\it Hybrids and Glueballs.}
 The existence of gluonic excitations in the hadron spectrum is 
 one of the most important unanswered questions in hadronic physics.
 Lattice  QCD  predicts a rich spectroscopy of charmonium hybrid mesons \cite{QWG}.
 There are three important decay modes for charmonium hybrids
 \cite{QWG}:
  (i)  the decays to $D$ mesons
  (ii) the cascade 
  to conventional $c\bar{c}$ states
  (iii) decays to light hadrons via intermediate gluons,
   $\psi_g$ hybrids with exotic $J^{PC}$ quantum numbers offer the most 
  unambiguous signal 
  since they do not mix with conventional quarkonia. 
 Gluonic excitations may be studied through radiative decay, i.e. 
 $J/\psi\to\gamma X$ where $X$ is a glueball ~\cite{Close:1996yc}.
 Also  $\psi(2S)$ radiative decays
 are expected to be a prime source for glue-rich final states. 
 Although one expects the 
 majority of this data to come from $J/\psi$  running, $\psi(2S)$ decay would 
 also allow flavour tagging through the hadronic decays where a low mass 
 vector meson (i.e. $\rho$, $\omega$, $\phi$) replaces the radiative photon. 
 The possibility of studying $J/\psi$ decay 
 using $\psi(2S)$ running and tagging the 
 $J/\psi$ from $\psi(2S)\to\pi\pi J/\psi$ is also promising.
\item{\it Physics at  $\psi(3770)$.}
 A run at $\psi(3770)$ energy, besides
  measuring $f_D$ from $\bar DD$ decays, can also look for rare
  radiative and hadronic decays to lower $c \bar c$ states. This study
  can give a unique insight into the $S$-$D$ mixing and coupling to
  decay channel effects.  It may also give clues to the understanding
  of the $\rho$-$\pi$ puzzle. 
%For a discussion of the puzzle see \cite{Bianco:2003vb}.
 \end{itemize}
 \subsubsection*{Search for new  states.}
  The new
 resonances recently observed at {\small BELLE}, {\small BABAR}, CLEO, BES and Fermilab
 [$X(3872) $, $Y(4260)$, $X(3940)$, $Y(3940)$, $Z(3939)$] \cite{QWG2},
 open a discovery
 potential for states of new type (molecular, multiquark, heavy hybrids)
 never seen before  and of great impact for acquiring insight in the strong interaction 
 dynamics.
 Studies of narrow resonances via the radiative return method
 will be  feasible at the Frascati \tc factory and will 
 greatly enhance such discovery potential. 
  The observation of the $X(3872)$ \cite{QWG2} has been  the start of challenging searches for
 non-vector states across the open flavour threshold. This is probably
 the richest experimental field of research on heavy quarkonia at
 present.   The nature of this new, 
 narrow state is not yet clear, and 
 speculation ranges from a $\rm D^0\overline{D^{0*}}$ molecule, a $\rm ^3D_2$
 state to diquark-antidiquark state \cite{X}. 
 There are theoretical problems with all these interpretations, 
 and further, more accurate measurements of its width and 
 particularly of its 
 decay modes are needed to shed light on this state.
 Studies on the nature of the $X(3872)$ 
 can benefit from data taking at \tc factories.
 The study of the energy region above the $\rm D\overline{D}$ threshold is 
 one of the most interesting open problems in charmonium physics 
 and will require high-statistics.
 \subsubsection*{Investigation of low energy QCD.}
 The accuracy with which 
 the fundamental parameters can be measured is
 at present limited by nonperturbative contributions whose form is in
 many cases known \cite{eft}, but whose size is not known with sufficient
 precision.  Therefore, the main theoretical challenge is the precise
 determination of these nonperturbative contributions.  
 We may  use the 
 lower quarkonium states as a theoretically clean environment to
 study the interplay of perturbative and nonperturbative effects in QCD
 and extract nonperturbative contributions by comparison with data. 
 Therefore, precise quarkonium data are
  important today more than ever.  They may check factorisation,
 and severely constrain theoretical determinations and predictions.
\subsubsection*{Searches for new physics with charmonium.}
 Heavy quarkonium offers an interesting place where probing
 new physics which would manifest experimentally as 
 slight but observable modifications of decay rates and
 branching fractions; unexpected topologies in decays; CP and lepton
 flavour violation \cite{QWG}.\par\noindent
 {\it CP tests with $J/\psi$ decays.}
By using the decay mode
 $J/\psi\to\gamma\phi\phi$, the
 electric- and chromo-dipole moment can be probed
 at order of $10^{-13}e~{\rm cm} \sim 10^{-14}e~{\rm cm}$.
 A nonzero electrical dipole moment (EDM) of a quark or a lepton
 implies that CP symmetry is violated, since
 EDM's of quarks and leptons are very small
 in the SM, and, more importantly, that a signal exists for the intervention of BSM
 physics.
   \par \noindent
 {\it Lepton flavour violation.}
 Lepton flavour is violated in many extensions of the SM, 
 such as grand unified theories, 
 supersymmetric models, 
 left-right symmetric models, 
and models
 where the electroweak symmetry is dynamically broken. 
 Recent results 
 indicate that neutrinos have nonzero masses and can mix with 
 each other pointing to the fact that  lepton flavour may be a  broken symmetry in nature. 
 Lepton flavour violation (LFV)  can be tested 
 via the two-body $J/\psi$ decay (which conserves total lepton number):
 $ J/\psi \to \ell\ell'$
 with $\ell$ and $\ell'$ denoting charged leptons of different species.
 This process could occur at tree-level 
 induced by leptoquarks, sleptons (both in the $t$-channel) or mediated
 by $Z'$ bosons (in the $s$-channel).  
 The large sample ($5.8 \times 10^7$ events) collected in
 leptonic decays of $J/\psi$ resonances at BEPC and analysed by 
 BES up to now makes this search especially interesting; in fact, upper limits 
 for different lepton combinations have
 already been set at 90\% C.L. \cite{Bai:2003sv,unknown:2004nn}.
 In the future, larger samples collected at a \tc factory at LNF
 could allow to test LFV at a 
 higher precision, severely constraining new physics models.
%
%%OPEN CHARM
%
\subsubsection{Open charm  physics}

Studying charm decays at the threshold process $e^+e^- \rarr
\psi(3770) \rarr D \bar D$ offers many  advantages.
Threshold production of charm hadrons leads to extremely clean events, 
with optimum signal/background ratios;    
the background due to not-$D \overline D$ processes can be directly measured 
running below the production threshold.  
It is possible to tag the events to obtain absolute branching 
fraction measurements; the $D \bar D$ pairs are produced in a 
coherent quantum state providing information on D mixing 
and CP violation. In the SM, CP-violating amplitudes arise from penguin 
or box diagrams with b-quarks; however they are strongly suppressed by the 
small V$_{cb}$ V$_{ub}^*$ combination of the CKM matrix elements. 
The SM does predict CPV in Singly-Cabibbo suppressed modes at
 the level of 0.001 or so. No CP violation is allowed in the SM for
  Cabibbo-favoured nor Doubly-Cabibbo suppressed modes. SM does 
  direct CP in partial widths and in
 final state distributions (Dalitz plots, T-odd moments) and CP violation involving
oscillations. The question is whether close to  threshold one has systematic
advantages, when the CP asymmetries are very small, i.e 0.001 or less.
The $D \overline D$ pairs are produced in a $C=-1$ 
initial state so that final states containing two CP eigenstates 
of the same parity are a manifestation of CP violation. With integrated 
luminosities of the order of 100 fb$^{-1}$ the discovery window in the 
analysis of these final states could be extended to the 10$^{-4}$ level.   

Moreover, it is possible to put stringent limits on the oscillations in the 
charm sector running above the $\psi(3770)$ and comparing the 
correlations in $D$ decays produced in $e^+e^- \to D^0\bar D^0$,  
 $e^+e^- \to D^0\bar D^0 \gamma$ and 
$e^+e^- \to D^0\bar D^0\pi ^0$~\cite{Bianco:2003vb}. Although $ D^0\bar D^0 $ 
oscillations are not the most suitable tool to look for new physics, their 
study at the \tc factory  is unique because of the absence 
of concurrent 
Doubly Cabibbo Suppressed decays in most of final states. 
At a \tc factory  is also 
possible the measurement of the strong phase shift by comparing oscillations 
in hadronic and semileptonic final states. 
Other channels of interest are the FCNC decays, in particular the 
$D \to \pi$ $l^{+} l^{-} $ and $D \to \rho$ $l^{+} l^{-} $ which in the SM 
are expected to have a BR of the order of 10$^{-6}$:  effects of 
new physics could show up in the region of low di-lepton 
masses \cite{burdmanshipsey}. 
\par
Integrated luminosities of the order of 100 fb$^{-1}$ would provide 
statistics for many important studies of the SM, summarised in 
Table~\ref{TAB:CHARMREACH}
\begin{itemize} 
 \item  
  Precision measurement of the  $D^+ \to \mu ^+ \nu$ branching ratio, 
  allowing the extraction of the charm decay constant $f_D$, 
  can be carried out with unrivalled control over 
  systematics. Extracting precise numbers for the decay constant $f_D$  
  represents an important test for lattice QCD calculations ~\cite{LQCD4}
 and will thus indirectly  support the lattice  
  calculations done in the beauty sector
where direct 
 precision measurements are not available. If the \tc factory could run at the
  $D_s$ production threshold, one would be allowed to study 
   also $D^+_s \to \mu ^+ \nu$ and $D^+_s \to \tau ^+ \nu$.
%, that are expected 
%  to become very precise in this sector in the near future~\cite{LQCD4}.   
 \item  
  Significant improvements in the measurements of  
  $|V(cd)|$ and $|V(cs)|$ from $D \to \ell \nu \pi$ and $D \to \ell \nu K$  
  modes could be obtained, providing also another sensitive test for 
  the attainable precision with lattice QCD.
% that is expected to reach  
%  the percent level in this sector~\cite{LQCD4}. 
 Accurate charm data are very 
  useful to understand the reliability of the description of  
  non-perturbative dynamics. 
%  and thus indirectly to support the lattice  
%  calculations done in the beauty sector.
 \item  
  The absolute branching ratios for non-leptonic decays like  
  $D^0 \to K\pi$ and $D^+ \to K \pi \pi$  could be 
  measured with  uncertainties of the order of per mil. 
   Absolute measurements of hadronic
  charm meson branching fractions are relevant  in the study of the
  weak interactions because they are needed  to normalise several branching
  fractions,    from which CKM matrix elements are extracted. Better
  results on modes like $D^0\to K^-\pi^+$ or $D^+\to K^-\pi^+\pi^+$ have
  already been obtained by the CLEO-c collaboration, and more data will be
  provided soon  by BES-III. 
 These  data will help to get further insight into the scalar
  meson sector and a better determination of the parameters of the
  already well established resonances. 
  \item 
  Accurate studies of low mas hadronic spectroscopy in charm decays would be possible.
\end{itemize} 
There is a deep interplay between precision 
measurements in the charm sector and the present/future physics 
programs in the beauty sector.  
Absolute charm branching ratios and decay chains represent important  
inputs for $B$ decays, and the present uncertainties are becoming a 
bottleneck in the analysis of the beauty decays.  
%Includere Tabella
%, as it is shown in Tab.TABELLACHARM
%
\par
\begin{table}
 \begin{center} 
  \begin{tabular}{|l|r|r|}
    \hline\hline
  Charmed Meson   &  Produced & Detected \\ 
\hline
 $D^0$   & $400 \times 10^6$ & $160 \times 10^6$ \\
 $D^+$   & $160 \times 10^6$ & $63 \times 10^6$ \\
 $D^+_S$ & $30 \times 10^6$  & $9 \times 10^6$ \\
\hline\hline
   Mode        &   Decay Constant   &   $\Delta f_{D_q}/f_{D_q}$     \\
\hline 
  $D^+\rightarrow \mu^+ \nu$  &  $f_D$  &   0.5\%         \\
  $D^+_S\rightarrow \mu^+ \nu$  &  $f_{D_S}$ &   0.4\%    \\
  $D^+_S\rightarrow \tau^+ \nu$  &  $f_{D_S}$  &   0.3\%  \\
\hline\hline
   $\Delta (V_{cd})$ &  $\Delta (V_{cs})$ & 
        $\Delta R/R ; R\equiv |V_{cd}|/|V_{cs}|$ \\
\hline
  0.3\%    &    0.3\%    &  0.2\%   \\
\hline\hline
  Abs. Hadronic BRs  & Num. Double Tags ($\times 10^3$)  &  Stat. Error \\
 \hline
 $D^0$      &    1500     &   0.1\% \\
 $D^+$      &    1800     &   0.1\% \\
 $D^+_S$    &      180    &   0.3\% \\
\hline
\hline 
\end{tabular} 
\caption{Physics reach for 
outstanding SM studies at \tc factory. 
Estimates are extrapolated from CLEO-c, 
the factory presently operating at the Cornell 
storage ring, and projected to  100 fb$^{-1}$ luminosity.   
\label{TAB:CHARMREACH} }
\end{center} 
\end{table}

%%TAU
%
\subsubsection{$\tau$ physics}
The $\tau$ lepton is
 an ideal laboratory \cite{davier} for precise studies of  the electroweak and strong sector of the SM.
In addition, searches for physics beyond the SM can be
performed, with  precision measurements or direct searches
for non-SM processes. 
$\tau$-data can also be exploited to reconsider the contribution of the 
hadronic vacuum polarisation to $\alpha_{QED}$ and the anomalous magnetic moment of the muon.
\par
An $e^{+}e^{-}$ collider reaching ${\cal L} \sim 10^{34}$ cm$^{-2}$ s$^{-1}$ 
at $\sqrt {s}$ just 
above 3.7 GeV could deliver 100 fb$^{-1}$ per year (1 y = 10$^7$ s), 
thus allowing for the study, with $\sim$3 years of data taking, 
of the lepton-number violating channel $\tau \to \mu \gamma $ on the 
basis of a sample of 10$^9$ events, i.e. in the region of interest for the 
SUSY theories~\cite{taugsusy}.
\begin{itemize}

\item[] \emph{$\tau$ spectral functions, hadronic cross sections and 
$\tau$ decays}

Hadronic $\tau$ decays provide one of the most powerful testing ground 
of QCD \cite{davier}. This is due both to the high statistics 
and high precision obtainable  in the data and to the fact that the theoretical description 
is found to be dominated by perturbative QCD. Because of its large mass 
the $\tau$ can decay into hadrons while it 
has the QCD vacuum as the initial state and thus can  provide a 
particularly clean tool to investigate  strong interactions and charged 
weak hadronic currents.
 $\tau$ decays  reveal a rich structure of resonances, while 
the leptonic environment provides a way to isolate  clean hadronic systems 
and measure  their parameters.
Observables based on the spectral functions of hadronic $\tau$ decays can be used 
to obtain precise determinations of $\alpha_s$,  $m_s$, $V_{us}$ and parameters of the 
chiral Lagrangian.
$\tau$ decay results are complementary to the $e^+ e^-$ data to perform 
detailed studies at the fundamental level through the determination 
of the spectral functions. 
The non-strange vector spectral function 
is related, vis isospin symmetry, to
   the corresponding $e^+e^-$ spectral function. 
The precision reached makes it necessary to
correct for isospin-symmetry breaking. As discussed in Sect.\ref{Hadro} these vector spectral 
functions are used 
to compute vacuum polarisations, which enter the evaluation of 
the running of $\alpha$ 
and the muon anomalous magnetic moment.
 At present, $\tau$ and $e^+e^-$ data sets produce a discrepancy
   at the 2-3 $\sigma$ level which has to be clarified.
\blankline
\item[] \emph{New physics searches with  $\tau$.}
Tests of charged-current lepton universality,
electro-weak dipole moments and lepton-flavour violating decays such as
$\tau\to\mu\gamma$ and $\tau^-\to l^+ l^- l^-$ will be possible with the
future high-precision and high-statistics $\tau$ decay data.
Data in the $\tau$ sector should reach  levels below the
per-mil level.  The observation of non-zero weak dipole
moments would signal CP-violation BSM.
%\note{inserire commento di Ikaros su CPV in Tau. Stefano}
The highest cross section for $\tau \overline \tau $ production occurs 
at the $\psi \prime$ resonance (3.69 GeV) where visible cross section 
is approximately proportional to   1/$\sigma$(E) ($\sigma$(E) is 
the machine beam energy spread) and it is $\sim$3 nb for machines adopting 
standard optics. 
The recent upper limit obtained by {\small BABAR} Collaboration for 
the $\tau \to \mu \gamma $~\cite{TMUGBabar}  
is BR( $\tau \to \mu \gamma $)$\leq$ 6.8$\times$ 10$^{-8}$ at 90\% C.L., 
based on the analysis of 
$\sim2 \times 10^8$ produced $\tau$ pairs.
Radiative $\tau$ decays such as $\tau \to \pi  \nu \gamma$ and 
$\tau \to \mu \pi \nu \nu \gamma$ are for the same reason best 
studied immediately above $\tau$ threshold.  
Moreover, the $\tau$ physics program includes  
precise measurement of the $\tau$ mass, sensitive studies of weak couplings 
and lepton universality via purely leptonic and 
semileptonic decays,  
%a new direct limit on the $\nu_{\tau}$ mass, 
%the study of the radiative decays 
and the measurement of other rare processes as those
involving kaons. As stated above, thanks to the kinematic constraints, 
the $\tau \to \pi \pi^{0} \nu$ channel can be well separated from 
the other decays giving the opportunity for a new test of the CVC 
rule via the comparison with the $e^{+}e^{-} \to \pi^{+} \pi^{-}$ data.  

\item[] \emph{CPV asymmetries in $\tau\rightarrow K_{S,L} \pi \nu$.}
 It has been pointed out recently\cite{Bigi:2005ts} how threshold production 
and decay of
 $\tau^+\tau^-$ pairs, such as $\tau\rightarrow K_{S,L} \pi \nu$ 
 would provide information on known SM sources of CP
 violations, complementary to
 $K_{S,L}$ semileptonic decays.
\end{itemize}

\subsection{\bf Comparison with present and future competitors}
\label{SECT:COMPETITORS}
CESR at Cornell, Ithaca (US), after almost twenty years of successful 
operation as a B-factory,
with a long lasting record in peak luminosity (1.3$\times$10$^{33}$
cm$^{-2}$ s$^{-1}$ at the $\Upsilon(4S)$ energy), has been recently modified in
order to run at 
lower center-of-mass energies (CESR-c) \cite{ref5qgwg,QWG,Cassel03vr,Heltsley04}. SC wigglers have been added
to increase the radiation damping and improve luminosity. CESR-c will
run until 2008 
%with two asymmetric energy beams, 
at the $\tau$, $D \overline D$ and $D_s \overline D_s$ thresholds. 
The goal peak luminosity is 3$\times$10$^{32}$ cm$^{-2}$ s$^{-1}$. 
   During first quarter of 2004, CLEO-c has been 
   running\cite{Heltsley04} at a peak luminosity of about 
   $5 \times 10^{31}$ cm$^{-2}$s$^{-1}$. Their plan from today to 2007
   is to integrate  3~fb$^{-1}$ at the $\psi(3700)$ (or $1.3 \times 10^9$ 
   $\jp$), then switch to the $D_sD_s$
   threshold and accumulate another 3~fb$^{-1}$ (or $3 \times 10^7$ events), 
   and finally to switch to the $\jp$ aiming to collect $10^9$ events, i.e., 
   twenty times the BES statistics \cite{QWG}.
\par
BEPC\cite{Liu04}, at Beijing (China), which reached a peak luminosity of
1.1$\times$10$^{31}$ cm$^{-2}$ s$^{-1}$, has been dismounted and is
being upgraded to become BEPC-II, the first completely dedicated
\tc factory, still maintaining synchrotron radiation
production. Its design is based on a double ring scheme, with energies
ranging between 1.5 and 2.5 GeV/beam, optimised at 1.89 GeV, with a
design luminosity of 10$^{33}$\cm2s1. A new inner
ring will be installed inside the old one, so that each beam will
travel in half outer and half inner ring. SC cavities will also be
installed. A by-pass will allow the electron ring to be used as a
synchrotron light source as well. Commissioning of the new rings is 
planned for summer 2007. Experiment BES-III, that is an upgraded version of
BES which provided  record samples of 
$J/\psi$'s and $\psi^\prime$'s in the last years, will run a new intensive 
 program at these energies from 2007 on. 
\cite{QWG}.
\par
CLEO-c and BES address topics in both open charm and quarkonium physics. The world scenario 
for quarkonium \cite{QWG}, however, is not limited to them but sees other competitors
\begin{itemize}
\item \tc  factories:
 besides BES and CLEO-c, KEDR which, exploiting 
the polarimeter in the VEPP-4 collider, 
has recently provided high precision 
measurements of $J/\psi$ and $\psi^\prime$ masses;
\item  B-factories: after CLEO,  {\small BABAR} and {\small BELLE},  have proved their large physics potential also 
as charmonium factories, through a rich variety of reactions(B decays to charmonium,  
$\gamma\gamma$ , ISR, double $c\bar{c}$);
\item $\bar{p}p$ charmonium factory: the Antiproton Accumulator of the 
Tevatron, at Fermilab, was exploited by the  E835 experiment, 
to scan all
known narrow charmonium states in formation from $p \bar{p}$ annihilation.
\end{itemize}
\par
In these last years, clean record samples of all the narrow vector resonances
have been accumulated. 
Table \ref{ch2:ccbars} (taken from Ref. \cite{QWG}) shows the record samples of charmonia 
produced (or formed) in
 one B-factory (via B decays, $\gamma\gamma$, radiative return) with 
250 $\mathrm{fb}^{-1}$ (such quantity is continuously increasing at present); 
 the highest statistics runs recently done by the \tc factory BES
(58 M $J\psi$'s and 14 M $\psi(2S)$); and
 the data samples formed in the $p \bar{p}$ charmonium experiment E835.
Another $p \bar{p}$ charmonium factory is going to start data taking 
at GSI in the next decade.

\begin{table}[htb]
\begin{center}
\label{ch2:ccbars}
\begin{tabular}{|l|lllllll|}
\hline
Particle & $\psi(2S)$ & $\eta_c(2S)$ & $\chi_{c2}$ & 
$\chi_{c1}$ & $\chi_{c0}$ & $J/\psi$ & $\eta_c(1s)$\\
\hline 
 B decays      & 0.8M & 0.4M & 0.3M & 0.9M & 0.75M & 2.5M & 0.75M \\
$\gamma\gamma$ &  -   & 1.6M & 1M &   -   & 1.2M &  -   & 8.0 M  \\
 ISR           &  4M  &    -  &   -  & -     &  -    & 9M & -    \\
 $\psi(2S)$ decays & 14M & ?  & 0.9M & 1.2M & 1.2M & 8.1M & 39K      \\
 $J/\psi$ decays &   -  &   -   &   -  &   -   &   -   & 58M  & 0.14 M \\
 $p \bar{p}$      & 2.8M &   ?   &  1M   &  1M   &  1.2M & 0.8M & 7M \\
\hline 
\end{tabular}
\caption{Charmonium states
produced in the B-factories and \tc factories,  or formed in $p \bar{p}$ 
(from Ref.\cite{QWG}).}
\end{center}
\end{table}
\subsubsection{Systematic limits of the present generation of high statistics experiments}
 Running at $\ddb$ threshold has obvious advantages in terms of number of charm  events over background,  ratio of events produced, 
systematics and rates. Similar considerations do apply for threshold production of $\tau$ leptons.
As an example, the sensitivity of the {\small BABAR} experiment  for 
the $\tau \to \mu \gamma $~\cite{TMUGBabar}  decay discussed above is 
limited by the background, mainly  
from $\mu \mu (\gamma)$ and $\tau \tau (\gamma)$ events.
Running near threshold, the radiative $\tau \tau \gamma$ component is  
suppressed, and the kinematics of the 2-body $\tau$ decays is 
quasi-monochromatic\cite{LNFF5}, giving further elements 
besides statistics to definitively improve the results from the B-factories. 
Radiative $\tau$ decays such as $\tau \to \pi  \nu \gamma$ and 
$\tau \to \mu \pi \nu \nu \gamma$ are for the same reason best 
studied immediately above $\tau$ threshold.  
Finally, 
results from all $e^+ e^-$ experiments crucially depend on the subtraction of 
radiative corrections on the initial state \cite{QWG}.
\subsection{\bf Conclusions}
When the LHC will have directly probed 
 the physics at the TeV scale, precision studies of
flavour physics will be necessary. Within flavour physics, the charm quark plays
an important role\cite{Bianco:2003vb,burdmanshipsey}. The physics
program at a \tc factory is very    
broad and robust\cite{snowmass01,QWG} with a very strong list of SM
issues such as charmonium spectroscopy and decays, absolute branching
ratios, decay constants and CKM matrix elements from semileptonic decays.
%makes a \tc factory an enterprise of much broader scope than the 2.5 GeV
%option described in the other sections of this report. 
Besides, relevant
BSM topics contribute to the understanding of the LHC scenario: CP
violation, mixing, charm rare decays, tau rare decays.   
\par
Some of these topics are unique to \tc factories with respect to their
direct competitors, the SuperB-factories \cite{SUPERBWG}: absolute
branching ratios, purely leptonic decays.
Other topics, both BSM- and SM-related, benefit from threshold running at
\tc factories, such as mixing, semileptonic decays for meson decay constants,
$\tau$ rare decays, searches for CP violation in the charm system, charmonium spectroscopy.      
\par
The interplay, compatibility, complementarity and overlap of the physics
programs of future \tc factory and SuperB factory were subject recently of
very detailed and specific discussion at DIF06 Workshop~\cite{DIF06PROC}. 
Main conclusions
of the Workshop were
\begin{itemize}
 \item 
   Flavour physics is crucial in the LHC era to constrain the flavour structure of the BSM
   physics found;
 \item
    If no BSM physics is found, flavour physics may be the only tool we shall have to
    acquire  clues on the features of BSM physics;
 \item
   A Super-B factory is also a powerful $\tau$-factory --- in the $\tau$ sector, very few
   measurements are performed with great advantage running at threshold rather than at
   the $\Upsilon (4S)$;
 \item 
   A Super-B factory is also a charm-factory, although in this case running at charm
   threshold preserves a few uniqueness, such as charm mixing and absolute branching
   ratios;
 \item
  A Super-Flavour factory (i.e., a collider running at the $\Upsilon (4S)$ at
  $10^{36}$\cm2s1 which retains the capability of lowering energy down to the $\jp$ with
  peak luminosity there of $10^{34}$\cm2s1) is a very attractive concept of a machine able
  to fulfil the need of a thorough investigation of the flavour sector;
 \item
 A Super-Flavour factory machine design could be based on new ideas
 ~\cite{DIF06PROC}, which heavily exploit the ongoing R\& D for the ILC;
 \item 
 Technical feasibility, cost and schedules for both machine and detector at a
 Super-Flavour factory need very careful insights.
\end{itemize}
\par
Even if one restricts the scope to a \tc factory, 
machine\cite{Raimondi05,notemacchina} and detector are expected to
be challenging. 
%Recent estimates for the cost of the collider are
%in the range of 40~M\textgreek{\euro} for accelerator and 20~M\textgreek{\euro} for Linac and
%dipoles \cite{Raimondi05}.   
 Expertise on both machine and detector does exist in Frascati, but it is
clear that such an enterprise must be an international joint
venture. The new \tc factory rings would not fit the existing building, 
and it is likely that a new   
 site close to Frascati should be selected, possibly in synergy with
facilities compatible in energy and luminosity, such as medical therapy,
FEL, material science and synchrotron radiation.   
\par
Finally, very strong competition does exist from the Beijing machine,
which is scheduled to start commissioning in 2007. A \tc factory in
Frascati with $10^{34}$\cm2s1 would provide a 10-fold
event sample with
respect to the Beijing collider starting from 2013, its first planned year of
operation. A Frascati \tc factory project indeed requires to meet a very
tight schedule to be a winner.    

\section{Detector considerations}
\label{DetAcc}

\newcommand{\bfor}{\begin{displaymath}}
\newcommand{\efor}{\end{displaymath}}
\newcommand{\kc}{$K^{\pm} \hspace{0.1cm}$}
\newcommand{\daf}{{\small DAFNE}-2\hspace{0.1cm}}
We discuss here the requirements posed by the {\small DAFNE}-2 physics
program outlined in Sect.\ref{DAFNE-2}, on the possible detectors that can
be used. This section complements the considerations done on each single
section.  

The variety of physics items described above puts forward a number of
detector requirements well met by a general purpose
detector, apart few special cases, namely the small angle tagger for
the $\gamma\gamma$ physics, the neutron detector and the polarimeter for
the nucleon form factor 
measurement. The small angle tagger for $\gamma\gamma$ physics has been partially
discussed in the section dedicated to such final states (see
Sect.\ref{Gamma}). 
Its design, which is
in principle very simple, depends very much on the accelerator choices and
would be impossible here to discuss it in more details. However it can be
realized with a moderate investment.\par
All the other requirements are easily summarised as follows:
\begin{enumerate}
\item full angular coverage;
\item efficient tracking, providing good momentum resolution down to low
  momentum values;
\item hermetic calorimetry with excellent photon detection capability;
\item good particle identification performances.
\end{enumerate}
The first design of {\small DAFNE-2} foresees one interaction region
only. This implies that either one single detector is used, or that the
possibility to move in and out different detectors in the same experimental
region has to be considered.
\par 
In any case, for the sake of minimising the cost it is wise to
consider the possibility of re-using large parts of the presently operating
detectors at the {\small DAFNE} machine, namely {\small KLOE} \cite{KLOE} 
and Finuda \cite{Finuda}.  
Out of the two, only {\small KLOE} has been designed to be a
general purpose detector, and thus matches broadly the requirements just 
stated. In fact, the {\small KLOE} detector provides full angular coverage, and
consists of a large Helium drift chamber\cite{kloe_dc} immersed in a 5 kG 
solenoidal magnetic field, providing a momentum resolution of 0.4\% for
track with polar angle above $45{}^{\circ}$, surrounded by a
lead-scintillating fibers calorimeter\cite{kloe_cal} able to detect photons
down to few MeV with good efficiency and providing an energy
resolution of 5.7\%/$\sqrt{E}$ and a time resolution of 54 ps/$\sqrt{E}\,
\oplus\, 50$ ps. The excellent time of flight measurement provides also good
particle identification.\par 
The Finuda detector\cite{Finuda}, designed and optimised for hypernuclear
physics, has been shown recently well-suited for the determination
of the nucleon form factors (both in the proton and in
the neutron final state). With
the implementation of a polarimeter in the upgraded detector, it is also able
to measure the normal polarisation of the outgoing baryon and hence
the phases of the form factors. We refer the
interested reader to the Letter of Intent recently submitted
\cite{LoI}.
 \par
In the following we will make the assumption that the {\small KLOE}
detector can be used as a starting point to realize a detector for \daf .
We will consider first the issues related to operating KLOE as it is, and secondly we 
will discuss the possible detector upgrades.\par  
%To be more explicit: the ``zeroth order approximation'' would be to plan
%to re-use {\small KLOE} as it is, while ``higher orders'' would imply the design
%and construction of a certain number of upgrades.

\subsection{\bf Operating the {\small KLOE} detector at \daf }
To run at \daf without major modifications, the {\small KLOE} detector must face
essentially 3 challenges:
\begin{enumerate}
\item the aging of the electronics and the obsolescence of all the
  computing systems, both online and off-line;
\item the increase in event rate, by a factor of $\sim 10$, due to the 
increased luminosity of the machine at the $\phi$ peak;
\item the increase in center of mass energy, by a factor of $\sim 2$,
  needed for some of the measurements described in the previous sections.
\end{enumerate}
The first point is of key importance, given the fact that the \daf machine
is not expected to start operating before year 2010, and will have to
run for at least 5 years. 

%Even if the failure rate of the {\small KLOE} electronics
%is still very low, it is very unlikely that it would keep at such level for
%other 10 years or more. Moreover, the spare modules are very limited and
%they cannot be duplicated because most of their components are not anymore
%available on the market. This has a strong impact on the front-end
%electronics, on the trigger
%and also on the data acquisition system.

\subsubsection*{Front-end electronics}
Aging is a common issue both for calorimeter and chamber electronics, and
the only reasonable solution can be to redo (and probably redesign) a
sizeable part of it. On the other hand, the
increased rate should not be a big problem, provided the on-board data
buffers are replaced with deeper ones. The energy increase instead will 
require to extend the dynamic range of the calorimeter ADCs. 
%a technical solution will have to be studied in order not to
%destroy the sensitivity at low energy, while allowing the detection of
%tougher showers. 

\subsubsection*{Trigger}
Apart from the common aging problem, the {\small KLOE} trigger
system\cite{kloe_trig} would not have
problems to run at an increased energy, while the rate increase would
certainly be an issue, particularly if we consider a possible increase also
of the machine background, at least in the first 1$\div$2 years of operation.
In more detail, it has been designed as "minimum bias" trigger, made of
two independent systems: one based on the calorimeter multiplicity and the
other on the drift chamber multiplicity. The system results in a rate of 2$\div$3 kHz with
the present {\small DAFNE} luminosity and background conditions.\par
At \daf minimum bias trigger with the same philosophy would produce a rate around 
20 kHz, putting severe constraints both on data acquisition and on
the data processing and storage. However a different philosophy (selective
triggers) seems really unaffordable given the variety of the physics channels
to be studied and the precision required for some of the measurements (for
example the hadronic cross section), where even small trigger bias could be
very important. But at least the minimum bias trigger could be fruitfully
completed by an effective third level stage, based on online event
reconstruction, able to reduce the machine background, the cosmic 
and noise events stored on tape.\par

\subsubsection*{Data acquisition}
Here again the aging issue is critical for the 
entire {\small KLOE} data acquisition system\cite{kloe_daq}, and applies not only
to custom boards but also to commercial CPUs, operating systems and data
transmission protocols. The construction of a brand new system is highly
advisable. However the architecture of the {\small KLOE} system can be safely
extended to the \daf environment, provided some modifications are
implemented in order to increase the maximum sustainable rate.
In fact the system was designed to sustain 50 Mbyte/s and has been tested 
up to 80 Mbyte/s, while the usual bandwidth 
during the present {\small KLOE} run does not exceed 10 Mbytes/s. If the
minimum bias philosophy is kept for the trigger at \daf , and if the event
size will increase (in case the {\small KLOE} detector is upgraded in some parts),
it is necessary to foresee a much higher data throughput.\par

\subsection{\bf Possible upgrades to the {\small KLOE} detector}
The {\small KLOE} detector was optimised to efficiently detect in-flight
decays of the $K_L$, thus leaving some important weak points from the point
of view of \daf physics. In fact all the channels discussed in this
document include prompt particles, with momentum spectra often extended to
very low values, whose detection could be inefficient in {\small KLOE} due to the
absence of any tracking device in a radius of 25 cm from the interaction point.
Also, the value of the magnetic field is such that many low momentum tracks
may escape detection by spiralising before entering any detector.\par
On the other hand, a number of the final states we are considering
contains many
photons, for which the readout granularity of the {\small KLOE} calorimeter has
proven to be too coarse, thus inducing accidental cluster splitting or
merging. A refinement of the calorimeter readout granularity would be also
of great help in $e/\mu/\pi$ separation, which is required to reject
background in various channels.\par 

\subsubsection*{Magnetic field}
\begin{figure}[htbp]
\begin{center}
\vspace{8 cm}
\includegraphics{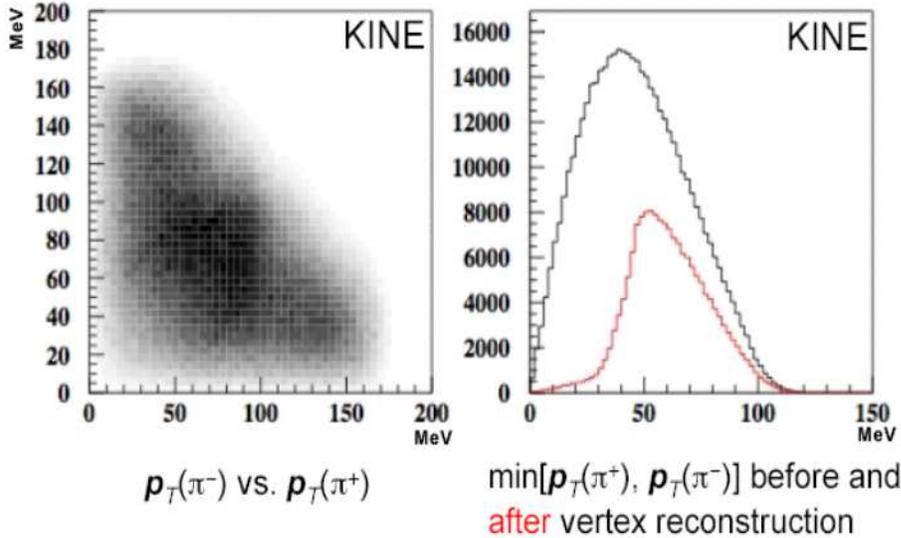}
\caption{ (Left) momentum distribution of the charged pions from 
 $K_S \rightarrow\pi^0\pi^+\pi^-$. 
Right) Momentum distribution of the 
same charged pions at generation level and 
after the reconstruction } 
\label{fig:ks3p}
\end{center}
\end{figure}

The magnetic field value used in {\small KLOE} ($\simeq 0.52 T$) was chosen to maximise
the kinematical separation between the $K_L \rightarrow \pi^+\pi^-$ and the
$K_L\rightarrow \pi \mu \nu$ decays in the drift chamber volume. 
Due to such high B field a sizeable fraction of the low momentum tracks 
coming from the interaction point falls out of the detector 
acceptance. Furthermore, the smaller the track curvature radius, the higher
the probability that the pattern recognition splits it in two or more segments.
An example of this effect is shown in Fig.\ref{fig:ks3p} for simulated $K_S
\rightarrow\pi^0\pi^+\pi^-$ decays, which are kinematically very similar to
$\eta \rightarrow\pi^0\pi^+\pi^-$. The momentum distribution of the two
charged pions at the generation level and after the reconstruction is shown
in the right plot, where it is clear the acceptance drop at low momenta.\par

However it must be kept into account that lowering the magnetic field would 
have an effect on momentum resolution and vertex efficiency, which would be
further worsened in the \daf high energy run. The prediction of such
effects and their convolution as a function of the B field value is not
trivial and depends very much on the specific final states. A careful
simulation is necessary, selecting a subset of benchmark channels.
However, it is clear that the run at the $\phi$ and the high energy run may
use a different magnetic field value.\par  
%
% vertex
%
\subsubsection*{Inner tracker}
There is no doubt that a tracking device as close as possible to the
interaction point, providing 3-dimensional spatial measurements with
moderate hit resolution (few hundreds $\mu m$), would greatly improve the
{\small KLOE} tracking and vertexing performances, and would allow to safely afford
all the measurements discussed in this document. However the choice of the
detector to be used and its design must be carefully tuned according to the
following considerations:
\begin{enumerate}
\item compatibility with the kaon physics program at \daf ;
\item amount of material required (for the detector, its supporting
  structure, electronics and cables) and its effects on multiple scattering and photon conversions;
\item sustainable rate and occupancy, mainly with respect to Bhabha events
  rate and machine background.
\end{enumerate}
The first point is due because it is clear that during the run at the
$\phi$ peak a kaon physics program will be also pursued, using exactly the
same detector we are considering here. Up to now we have not considered the
(more stringent) detector requirements coming from kaon physics, but in
this case we cannot avoid it, because an inner tracker too close to the
interaction region would definitely spoil any study of kaon
interferometry. 
The minimal distance required amounts to 
20 $K_S$ lifetimes, i.e. about 12 cm. This essentially means that the
spherical beam pipe structure used by {\small KLOE} will have to be kept at \daf ,
and that a possible cylindrical inner tracker must have inner radius
$\geq $ 12 cm and outer radius $\leq$ 25 cm (which is the {\small KLOE} chamber
inner wall radius).\par
\begin{figure}[htbp]
\begin{center}
\vspace{7 cm}
\includegraphics{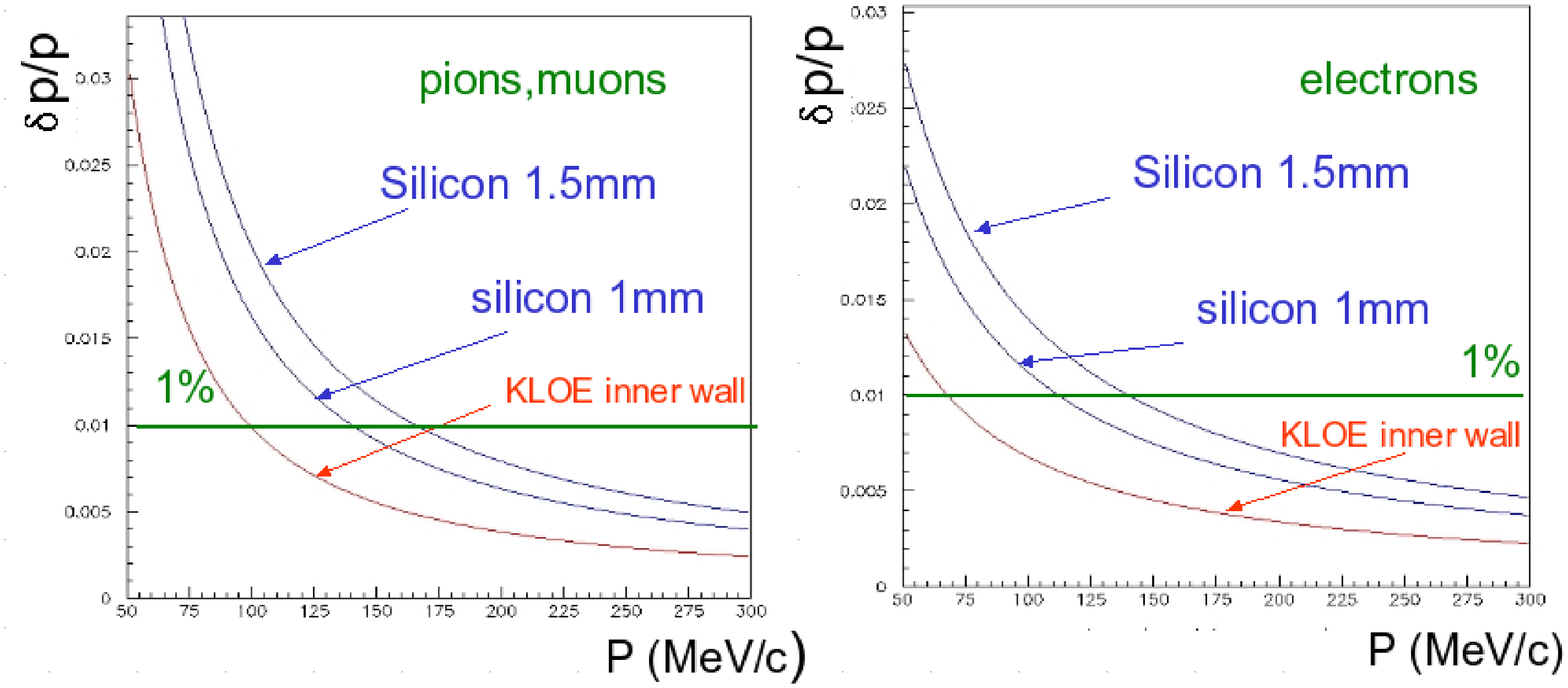}
\caption{ Relative uncertainty on the momentum measurement as a function
of the particle momentum for different thickness of silicon equivalent compared
to the uncertainty due to the {\small KLOE} drift chamber inner wall}
\label{fig:mate}
\end{center}
\end{figure}

For low momentum tracks the multiple scattering contribution to the momentum
and vertex resolution is dominant. Vertex resolution is not crucial for
most of the measurements we are considering here, but the momentum
resolution must be
kept very good to allow the kinematic closure of events and reduce
selection systematics. Moreover, when searching for rare decays, small non 
Gaussian 
tails in the measured distributions should be avoided in order to keep the
background under control. A first study of the multiple
scattering influence on the momentum resolution has been performed, and is 
shown in 
Fig.\ref{fig:mate}, where the effect of {\small KLOE} drift chamber inner wall is
compared with what would happen adding 1 mm or 1.5 mm of equivalent silicon
thickness. 
The reference line of $\frac{\delta p}{p} = 1 \%$ is also drawn: it is
clear that below 100 MeV the resolution worsens very rapidly, and it would
be a good choice not to exceed 1 mm of silicon equivalent thickness 
($\sim$ 1\% $X_0$) for the whole inner tracker.\par

The event rate coming from $\phi$ events at \daf is expected to be of the
order of 10 kHz, with low particle multiplicity, and then
should be easily tolerable for any type of inner tracker. Bhabha events
may be a problem instead, but they are peaked in the forward region: the
event rate can be kept under control by
simply limiting the inner tracker acceptance at polar angles above 
$30^{\circ}$. On the other hand it is very difficult to estimate the machine 
background contribution,
because its rate, composition (electron vs photons), and spatial
and momentum distribution are highly dependent on the machine optics and 
operation conditions. However if we scale 
the single counting rate observed on the {\small KLOE} drift chamber  
wires closer to the IP (at a radius of $\simeq $ 28 cm ), we obtain that 
a cilyndrical active layer at a radius $ 12\div 20 cm$  from IP would see a
rate of the order of 10$\div$20 MHz. Such rough estimate is consistent with the
one based on the counting rate of the Finuda vertex detector: at a radius
of 6 cm from IP the 
silicon detector collected 7/8 hits/planes in $2 \mu s$  of integration time 
at a luminosity near $10^{32}cm^{-2}s^{-1}$, which scaled to \daf expected
luminosity yields a rate around 30 MHz. To cope with such environment, the
tracker should be designed as fast as possible, with 
a short integration time to minimise the pile-up of machine 
background.\par

Obviously a detailed simulation study is needed before the choice of a specific
technical solution, however, bearing in mind the considerations reported above, 
we suggest to discuss 3 possible choices:
\begin{enumerate}  
\item Light drift chamber : designed as an extrapolation toward IP of the 
{\small KLOE} drift chamber.The longitudinal coordinate can be given both by
 stereo geometry wires and by charge division. It would have the advantage
 to be very light and to require a limited number of channels (of the order
 of few hundreds), but the disadvantage of being slow due to the drift.
However a suitable choice of the cell size (for example $0.5\times 0.5
cm^2$) could solve the problem.
\item Silicon detector: organised in 4 or 5 cylindrical planes of the well
  established double sided strip readout, which would provide 3-dimensional
  space measurements with very good hit resolution, and maybe also $dE/dx$
  measurement. If the strip pitch is not chosen too small (also 1 mm could
  be enough for many measurements) the number of channels could be limited
  around $10^4$, depending on the number of layers. However the big
  disadvantage of this solution if the material thickness required, not only
  for the detector but also for the electronics and support structure. 
\item Cylindrical GEM : this very promising new technology has been widely
  developed and tested at LNF\cite{gem}, where a valuable expertise is present.
 Three dimensional spatial measurement is easily implemented, and the
 overall material thickness can be kept very low. However if 1 cm${}^2$ readout
 pads are implemented, around $3\times 10^4$ channels are required.
\end{enumerate}

\subsubsection*{Calorimeter}
The {\small KLOE} calorimeter read-out cell size is $4.4\times4.4 cm^2$, for a 
total of 5000 read-out cells, equally divided between barrel and endcaps.
Such a segmentation has to be compared with a Moliere radius of the order 
of 1.5 cm. A finer granularity is then suggested by the natural transverse 
size of the showers in the calorimeter, and could be of great help to minimise 
cluster splitting and/or merging, and would allow a better cluster 
shape analysis for particle identification. \par
\begin{figure}
\begin{center}
\vspace{8 cm}
\includegraphics{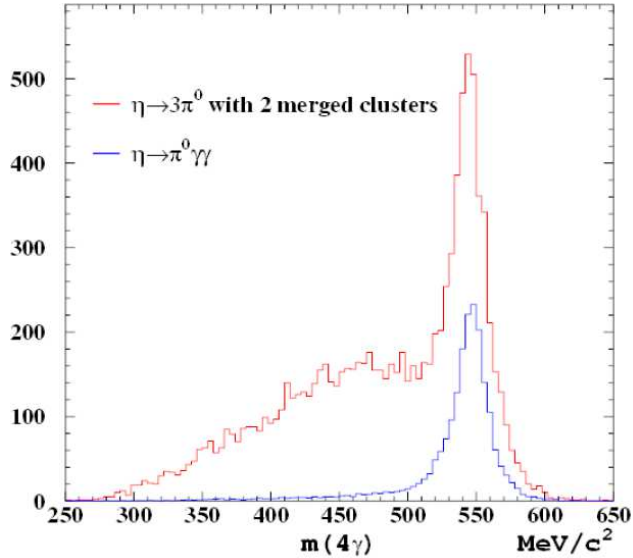}
\caption{ Invariant mass of the four photons for the 
$\eta \rightarrow \pi^0 \gamma \gamma$ events and for the  
$\eta \rightarrow 3\pi^0$ background with two cluster merging}
\label{fig:eta}
\end{center}
\end{figure}

The effect of cluster splitting/merging affects any analysis with cluster
counting, and in particular the search for rare decays into neutrals.
%As an example of this effect we refer to $K_S \rightarrow 3\pi^0$ {\small KLOE} 
%analysis where one of the main background is due to the events made of
%a $K_S \rightarrow 2\pi^0$ decay with double splitting of the photon clusters.
%In fig. \ref{fig:3pi0} we show the different population of the $\chi^2$ 
%variable in $3\pi^0$ and $2\pi^0$ hypothesis for the MC signal and for the
%data events.
In fig \ref{fig:eta} is shown a striking example of cluster merging
in the search for $\eta \rightarrow \pi^0 \gamma \gamma$ events, where a 
huge background survives the kinematical fit coming from
$\eta \rightarrow 3\pi^0$  decays with double merged clusters. 
Due to the merging of two couple of photons the topology of the 
background  becomes equal to the signal, since the invariant mass of the 
four photons peaks anyway at the $\eta$ mass.\par

%\begin{figure}
%\vspace{8 cm}
%\special{psfile=3pi0.eps hscale=50 vscale=50}
%\caption{ Distribution of the $\chi^2$ variable in $3\pi^0$ and $2\pi^0$ 
%hypothesys of MC $K_S\rightarrow3\pi^0$ signal and of the data}
%\label{fig:3pi0}
%\end{figure}

The best readout granularity choice can be studied by a detailed simulation
of the calorimeter, using a package, like FLUKA or GEANT4, which implements
an accurate description of low energy processes. Such simulations should
however be verified by test beams, where also the technical solution to
refine granularity could be tested.\par

The read-out device to be used with the new granularity must not spoil the
excellent energy and time resolution of the calorimeter, i.e. it should have
high quantum efficiency and fast timing performances. At the moment the
most interesting solutions on the market are the multi-anode photomultipliers 
(R760000 M4 or M16 or M64 by Hamamatsu) which have rise-time and quantum 
efficiency similar to the {\small KLOE} photomultipliers.\par 
The increase in the number of channels can be really sizeable: a simple
reduction by a factor of 2 in the cell linear size results in a factor of 4
increase of the number of channels. However it is also possible to
implement the granularity refinement only on the first 2 planes of the
calorimeter, provided the simulation and the tests show this is sufficient
to reduce clusterization errors. To cope with a high number of channels, it
would be advisable to process the signal as near as possible to the
detector, and send out only the digitised information.\par

\subsection{\bf Detector requirements for nucleon form factor measurements}
The {\small KLOE} detector is in principle well-suited for the measurement
of the proton form factor, apart from the proton polarisation and also for
$\Lambda$ and $\Sigma$ form factors, through the measurement of their decay
products. On the other hand, concerning the
neutron form factor, the key point is to understand the efficiency for
neutron detection. We discuss here the two items of neutron detection
efficiency and of proton polarisation measurement.
\par  
The neutron detection capability of the {\small KLOE} detector 
is not known up to now. Although neutron
detection with bulk organic scintillators is known to have roughly an efficiency
of 1\% per each cm of scintillator thickness, it is not clear how this
figure can be applied to the peculiar lead-scintillator structure of the
{\small KLOE} calorimeter, where an equivalent thickness of about 11 cm is
embedded in a fine grain sampling structure.
Studies on this subject are in progress.\par
%both using the present {\small KLOE}
%data, and also using a detailed simulation of nucleon interaction in the
%calorimeter. Moreover a test beam at a neutron facility will be done to
%clarify the situation. \par
A proton polarimeter is also required for the form factor measurement. Such
a device normally consists of a layer of carbon placed between two precise
tracking devices, typically silicon detectors. This object cannot be easily
incorporated in the {\small KLOE} structure and would spoil the tracking resolution
of the detector. It should then be inserted only for a dedicated
run,  maybe replacing part of the beam pipe or of the inner tracker.\par
Finally the wide program of measurements of the $KN$ interactions in the
$p_K\sim 100MeV/c$ momentum region, requires in principle several different
gaseous targets around the interaction region. On this respect, as pointed
out in Sect.\ref{Kaons}, the present {\small KLOE} Drift Chamber, filled with
a helium based gas mixture provides a good starting point for a complete
measurement of kaon interactions on $^4He$ nuclei.
\section{Summary}
\label{Concl}

The {\small DAFNE}-2 project,
starting around the year 2011, will have a relevant impact on a wide variety of
physics topics, ranging from
precision tests of the Standard Model to several controversial subjects in
the field of hadronic
physics. 
For each single item, we have compared the physics potential of the project
with the possible present and future competitors.
In Tab.\ref{competitors} we summarise the most relevant competitors that we
have considered.
\begin{table}[ht]
  \centering
  \begin{tabular}{| c | c | c |}
    \hline
 Experiment & Measurements & Time scale \\
    \hline
 VEPP-2000 & Hadronic cross section 0.4$\div$2.0 GeV & Start in 2007 \\
           & Proton and neutron FF at threshold & \\
\hline
 BES-III   &Hadronic cross section $\geq2.4$ GeV & Start in 2007 \\
           & Proton FF for $q^2\geq6$ GeV$^2$ & \\
\hline
{\small BABAR} & Hadronic cross section 0.3$\div$5 GeV & Data taking up to
 2008 \\
               & Proton FF for $q^2$ up to $\sim 10$ GeV$^2$ & \\
               & $\gamma\gamma$ physics (?) & \\
\hline
{\small BELLE} & Hadronic cross section 0.3$\div$5 GeV (?)& Data taking  \\
               & Proton FF for $q^2$ up to $\sim 10$ GeV$^2$ (?) & \\
               & $\gamma\gamma$ physics: ($\gamma\gamma\to\pi^+\pi^-$)
 $W_{\gamma\gamma}\geq 700$ MeV  & \\
\hline
{\small PANDA} & Proton FF up to high $q^2$ & Start in 2013 \\
\hline
{\small PAX}   & Proton FF (including polarisation) up to high $q^2$ &
 Beyond 2015 \\
\hline
CrystalBall @ MAMI & $\eta$ and $\eta'$ physics & Data taking \\
\hline 
WASA @ COSY & $\eta$ and $\eta'$ physics & Start in 2007 \\
\hline
  \end{tabular}
  \caption{List of the competitors for the {\small DAFNE}-2 project. For
    each experiment we indicate which measurement among those of the
    {\small DAFNE}-2 program can be done. Moreover we indicate the time scale 
    of the project. With the symbol
    (?) we indicate those measurements that in principle can be done by the
    experiment but have not yet been done. For discussions and comparisons
    we refer to the single paragraphs of Sect.2.}
  \protect\label{competitors}
\end{table}

The present {\small KLOE} detector with some upgrades appears well-suited
for the measurements discussed here. Among the upgrades we remark the
relevance of the inner tracker, of 
the small angle tagger for the $\gamma\gamma$ physics, and of
the proton polarimeter to access the phases of the nucleon form factors. 
For the latter
subject, the possibility to use the Finuda
detector should also be considered. 

We remark that
a large part of the {\small DAFNE}-2 program
is
based on the assumption that the center of mass energy will be increased up to
at least 2.5 GeV, and will not be possible in case {\small DAFNE}-2 will work at
the $\phi$ only. 
Moreover we stress that, since many of the measurements discussed here are
precision refinements of measurements already done, the project will be 
significant
only if it will provide ``ultimate'' and complete
measurements. This has to be taken into account in the design of the experiments.

We have also considered the physics potential of a higher energy $e^+e^-$
collider,  
a $\tau-$ charm factory with a luminosity of order $10^{34}cm^{-2}s^{-1}$.
It offers interesting possibilities for flavour physics. Among the
relevant 
topics we mention charmonium spectroscopy, the study of 
charmed mesons decays and
$\tau$ physics. It should be considered as an option in the framework of
the Super - Flavour factory.

\ack
We warmly acknowledge the colleagues who 
contributed to the analyses
presented here: A.Bianconi, S.Eidelman, R.Escribano,
F.Jegerlehner, B.Pasquini, E.Tomasi-Gustafsson, N.Tornqvist, L.Trentadue. 
We thank the Frascati Laboratories and the INFN units of Milano and Roma,
for hospitality during the preparation of the document.
We thank the
Director of the Frascati Laboratories M. Calvetti and the President of the
INFN CSN-1 F. Ferroni, who supported this work. Finally we
express our gratitude to L.Maiani for many
discussions especially on the
subject of low mass scalar spectroscopy, to I.Bigi for his suggestions 
concerning the $\tau$-charm physics, and to R.Baldini for his support
on all the topics discussed in this report.

\end{document}